\documentclass[a4paper,11pt,nohyper]{article}
\pdfoutput=1
\usepackage{jheppub}

\usepackage[T1]{fontenc}
\usepackage[mathletters]{ucs} 
\usepackage[utf8x]{inputenc}
\usepackage{multirow,booktabs,array}
\usepackage{amssymb}
\usepackage{amsmath}
\usepackage{lmodern}
\usepackage{braket}
\usepackage{verbatim}
\usepackage{multirow}
\usepackage{graphics,graphicx}
\usepackage{psfrag}
\usepackage{physics}

\usepackage{mathtools}

 \def\bd{\begin{document}} \def\ed{\end{document}}
\def\ds{\documentstyle} \let\fr=\frac \let\bl=\bigl \let\br=\bigr
\let\Br=\Bigr \let\Bl=\Bigl
\let\bm=\bibitem
\let\na=\nabla

\newcommand{\be}{\begin{equation}}
	\newcommand{\ee}{\end{equation}}

\newcommand{\bea}{\begin{eqnarray}}
	\newcommand{\eea}{\end{eqnarray}}

\usepackage{amsmath,latexsym,amssymb,slashed}

\title{Generalised proofs of the first law of entanglement entropy}
\author{Marika Taylor and Linus Too} 
\affiliation{School of Mathematical Sciences and STAG Research Centre, University of Southampton, Highfield, Southampton, SO17 1BJ, UK.}

\bigskip
\emailAdd{m.m.taylor@soton.ac.uk, l.h.y.too@soton.ac.uk}

\abstract{In this paper we develop generalised proofs of the holographic first law of entanglement entropy using holographic renormalisation. These proofs establish the holographic first law for non-normalizable variations of the bulk metric, hence relaxing the boundary conditions imposed on variations in earlier works. Boundary and counterterm contributions to conserved charges computed via covariant phase space analysis have been explored previously. Here we discuss in detail how counterterm contributions are treated in the covariant phase approach to proving the first law. Our methodology would be applicable to generalizing other holographic information analyses to wider classes of gravitational backgrounds. 
}

\preprint{}

\begin{document}
	\flushbottom
	\maketitle
	
\section{Introduction}

The first law of entanglement entropy states that the variation of the entanglement entropy $S_B$ is equal to the variation of the modular energy $\langle H_B \rangle$,
\begin{align}
	\delta S_B=\delta \expval{H_{B}}.\label{eq:first1st}
\end{align} 	
The holographic first law of entanglement entropy first demonstrated in \cite{Faulkner:2013ica} is applicable to field theories that admit a holographic description. In the analysis of \cite{Faulkner:2013ica} the righthandside of (\ref{eq:first1st}) is by construction finite, as it derives from the standard holographic renormalization expressions for energy \cite{deHaro2001}. The authors of \cite{Faulkner:2013ica} work with a regulated entanglement entropy and restrict variations such that the left hand side of (\ref{eq:first1st}) has no ultraviolet divergences. The goal of this paper is to demonstrate the holographic first law in general situations, without imposing restrictions on variations, making use of the consistent renormalization procedure for the entanglement entropy developed in \cite{Taylor:2016aoi}. Holographic renormalization of entanglement entropy has been discussed  in a number of other works, including \cite{Taylor:2017zzo,Anastasiou:2018mfk,2018aao2,Anastasiou:2019ldc,Anastasiou:2020smm}. 
	
Entanglement entropy in quantum field theory is divergent due to the correlations across the boundary of the entangling region. Holographically, the entanglement entropy is captured by the area of the Ryu-Takayangi surface \cite{Ryu:2006bv}, which is also divergent due to the infinite volume of the entangling surface in the bulk spacetime. In both situations the entropy can be systematically renormalized, inheriting its renormalization scheme from that for the partition function of the theory. The renormalized holographic entanglement entropy in \cite{Taylor:2016aoi} can be derived from the holographically renormalized action \cite{deHaro2001} using the replica trick. We will show it is necessary to use the renormalized entanglement entropy on the left hand side of $(\ref{eq:first1st})$ to obtain the correct finite contributions when one considers general linear perturbations. 
	

The covariant charge formalism can be used to give an elegant discussion of the holographic first law \cite{Faulkner:2013ica}. In the covariant formalism both sides of $(\ref{eq:first1st})$ can be expressed as integrals of charge densities over entangling surfaces. We will review this approach in the following section. The charge associated with the change in modular energy used in \cite{Faulkner:2013ica} was renormalized, following the earlier works of  \cite{Papadimitriou:2004ap, Papadimitriou:2005ii}. However, the charge associated with the change in entropy was not renormalized; its variation was finite in \cite{Faulkner:2013ica} due to constraints on the asymptotic falloff of metric perturbations. In this paper we construct a renormalized charge corresponding to the change in entropy such that the integral version of the holographic first law applies to generic metric perturbations. 

At a technical level, one can understand the construction of this charge as follows. Onshell, the density of the conserved charge is defined in terms of the current density as
\begin{align}
	\boldsymbol{J}=d\boldsymbol{Q}
\end{align}
where $\boldsymbol{J}$ and $\boldsymbol{Q}$ are differential forms. The charge density clearly has an intrinsic ambiguity: additional exact terms in $\boldsymbol{Q}$ will not change the current. In our context, the exact term ambiguity in the density of the conserved charge contributes to the entanglement entropy (and modular energy) at the boundary of the entangling surface. The holographic counterterms fix the ambiguity in the density of the conserved charges, with the matching of renormalization schemes for energy and entropy ensuring that the first law holds. Relative to the expressions given for the entropy in \cite{Faulkner:2013ica}, our expressions have additional boundary terms. Our general expressions are applicable to variations of the entanglement entropy associated with generic variations of the bulk metric i.e. perturbations of the non-normalizable terms in the metric. 

Boundary terms in the construction of charges using the covariant phase space formalism have been discussed recently in \cite{2020HarlowWu}. The boundary counterterms associated with holographic charges were constructed using Hamiltonian renormalisation methods in  \cite{Papadimitriou:2005ii}. 
There are key conceptual differences in the entropy variation that require us to generalize relative to both of these works. The vector used to construct the entropy variation is no longer Killing. In \cite{Papadimitriou:2005ii} the goal was to compute conserved charges for black holes and accordingly any variations considered would preserve the non-normalizable modes of the background. In our case the non-normalizable modes are not held fixed: the metric perturbations can be such that the non-normalizable modes vary, corresponding to deforming not just the state of the dual field theory, but the theory itself. This different physical setup leads to differences in the counterterms arising in the analysis of the covariant phase space construction, which are explained in detail in Appendix C. 


\bigskip
The structure of the paper is the following. In section $\ref{section:REE}$ we review the holographic renormalized entanglement entropy, introducing the notion of renormalized area integral for codimension two minimal surfaces in $AlAdS$ that allows us to express the renormalized entanglement entropy functional in terms of certain conformal invariants. In section $\ref{section:Ham}$ we summarise the covariant formalism or Hamiltonian formalism for holographic renormalization and conserved charges and explain in $\ref{section:1stLawIntro}$ the first law of holographic entanglement entropy, explaining the constraints imposed on variations in previous works.  In section \ref{section:InfRFL} we explore the infinitesimal version of the first law i.e. the radius of the entangling region is infinitesimal, for general variations, explaining the differences between odd and even dimensions. In odd dimensions the variation of the renormalized entropy can be expressed elegantly in terms of the pullback of the Weyl tensor variation. 

We demonstrate the integral version of the renormalized first law in section $\ref{section:IRFL}$. We first need to introduce in $\ref{section:CER}$  the proper definition of the conserved charges and their integrals: we demonstrate how the equivalence relations between conserved charges need to be generalized to include appropriate counterterms once one allows for generic variations. In section $\ref{section:VoC}$, we use these conserved charges and their equivalence relations to derive the renormalized first law of entanglement entropy. This general proof is illustrated using two examples in $d=3$, $4$ and $5$. We end the paper with discussion of implications and applications of our results. 

\section{Review of renormalized entanglement entropy, holographic charges and first law}

In this section we briefly review the definition of renormalized entanglement entropy and holographic charges, and describe the first law of entanglement entropy. 

\subsection{Renormalized Entanglement Entropy}
\label{section:REE}

One of the main goals of this work is to generalise first laws for holographic entanglement entropy, relaxing assumptions on boundary conditions for bulk metric perturbations. One of the tools that will be used in our analysis is renormalized entanglement entropy; this is relevant as general boundary conditions for bulk metric perturbations are associated with UV divergences in the regulated entanglement entropy. Working with quantities that are consistently renormalized allows us to work systematically with such setups. 

Renormalized entanglement entropy was discussed extensively in \cite{Taylor:2016aoi}, with explicit formulae for holographic renormalized entanglement entropy being derived. A convenient way to construct expressions for renormalized entanglement entropy is via the replica trick. 
Using the replica trick, entanglement entropy can be derived as the limit of R\'{e}nyi entropy	 
\begin{equation}
	S=-\alpha\partial_{\alpha}[\log\mathcal{Z}(\alpha)-\alpha\log\mathcal{Z}(1)]_{\alpha=1}
\end{equation}
where ${\mathcal{Z}}(\alpha)$ is the partition function on the $\alpha$ fold cover manifold. In much of the condensed matter literature this approach is applied to UV regulated quantities, with the UV regulator being interpreted in terms of the lattice scale of the discrete condensed matter system of interest. However, from a quantum field theory perspective, it is much more natural to work directly with renormalized quantities, ie.e. ${\mathcal{Z}}$ is the renormalized partition function. 

In holography the partition function is computed to leading order from the onshell bulk action i.e. 
\begin{equation}
	\mathcal{Z}_{grav}=e^{-I_{grav}},
\end{equation}
where $I_{grav}$ denotes the onshell gravitational action. Applying the holographic dictionary and the replica trick to the {\it renormalized} gravity action one obtains a formal definition of the renormalized holographic entanglement entropy. 
\begin{equation}
	S = \alpha\partial_{\alpha} [I_{ren}(\alpha)-\alpha I_{ren}(1)]_{\alpha=1} \label{eq:repEE}
\end{equation}
This approach thus directly relates the renormalization scheme for the partition function (gravitational action) to the scheme for the entanglement entropy. 

\bigskip

To obtain a finite value for the gravitational action, one needs to use holographic 
renormalization. The renormalized action can then be obtained by the procedure of regularization and the introduction of appropriate covariant boundary counterterms
\begin{equation}
	I_{ren}=I_{reg}- I_{ct}
\end{equation}
For pure gravity with negative cosmological constant, the renormalized action in $(d+1)$ dimensions takes the form \cite{deHaro2001}
\begin{align}
	I_{ren}=\frac{1}{16\pi G_{d+1}} &\int_{\mathcal{M}_{z\geq\epsilon}}d^{d+1}x\sqrt{g}(R + d (d-1) ) \\
	-\frac{1}{16\pi G_{d+1}}&\int_{\tilde{\mathcal{M}_{\epsilon}}} d^dx\sqrt{\tilde{g}}\big[ K +2(1-d)\mathcal{R}+\frac{1}{2-d}\mathcal{R}\nonumber\\
	&-\frac{1}{(d-4)(d-2)^2}(\mathcal{R}_{\mu\nu}\mathcal{R}^{\mu\nu}-\frac{d}{4(d-1)}\mathcal{R}^2) -\log\epsilon a_{(d)}+ \cdots\big] \nonumber
\end{align}
In these expressions the bulk manifold is regulated using a radial coordinate $z \geq \epsilon$; $R$ denotes the curvature of the bulk manifold while
$K$ and $\mathcal{R}$ refer to the extrinsic and intrinsic curvature of the boundary manifold respectively. 
Here the given counterterms suffice for $d \le 5$; expressions for the additional counterterms required for $d > 5$ can be found in \cite{deHaro2001}. Logarithmic counterterms associated with conformal anomalies arise for $d$ even, and explicit expressions for these can also be found in \cite{deHaro2001}. 

One can then derive the renormalized entanglement entropy from the renormalized action, making use of the following expressions for the integrals of curvature invariants, expressed as series in powers of $(1-\alpha)$ \cite{Fursaev_1995,Fursaev_2013}:
\begin{align}
	&\int_{\mathcal{M}_{\alpha}}d^{d+1}x\sqrt{g}\mathcal{R}_{\alpha}=\alpha\int_{\mathcal{M}}d^{d+1}x\sqrt{g}\mathcal{R}+4\pi(1-\alpha)\int_{\tilde{B}}d^{d-1}x\sqrt{\gamma}     \label{eq:repRint}    \\
	&\int_{\mathcal{M}_{\alpha}}d^{d+1}x\sqrt{g}\mathcal{R}_{\alpha}^2=\alpha\int_{\mathcal{M}}d^{d+1}x\sqrt{g}\mathcal{R}^2+8\pi(1-\alpha)\int_{\tilde{B}}d^{d-1}x\sqrt{\gamma}\mathcal{R}\nonumber\\
	&\int_{\mathcal{M}_{\alpha}}d^{d+1}x\sqrt{g}{\mathcal{R}_{\alpha}}_{\mu\nu}\mathcal{R}_{\alpha}^{\mu\nu}=\alpha\int_{\mathcal{M}}d^{d+1}x\sqrt{g}\mathcal{R}_{\mu\nu}\mathcal{R}^{\mu\nu}\nonumber\\
	&\;\;\;\;\;\;\;\;\;\;\;\;\;\;\;\;\;\;\;\;\;\;\;\;\;\;\;\;\;\;+4\pi(1-\alpha)\int_{\tilde{B}}d^{d-1}x\sqrt{\gamma}\big(\mathcal{R}_{\mu\nu}n^{\mu}\cdot n^{\nu}-\frac{1}{2}(TrK)^2\big)   \nonumber 
\end{align}

Using these replica curvature integrals the explicit expression for the holographic renormalized entanglement entropy becomes \cite{Taylor:2016aoi}
\begin{align}
	&S_{ren}= \frac{1}{4G_{d+1}}\int_{\tilde{B}}d^{d-1}x\sqrt{\gamma}-\frac{1}{4(d-2)G_{d+1}}\int_{\partial\tilde{B}}d^{d-2}x\sqrt{\tilde{\gamma}}  \label{eq:renhee}\\
	&-\frac{1}{4(d-2)(d-4)G_{d+1}}\int_{\partial\tilde{B}}d^{d-2}x\sqrt{\tilde{\gamma}}\bigg(\mathcal{R}_{\mu\nu}n^{\mu}\cdot n^{\nu}-\frac{1}{2}(Tr K)^2-\frac{d}{2(d-1)}\mathcal{R}\bigg) \nonumber
\end{align}
where $\mathcal{R}$ is the Ricci scalar of the metric $g_{\mu\nu}$, $\mathcal{R}_{aa}=\sum_{a}(-1)^a\mathcal{R}_{\mu\nu}n^{\mu}_an^{\nu}_a$ is the projection of the Ricci tensor on the subspace orthogonal to $\partial \tilde{B}$ with temporal and spatial normals $n_a^{\mu}$, $a=1,2$, $\widetilde\gamma$ is the determinant of the induced metric on $\partial\tilde{B}$ and $k^2=\sum_{a}K_aK_a$ with $K_a$ trace of the extrinsic curvature corresponding to the two normals $n_a^{\mu}$. Here $\tilde{B}$ denotes the entangling surface with boundary $\partial \tilde{B}$. The counterterms given here are sufficient for $d < 6$, but can straightforwardly be computed for $d \ge 6$. For $d$ even there are logarithmic counterterms related to conformal anomalies, see \cite{Taylor:2016aoi} for details. 
\\
\\
When the CFT dimension $d$ is odd, the renormalized entanglement entropy can be written in terms of the Euler characteristic and other renormalized curvature invariants of the bulk entangling surface \cite{Taylor:2020uwf},
\begin{align}
	S _{ren}(\tilde{B}) \sim(-1)^{n+1}  {\cal F}_{n} \; \chi (\tilde{B}) - \sum_{r} {\cal W}_{r} (\tilde{B}) - \sum_{p} {\cal H}_{p} (\tilde{B})- \sum_{q} {\cal I}_{q} (\tilde{B}). \label{structure}
\end{align}
where $\mathcal{W}_r$ are renormalized integral of projections of the Weyl curvature, $\mathcal{H}_p$ are renormalized integral of even powers of the extrinsic curvature and for $d>5$ there are ${\cal I}_{q}$ renormalized integrals containing products of Weyl and extrinsic curvature. More explicitly the renormalized entanglement entropy proportional to the renormalized area of the bulk entangling surface $\tilde{B}$
\begin{align}
		S _{ren}(\tilde{B}) =\frac{\mathcal{A}(\tilde{B})}{4G_{d+1}}
\end{align}
and renormalized area integral in $d=3$ is
\begin{align}
	{\cal A}  (\tilde{B}) = - 2 \pi \chi(\tilde{B}) - \frac{1}{2} \int_{\tilde{B}} d^2 x \sqrt{g} | {K} |^2 - \int_{\tilde{B}} d^2 x \sqrt{g} {W}_{1212} \label{simp1},
\end{align}
and in $d=5$ is
\begin{eqnarray}
	{\cal A}(\tilde{B}) &=& \frac{4 \pi^2}{3} \chi (\tilde{B}) - \frac{1}{6} {\cal H}(\tilde{B}) - \frac{1}{3} {\cal W}(\tilde{B}) \label{eq:A6} \\
	&& 
	- \frac{1}{24}  \int_{\tilde{B}} d^4 x \sqrt{g} \left ( H^2 - 4 H_{\mu \nu} H^{\mu \nu} +  H_{\mu\nu\rho\sigma}H^{\mu\nu\rho\sigma} \right .  \nonumber \\
	&& \left . +4{W}_{1212}^2
	-4{W}_{\mu n \nu n}\widetilde{W}^{\mu n \nu n}+{W}_{\mu \nu\rho\sigma}{W}^{\mu \nu\rho\sigma} \right ). \nonumber 
\end{eqnarray}
In what follows we will find these geometric expressions for renormalized entanglement entropy useful. Note in particular that these will simplify considerably in the context of first variations around AdS backgrounds.

\subsection{Hamiltonian Formalism and Charges in AdS}
\label{section:Ham}

In this section we review the description of Wald Hamiltonians \cite{Lee:1990nz,Wald:1993nt,Iyer:1994ys,Iyer:1995kg,Wald:1999wa} and charges in anti-de Sitter spacetimes. Our review follows closely the work of \cite{Papadimitriou:2004ap, Papadimitriou:2005ii}, and more details may be found in these references. The Wald approach assumes that the gravitational theory is described by a diffeomorphism covariant {Lagrangian d-form} 
${\bf{L}}(\psi)$, where ${\bf{L}}(\psi)$ will depend both on the metric and other fields, denoted collectively as $\psi$. In the anti-de Sitter context we work with a renormalized Lagrangian  
\begin{align}
	\boldsymbol{L}^{ren}=\boldsymbol{L}-{d}\boldsymbol{B}
\end{align}
where $\boldsymbol{L}$ is the bulk Lagrangian form and $\boldsymbol{B}$ may be viewed as the combination of the Gibbon-Hawking term and boundary counterterms. The onshell regular Lagrangian is exact i.e.
\begin{align}
	\boldsymbol{L}^{onshell}= - \frac{{d}(\boldsymbol{\varepsilon}_an^a\lambda)}{16\pi G_{d+1}}.
\end{align}
where $n^a$ is the outward normal in the asymptotic radial direction,
\begin{align}
	n=-\frac{dz}{z},
\end{align}
 $\boldsymbol{\varepsilon}$ is the volume form and $\boldsymbol{\varepsilon}_{a_1\dots a_n}$ is a $(d-n+1)$ form
\begin{align}
	\boldsymbol{\varepsilon}_{a_1\dots a_n}=\frac{1}{(d-n+1)!}\varepsilon_{a_1\dots a_n b_1\dots b_{d-n+1}}dx^{b_1}\wedge\cdots\wedge dx^{b_{d-n+1}},
\end{align}
with the orientation
\begin{align}
	\varepsilon_{ztx^1\cdots}=+\sqrt{-g}.
\end{align}
In the Hamiltonian formalism, fields can be expanded asymptotically near the conformal boundary in series of dilatation eigenfunctions with ascending weight, see appendix \ref{section:DEE} for more detailed explanation. The general structure of the boundary term is then
{\begin{align}
		\boldsymbol{B}&=\boldsymbol{B}^{GH}-\boldsymbol{B}^{ct}\nonumber\\
	&= - \frac{\boldsymbol{\varepsilon}_an^a}{16\pi G_{d+1}}\left(K-(K_{ct}-\lambda_{ct})\right)\nonumber\\
	&=\frac{-\boldsymbol{\varepsilon}_an^a}{16\pi G_{d+1}}\left(K_{(d)}+\lambda_{ct}\right).\label{eq:Bform}
\end{align}}
where typically counterterms contribute up to terms at the $d^{th}$ order i.e. 
\begin{align}
	O_{ct}\sim\sum_{i<d}O_(i).
\end{align}
Variations can then be expressed as 
\begin{align}
	\delta\boldsymbol{L}^{ren}&=\delta \boldsymbol{L} -d\delta\boldsymbol{B}\\
	&=\boldsymbol{E}^{\psi} \delta \psi+d\boldsymbol{\Theta}[\delta\psi]-d\delta\boldsymbol{B}
\end{align}	
where $\boldsymbol{E}^{\psi} $ denotes the equations of motion and $\boldsymbol{\Theta}$ is the symplectic potential. This expression can be rewritten as 
	\begin{align}
	\delta \boldsymbol{L}^{ren}=\boldsymbol{E}^{\psi} \delta \psi+d\boldsymbol{\Theta}^{ren}[\delta\psi]
\end{align}
where the renormalized symplectic potential form can be expressed as
\begin{align}
	\boldsymbol{\Theta}^{ren}[\delta\psi]=	\boldsymbol{\varepsilon}_an^a\pi_{(d)}^{\mu\nu}\delta\gamma_{\mu\nu} \label{SymPot}
\end{align}
 The canonical momentum $\pi_{\mu \nu}$ can be expressed in terms of the extrinsic curvature as 
\begin{align}
	\pi^{\mu\nu}= - \frac{1}{16\pi G_{d+1}}\sqrt{\gamma}\left(K^{\mu\nu}-K\gamma^{\mu\nu}\right).\label{eq:ConMom}
\end{align}
If we expand both sides of the equality in the dilatation eigenfunction expansion, we can match the dilatation weights and obtain,
\begin{align}
	\pi_{(n)}^{\mu\nu}=- \frac{1}{16\pi G_{d+1}}\left(K^{\mu\nu}_{(n)}-K_{(n)}\gamma^{\mu\nu}\right)\label{eq:pin}.
\end{align}
In \eqref{SymPot} $\pi_{(d)}^{\mu\nu}$ is the $d^{th}$-term in the dilatation eigenfunction series of the conjugate momentum with respect to the metric and this is in turn related to the renormalized CFT stress tensor as
\begin{align}
	2\pi_{(d)}^{\mu\nu}=- \frac{1}{16\pi G_{d+1}}T^{\mu\nu}_{ren},\label{eq:pidTren}
\end{align}
i.e. the first variation of the renormalized action is 
\begin{align}
	\delta I^{onshell}_{ren}=\frac{-1}{32\pi G_{d+1}}\int_{\partial\mathcal{M}}d^dx\sqrt{-\gamma}T^{\mu\nu}_{ren}\delta\gamma_{\mu\nu}.  \label{eq:dITdg}
\end{align}
Now let us consider the asymptotic behaviour of metric perturbations. Expressing the AdS$_{d+1}$ metric as 
\begin{align}
ds^2 = \frac{dz^2}{z^2} + \frac{1}{z^2} \eta_{\mu \nu} dx^{\mu \nu}
\end{align}
where $\eta$ is the Minkowski metric, only the normalizable mode is allowed to vary under a Dirichlet condition and  
\begin{align}
	\delta\gamma_{\mu\nu}=z^{d-2}\delta\gamma_{(d)\;\mu\nu}+O(z^{d-1}),\quad\quad z\rightarrow0.
\end{align}
For $d$ odd, using the tracelessness of the stress tensor and absence of trace anomaly, the Dirichlet boundary condition can be automatically generalized to a conformal Dirichlet boundary condition which fixes the conformal class only. In the present of  a conformal anomaly, for the onshell action to be stationary under perturbations a representative of the conformal class has to be fixed.

\bigskip

Now let us turn to Noether charges. If the field variation is induced by a vector field $\xi$, we can define the Noether current form as
\begin{align}
	\boldsymbol{J}[\xi]=\boldsymbol{\Theta}[\delta_{\xi}\psi]-\iota_{\xi}\boldsymbol{L}\label{eq:JTL}
\end{align}
where $\iota_{\xi}$ contracts $\xi$ with the first index of $\boldsymbol{L}$. The exterior derivative of the Noether current is proportional to the equation of motion
\begin{align}
	d\boldsymbol{J}[\xi]=-\boldsymbol{E}^{\psi}\delta_{\xi}\psi,
\end{align}
and thus vanishes onshell.  Hence we can define the Noether charge form $\boldsymbol{Q}[\xi]$ as the exact term in the Noether current
\begin{align}
	\boldsymbol{J}[\xi]=d\boldsymbol{Q}[\xi]-\boldsymbol{N}[\xi]\label{eq:dQ}
\end{align}
where
\begin{align}
	d\boldsymbol{N}[\xi]=\boldsymbol{E}^{\psi}\delta_{\xi}\psi.
\end{align}
There is another conserved charge in $AlAdS$ induced by $\xi$ called the holographic charge. 
{Using $(\ref{eq:pidTren})$ and the fact that the renormalized CFT stress tensor is conserved, we can construct the total relativistic momentum of the boundary system.} The holographic charge form $\boldsymbol{\mathcal{Q}}[\xi]$ is defined by
\begin{align}
	\boldsymbol{\mathcal{Q}}[\xi]=-\boldsymbol{\varepsilon}_{ab}n^a2\pi_{(d)}^{bc}\xi_c.
\end{align}
{and this form is integrated over a timeslice at the boundary to obtain the holographic charge. This can be interpreted as the renormalized relativistic momentum along the $\xi$ direction.} 

In Lemma 4.1 in \cite{Papadimitriou:2005ii} it was proved that for any asymptotically locally anti-de Sitter space ${\mathcal{M}}$ 
the two definitions of charges corresponding to asymptotic conformal Killing vector, $\xi$, on a spatial slice on the conformal boundary, $\partial\mathcal{M}\cap C$, are equivalent i.e.
\begin{align}
-	\int_{\partial\mathcal{M}\cap C} \boldsymbol{\mathcal{Q}}[\xi]=\int_{\partial\mathcal{M}\cap C} \textbf{Q}^{full}[\xi]. \label{eq:Lemma4.1}
\end{align}
where
\begin{align}
\textbf{Q}^{full}[\xi]=\textbf{Q}[\xi]-\iota_{\xi}\textbf{B}.
\end{align}
Note that this equivalence is defined up to exact terms since $\partial\mathcal{M}\cap C$ is a cycle and the asymptotic conformal Killing vector $\xi$ has the following fall off condition:
\begin{align}
	\xi^z=O(z^d), && \xi^{\mu}=\zeta^{\mu}(1+O(z^{d+2})) \label{eq:falloff}
\end{align}
where $\zeta$ is a boundary conformal Killing vector. We will later need to generalise this equivalence to less restrictive fall off conditions on the vector field. 

\subsection{Holographic First Law of Entanglement Entropy}
\label{section:1stLawIntro}

In this section we briefly review the first law of entanglement entropy.  Given a reduced density matrix $\rho_B$ the modular Hamiltonian $H_B$ is given by
\begin{equation}
	\rho_B=e^{-H_B}
\end{equation}
Under a small variation of the entanglement entropy, we obtain the relation
\begin{align}
	\delta S_B=\delta\expval{H_B}
\end{align}
and the equivalence between $\delta\expval{H_B}$ and the change in energy $\delta E$ gives the first law of entanglement entropy.

Following \cite{Casini:2011kv}, we now review relevant properties of the modular Hamiltonian and modular flow for CFTs on Minkowski space. There is a symmetry group associated with the modular Hamiltonian: the modular group, a group of one-parameter transformations of the form
\begin{align}
	U_B(s)=e^{-isH_B}
\end{align}
where $\partial_s$ is called the modular flow. For QFTs on Minkowski space, the modular flow generates a boost. In null coordinates $X^{\pm}$ this is given by 
\begin{align}
	X^{\pm}(s)=X^{\pm}e^{\pm2\pi s}.\label{eq:modflowR}
\end{align}
For an accelerated observer in Rindler coordinates, the state is thermal in $\tau$ where the longitudinal part of the metric is given by 
\begin{align}
	dX^+dX^-=-\frac{\rho^2}{R^2}d\tau^2+d\rho^2,
\end{align}
where $R$ relates to the imaginary time periodicity i.e. $\beta = 2 \pi R$. The thermal density matrix of the state is
\begin{align}
	\rho_{\mathcal{R}}=\frac{e^{\beta H_{\tau}}}{Tr(e^{\beta H_{\tau}})}.
\end{align}
The modular flow generator is $2\pi R\partial_\tau$ and the modular Hamiltonian is given by $2\pi RH_{\tau}+\log Tr(e^{\beta H_{\tau}})$. 

\bigskip

For a spatial ball $B$ of radius $R$ centred at $x^i=0, \,t=0$ on $d$-dimensional Minkowski space, we can conformally map the causal development of the spatial ball $D(B)$ to the Rindler wedge. This conformal map $X^{\mu}\rightarrow x^{\mu}$ can also map the modular flow $(\ref{eq:modflowR})$ to 
\begin{align}
	x^{\pm}(s)=R\frac{(R+x^{\pm})-e^{\mp 2\pi s}(R-x^{\pm})}{(R+x^{\pm})+e^{\mp 2\pi s}(R-x^{\pm})}
\end{align}
and hence the modular flow generator $\partial_s$ as $\zeta_B$
\begin{align}
	\partial_s&=\frac{\pi}{R}\left((R^2-t^2-\vec{x}^2)\partial_t-2tx^i\partial_i\right)\\
	\zeta_B&=\frac{i\pi}{R}(R^2P_t+K_t)
\end{align}
where $P_t$ and $K_t$ are the time translation and special conformal transformation generators, respectively.  

Since the modular Hamiltonian is the translation operator in $s$, on $B$ we get
\begin{align}
	H_B = \int_B d^{d-1}xT^{ts}. \label{eq:HBs}
\end{align}
We can write $(\ref{eq:HBs})$ in covariant form as
\begin{align}
	H_B=\int_{B} d\sigma^{\mu}T_{\mu\nu}\zeta_B^{\nu}
\end{align}
and the modular energy as
\begin{align}
	E_B=\int_{B}d\sigma^{\mu}\expval{T_{\mu\nu}}\zeta_B^{\nu}.
\end{align}
The entanglement entropy of region $B$ can be calculated holographically by the area of the corresponding bulk entangling surface $\tilde{B}$.

A CFT in the vacuum state on the causal wedge $D(B)$ can also be mapped conformally to a CFT in a thermal state on the hyperbolic cylinder. This can be easily seen from writing the Rindler metric as: 
\begin{align}
	ds^2=\frac{\rho^2}{R^2}\left(-d\tau^2+\frac{R^2}{\rho^2}(d\rho^2+dX^idX^i)\right) =\frac{\rho^2}{R^2}ds^2_{\mathbb{R}\times \mathbb{H}^{d-1} }.
\end{align}
As discussed in \cite{Faulkner:2013ica}, the first law of entanglement entropy of the CFT thermal state on hyperbolic cylinder can be related to the first law of black hole dynamics via holography. Essentially, the CFT on hyperbolic cylinder $\mathbb{R}\times \mathbb{H}^{d-1}$ is dual to the Rindler $AdS_{d+1}$ black hole exterior and the bulk entangling surface $\tilde{B}$ can be viewed as the black hole horizon. The perturbation of entanglement entropy $\delta S_B$ is equal to the perturbation of black hole entropy calculated from the Wald functional $\delta S_{Wald}$ where
\begin{align}
	S_{Wald}=-2\pi \int_{\mathcal{H}}d\sigma \frac{\delta \mathcal{L}}{\delta R^{ab}_{\;\;cd}}n^{ab}n_{cd}.\label{eq:Swald}
\end{align}
Changing back to the interpretation in terms of a Minkowski boundary, we label $\Sigma$ as the bulk region enclosed by $B$ and $\tilde{B}$ and the bulk causal wedge of $\Sigma$ as $D(\Sigma)$. We extend the boundary modular flow to the bulk as the Killing vector
\begin{align}
	\xi_B=-\frac{2\pi}{R}(t-t_0)[z\partial_z+(x^i-x^i_0)\partial_i]+\frac{\pi}{R}[R^2-z^2-(t-t_0)^2-(\vec{x}-\vec{x}_0)^2]\partial_t \label{eq:xiB}.
\end{align}
Note that this Killing vector does not satisfy $(\ref{eq:falloff})$ but instead has the weaker fall-off behaviour
\begin{align}
	\xi^z_B=O(z), && \xi^{\mu}_B=\zeta^{\mu}_B(1+O(z^{2})). \label{eq:newfalloff}
\end{align}
One can check $\xi_B$ vanishes on $\tilde{B}$.

It was shown in \cite{Faulkner:2013ica} that for metric perturbations limited to normalizable modes, $\delta g_{\mu\nu}=z^{d-2}h^{(d)}_{\mu\nu}$, one get the infinitesimal first law of entanglement entropy as{
\begin{align}
	\delta \expval{T_{tt}\ }&=\frac{d^2-1}{2\pi \Omega_{d-2}}\lim_{R\rightarrow 0}\left(\frac{1}{R^d}\delta S_B\right)\nonumber\\
	\delta \expval{T_{tt}}&=\frac{d}{16\pi G_{d+1}}h^{(d)}_{tt}\\
\delta\expval{T_{\mu\nu}}&=\frac{d}{16\pi G_{d+1}}h^{(d)}_{\mu\nu} \nonumber
\end{align}
where the tracelessness condition of $h^{(d)}_{\mu\nu}$ is used to go from the second line to the final covariant expression. The final expression }matches the holographic dictionary between stress tensor and normalizable metric coefficient found in \cite{deHaro2001}. 

The covariant first law of entanglement entropy utilises the charges associated with with energy and entropy corresponding to the bulk Killing vector $\xi_B$ introduced in section \ref{section:Ham}. The entanglement entropy is 
\begin{align}
	S_B^{grav}=\int_{\tilde{B}} \textbf{Q}^{full}[\xi_B]\label{eq:EntropyInt}
\end{align}
and the modular energy is 
\begin{align}
	E_B^{grav}= - \int_B \boldsymbol{\mathcal{Q}}[\xi_B]\label{eq:ModEInt}.
\end{align}
Limiting to the variation involving only the normalizable mode with no boundary variation on $\partial B=\partial \tilde{B}$ and using $(\ref{eq:Lemma4.1})$,
\begin{align}
	-\int_B \delta\boldsymbol{\mathcal{Q}}[\xi_B] = \int_{B} \delta\textbf{Q}^{full}[\xi_B].
\end{align}
The off-shell difference is expressed in terms of the Einstein equations
\begin{align}
	\delta E_B^{grav}-\delta S_B^{grav}&=\int_{{B}}\delta\boldsymbol{Q}^{full}[\xi_{B}]-\int_{\tilde{B}}\delta\boldsymbol{Q}^{full}[\xi_{B}]\\
	&=\int_{\Sigma}d \delta\textbf{Q}^{full}[\xi_B]  =\int_{\Sigma} \delta\textbf{J}^{full}[\xi_B]   =\int_{\Sigma} -2\boldsymbol{\varepsilon}^a\delta E_{ab}\xi^a_B,  \nonumber
\end{align}
hence recovering the first law of entanglement entropy onshell. We also obtain the version of $(\ref{eq:Lemma4.1})$
\begin{align}
-\int_B\delta\boldsymbol{\mathcal{Q}}[\xi_B] =	\int_{\tilde{B}}\delta\boldsymbol{Q}^{full}[\xi_{B}]\label{eq:dLemma4.1}.
\end{align}
Note there are many caveats regarding boundary terms and fall-off condition when we allow the variation of the non-normalizable modes. We shall address them in the following sections.

\section{Infinitesimal Renormalized First Law}\label{section:InfRFL}
	In this section we will discuss the renormalized version of the first law of entanglement entropy in the infinitesimal limit, for $\mathcal{M}=AdS_{d+1}$ with spherical boundary entangling surfaces $\partial\tilde{B}=S^{d-2}$ in $d \le 6$. We begin by collecting together expressions for the renormalized entanglement entropy of such spherical regions. We derive the infinitesimal renormalized first law of entanglement entropy in $AlAdS_{d+1}$ for odd $d$ and explain its connection with the variation of the renormalized integral of a curvature invariant. Since the renormalized entanglement entropy in even dimensions is scheme dependent, we postpone the proof of the generalized first law in even $d$ to section $\ref{section:AlAdS5ex}$ to avoid repetitions.

\subsection{Spherical Entangling Region in AdS}
\label{section:SERAdS}
The metric of $AdS_{d+1}$ on the Poincare patch may be parameterized as
\begin{equation}
	ds^2=\frac{dz^2}{z^2}+\frac{1}{z^2}g_{\mu\nu}dx^{\mu}dx^{\nu},
\end{equation}
where $g_{\mu\nu}=\eta_{\mu\nu}$ is flat Minkowski metric with signature ($-$,$+$,$\cdots$,$+$). In the case of spherical entangling regions, the $(d-1)$-dimensional bulk extending entangling surface $\tilde{B}$ with boundary $\partial\tilde{B}$ as the entangling surface of the boundary CFT can be described by 
\begin{equation}
	r^2+z^2=R^2
\end{equation}
where $r$ is the radial coordinate on the boundary and $R$ is the radius of the spherical entangling region. The induced metric on the entangling surface in $AdS_{d+1}$ is then
\begin{equation}
	ds^2=\frac{R^2}{z^2r^2}dz^2+\frac{r^2 }{z^2}d\Omega_{d-2}^2. 
\end{equation}
where $d\Omega_{d-2}^2$ is the standard unit sphere metric. Another convenient choice of coordinates $(w,\,u)$ are defined by 
\begin{align}
	r=w\cos u, && z=w\sin u \label{eq:wcoord}
\end{align}
so the $AdS_{d+1}$ metric can be written as
\begin{align}
	ds^2=\frac{1}{w^2\sin^2u}(dw^2-dt^2)+\frac{1}{\sin^2u}dz^2+\frac{\cos^2u }{\sin^2u}d\Omega_{d-2}^2
\end{align}
and the induced metric on $\tilde{B}$ becomes
\begin{align}
	ds^2=\frac{1}{\sin^2u}dz^2+\frac{\cos^2u }{\sin^2u}d\Omega_{d-2}^2.
\end{align}
The regularised bulk contribution to the entanglement entropy for such an entangling surface is then
\begin{align}
	S^{reg}_B&=\frac{1}{4G_{d+1}}\int_{\tilde{B}_{\epsilon}}d^{d-1}x\sqrt{\gamma} \nonumber \\
	&=\frac{\Omega_{d-2}}{4G_{d+1}}\int_{\epsilon}^{R}dz\, \frac{Rr^{d-3}}{z^{d-1}}\label{eq:IntzSreg}\\
	&=\frac{\Omega_{d-2}}{4G_{d+1}}\int_{\xi}^{\frac{\pi}{2}}du\, \frac{\cos^{d-2}u}{\sin^{d-1}u}\label{eq:SIntu}
\end{align} 
where $\Omega_{d-2}$ is the area of ($d-2$)-dimensional unit sphere and $R\sin\xi \equiv \epsilon$. 

The divergent contributions are of the form $\epsilon^{-n}$ except in even $d$ where there are extra logarithmic terms. Focussing first on odd $d$, from $(\ref{eq:renhee})$ we know the counterterms for $d<6$ are
\begin{align}
	S^{ct}_B&=\frac{1}{4(d-2)G_{d+1}}\int_{\partial\tilde{B}_{\epsilon}}d^{d-2}x \sqrt{\widetilde\gamma}\left[1+\frac{1}{(d-2)(d-4)}(\mathcal{R}_{aa}-\frac{1}{2}k^2-\frac{d}{2(d-1)}\mathcal{R})\right]\label{eq:IntSct}\\
	&=\frac{\Omega_{d-2}}{4(d-2)G_{d+1}}\frac{r^{d-2}}{\epsilon^{d-2}}\left[1-\frac{(d-2)\epsilon^2}{2(d-4)r^2}\right]. \nonumber
\end{align}
Note that the intrinsic curvature terms do not contribute here since the boundary metric is flat, but they will contribute to the variation of the entanglement entropy under metric perturbations later. For odd $d$, using the definition of the entangling surface one obtains counterterm contributions \begin{align}	
	S^{ct}_B
	=\frac{\Omega_{d-2}}{4G_{d+1}}\left[\frac{R^{d-2}}{(d-2)\epsilon^{d-2}}-\frac{R^{d-4}}{(d-4)\epsilon^{d-4}}+\cdots\right].\label{eq:Sctoddd}
\end{align}
Combing $(\ref{eq:Sctoddd})$ with $(\ref{eq:IntzSreg})$ we get
\begin{align}
	&d=3:\quad\quad\quad\quad S_{ren}=-\frac{\pi}{2G_4}\\
	&d=5:\quad\quad\quad\quad S_{ren}=\frac{\pi^2}{3 G_6}. \nonumber
\end{align}
For $d=4$, the regularized entanglement entropy from $(\ref{eq:IntzSreg})$ is 
\begin{align}
	S^{reg}_B
	=\frac{\Omega_{2}}{4G_5}\left[-\frac{\ln R}{2}+\frac{\ln(R^2)}{2}+\frac{R(R^2-\epsilon^2)^{\frac{1}{2}}}{2\epsilon^2}+\frac{\ln \epsilon}{2}-\frac{\ln(R^2+R(R^2-\epsilon^2)^{\frac{1}{2}})}{2}\right]\label{eq:Srefd4}
\end{align}
and the corresponding the full set of counterterms, including the logarithmic counterterm, gives
\begin{align}
	S^{ct}_B&=\frac{1}{8G_5}\int_{\partial\tilde{B}}d^2x \sqrt{\widetilde\gamma}-\frac{\ln \epsilon}{16G_5}\int_{\partial\tilde{B}} d^2x \sqrt{\widetilde\gamma}(\mathcal{R}_{aa}-\frac{1}{2}k^2-\frac{2}{3}\mathcal{R})\nonumber\\
	&=\frac{\Omega_{2}}{8G_5}\left[\frac{r^2}{\epsilon^2}+ {\ln\epsilon}\right].
\end{align}
Combining these we obtain the renormalized entanglement entropy in $d=4$
\begin{align}
	S^{ren}_B=\frac{\Omega_{2}}{4G_5}\left[-\frac{\ln 2R}{2}+\frac{1}{4R}\right].
\end{align}
Since the action in this case has logarithmic counterterms there is an intrinsic scheme dependence in the renormalised entanglement entropy, which is completely determined by the scheme chosen for the renormalization of the action.

\subsection{First Law and Variation of Modular Energy}\label{section:1stlawVME}

We now consider the variation of the entanglement entropy under a linear perturbation of the bulk metric. We will express the perturbed metric in radial gauge so that 
\begin{equation}
	ds^2=\frac{dz^2}{z^2}+\frac{1}{z^2} (\eta_{\mu\nu} + h_{\mu \nu}) dx^{\mu}dx^{\nu},
\end{equation}
A general perturbation $h_{\mu \nu}$ can be expanded near the conformal boundary as 
\begin{equation}
h_{\mu \nu} = h^{(0)}_{\mu \nu} + z^2 h^{(2)}_{\mu \nu} + \cdots + z^d h^{(d)}_{\mu \nu} + z^{d} \log z \tilde{h}^{(d)}_{\mu \nu} + \cdots
\end{equation}
where the logarithmic terms arise in even $d$ and all coefficients in the expansion can be expressed in terms of the pair of data $(h^{(0)}_{\mu \nu},h^{(d)}_{\mu \nu})$ using the Einstein equations. 

The goal of this section is to show that the change in the renormalized entropy $\delta S_{B}^{ren}$ under such metric perturbations is equal to the change in modular energy i.e. 
\begin{align}
	\delta E_B=\delta S_B^{ren}.
\end{align}
In the previous literature \cite{Faulkner:2013ica}, the first law was derived by restricting the variation of metric to only normalizable modes i.e. imposing $h_{\mu\nu}^{(0)} = 0$ with $h_{\mu \nu}^{(d)} \neq 0$. Accordingly, the change in the entanglement entropy $\delta S_B$ is finite even without including the counterterms. Here we will derive the first law for general perturbations for which $h^{(0)}_{\mu \nu}$ is not 
necessarily zero; from a QFT perspective a general bulk metric perturbation corresponds to changing the background for the dual QFT as 
well as the state in the theory.

We will first demonstrate the renormalized first law in the infinitesimal limit where the radius of the boundary entangling region $B$ tends to zero $R\rightarrow 0$. The modular energy may be approximated by
\begin{align}
	\delta E_B&=\int_{B} d\sigma^{\mu }\delta T_{\mu\nu}^{ren}\xi_B^{\nu} \\
	\delta E_B&\stackrel{\scalebox{0.5}{$R\rightarrow 0$}}{\scalebox{1.3}{$\longrightarrow$}}\frac{2\pi R^d\Omega_{d-2}}{d^2-1}\delta T_{tt}^{ren}. \nonumber
\end{align}
From holographic renormalization \cite{deHaro2001}, the variation of the renormalized energy momentum tensor for odd $d$ is
\begin{align}
	\delta T_{\mu\nu}^{ren}=\frac{d}{16\pi G_{d+1}}h^{(d)}_{\mu\nu}.
\end{align}
In even dimensions the relation between the renormalized stress tensor and the coefficients of the asymptotic expansion is more complicated, capturing the conformal anomalies. For example, in $d=4$ 
\begin{align}
	\delta T_{\mu\nu}^{ren}=\frac{1}{16\pi G_{5}}\left(4h^{(4)}_{\mu\nu}+6\tilde{h}^{(4)}_{\mu\nu}\right),\label{eq:dTrend4}
\end{align}
i.e. there is an additional contribution associated with the coefficient of the logarithmic term $\tilde{h}^{(4)}$. At linearized order we can express $\tilde{h}^{(4)}$ in terms of the curvature $R^{(0)}$ of the perturbation of the QFT metric $h^{(0)}$ as
\begin{equation}
\tilde{h}^{(4)}_{\mu \nu} = - \frac{1}{48} \partial_{\mu} \partial_{\nu} R^{(0)} + \frac{1}{16} \partial^{\rho} \partial_{\rho} R^{(0)}_{\mu \nu} 
- \frac{1}{96} (\partial^{\rho} \partial_{\rho} R^{(0)}) \eta_{\mu \nu}.
\end{equation}
%
The infinitesimal first law of entanglement entropy for general variation is thus equivalent to showing that the variation of renormalized entanglement entropy can be expressed in terms of the renormalized stress tensor as 
\begin{align}
	\delta S^{ren}_B = \frac{2\pi R^d\Omega_{d-2}}{d^2-1}\delta T_{tt}^{ren}
	\label{eq:hddSren}
\end{align}

\subsection{Infinitesimal First Law for odd $d$}
\label{section:I1stod}
We shall focus on odd $d$. The linearized variation of regularized entanglement entropy can be expressed in Cartesian spatial coordinates as
\begin{align}
	\delta S^{reg}_B=\frac{R}{8G_{d+1}}\int_{\tilde{B}_{\epsilon}}d^{d-1}x\frac{1}{z^d}(h_{ii}-\hat{x}^i\hat{x}^jh_{ij}),
\end{align}
where $i$ runs over the spatial indices of the $d$-dimensional Minkowski space. 

To obtain the infinitesimal version of the first law, we consider the limit $R\rightarrow 0$. To evaluate the integrals explicitly it is more convenient to use the $(w,\,u)$ coordinates in $(\ref{eq:wcoord})$, in terms of which the variation of regularized entanglement entropy is
\begin{equation}
	\delta S^{reg}_B=\frac{1}{8G_{d+1}}\int_{\xi}^{\pi/2}du\int_{S^{d-2}}d\Omega_{d-2} \frac{\cos^{d-2}{u}}{\sin^{d-1}{u}}(\delta^{ij}-\cos^2{u}\hat{x}^i\hat{x}^j)h_{ij}.\label{eq:IntSregwcoord}
\end{equation}
For the variation of the counterterms we need the linearized variation of the spatial extrinsic curvature, which can be expressed as 
\begin{align}
	\delta K_2=-\frac{(d-2)z}{2r}\hat{x}^i\hat{x}^jh_{ij}+\frac{z}{2}\partial_r(h_{ii}-\hat{x}^i\hat{x}^jh_{ij})
\end{align}
and the variation of a specific combination of Ricci tensors, 
\begin{align}
	-\delta\mathcal{R}_{tt}+\delta\mathcal{R}_{rr}-\frac{d}{2(d-1)}\delta\mathcal{R}=(d-2)\left(h^{(2)}_{ii}-\hat{x}^i\hat{x}^jh^{(2)}_{ij}\right).
\end{align}
The latter equality holds at linearized level, see equation (\ref{fg2}) below. 

Substituting the above expressions into the variation of $(\ref{eq:IntSct})$ we get the following expression for the counterterms in general $d\leq 6$ to first order:
\begin{align}
	\delta S^{ct}_B&=\frac{1}{4(d-2)G_{d+1}}\int_{S^{d-2}}d\Omega_{d-2}\bigg[\frac{1}{2}\frac{r^{d-2}}{\epsilon^{d-2}}\Big(h_{ii}-\hat{x}^i\hat{x}^jh_{ij}\Big) -\frac{(d-2)}{4(d-4)}\frac{r^{d-4}}{\epsilon^{d-4}}\Big(h_{ii}-3\hat{x}^i\hat{x}^jh_{ij}\Big)\nonumber\\
	&+\frac{1}{(d-4)}\frac{r^{d-2}}{\epsilon^{d-4}}\Big(h^{(2)}_{ii}-\hat{x}^i\hat{x}^jh^{(2)}_{ij}\Big) -\frac{1}{2(d-4)}\frac{r^{d-3}}{\epsilon^{d-4}}\Big(\hat{x}^k\partial_kh_{ii}-\hat{x}^i\hat{x}^j\hat{x}^k\partial_kh_{ij}\Big)\bigg].\label{eq:ExdSCt} 
\end{align} 
In the $(w,u)$ coordinate system, the area integral for the bulk entangling surface $\tilde{B}_{\epsilon}$ is expressed in terms of an integral over the asymptotic angular coordinate $u$ and the spatial angular coordinates. We can thus evaluate the integral up to the upper limit $u=\frac{\pi}{2}$ even when expanding around $R=0$. 

The Taylor expansion around $x^i=0$ at each order of the Fefferman-Graham expansion can be written as: 
\begin{align}
	h^{(n)}_{\mu\nu}(x^i)=&\;h^{(n)}_{\mu\nu}(0)+R\hat{x}^i\partial_ih^{(n)}_{\mu\nu}(0)+\frac{R^2\hat{x}^i\hat{x}^j}{2!}\partial_i\partial_jh^{(n)}_{\mu\nu}(0)+\frac{R^3\hat{x}^i\hat{x}^j\hat{x}^k}{3!}\partial_i\partial_j\partial_kh^{(n)}_{\mu\nu}(0)\nonumber\\
	&+\frac{R^4\hat{x}^i\hat{x}^j\hat{x}^k\hat{x}^l}{4!}\partial_i\partial_j\partial_k\partial_lh^{(n)}_{\mu\nu}(0)+\dotsc\label{eq:xi0exp}
\end{align}
The Fefferman-Graham expansion also becomes an expansion in $R$,
\begin{align}
	h_{\mu\nu}=h_{\mu\nu}^{(0)}+R^2\sin^2{u} h_{\mu\nu}^{(2)}+\cdots\label{eq:FGu}
\end{align}
We can now expand $(\ref{eq:IntSregwcoord})$ using $(\ref{eq:FGu})$ and $(\ref{eq:xi0exp})$ up to $R^d$. The two angular integrals $du,\,d\Omega_{d-2}$ can be evaluated independently for each term in the expansion. In the appendix we give a general formula $(\ref{eq:OmegaxiInt})$ for integrating over products of unit vectors over $S^n$. Together with $(\ref{eq:deltaspartialsh})$, generic terms in the expansion after the spatial angular $d\Omega_{d-2}$ integral take the form
\begin{align}
	(\partial^2)^mh^{(n)}_{ii},\quad\quad\quad\quad (\partial^2)^m\partial_i\partial_jh^{(n)}_{ij}.
\end{align}
All the non-normalizable modes are related to the first term in the Fefferman-Graham expansion $h^{(0)}$ through the Einstein equation \cite{deHaro2001}. For $h^{(2)}$ we have
\begin{align}
h^{(2)}_{\mu\nu}=-\frac{1}{d-2}\Big(\delta\mathcal{R}_{\mu\nu}-\frac{1}{2(d-1)}\delta\mathcal{R}\eta_{\mu\nu}\Big) \label{fg2}
\end{align}
The linear variation of the Ricci tensor is,
\begin{equation}
	\delta\mathcal{R}_{\mu\nu}=-\frac{1}{2}\partial_{\mu}\partial_{\nu}{h^{(0)}}^{\sigma}_{\sigma}+\partial_{\sigma}\partial_{(\mu}{h^{(0)}}^{\sigma}_{\nu)}-\frac{1}{2}\partial^{\sigma}\partial_{\sigma}h^{(0)}_{\mu\nu}
\end{equation}
We can use the above information to express $h^{(2)}$ in terms of derivatives of $h^{(0)}$ and $h^{(4)}$ in terms of derivatives of $h^{(2)}$ as
\begin{align}
	h^{(2)}_{ii}=\frac{1}{2(d-2)}\left(\partial_i\partial_ih^{(0)}_{jj}-\partial_i\partial_jh^{(0)}_{ij}\right).\label{eq:h2toh0}
\end{align}
In $d=5$ we will also require the following relation between $h^{(4)}$ and $h^{(2)}$,
\begin{align}
	h^{(4)}_{ii}=\frac{1}{4}\partial_j\partial_jh^{(2)}_{ii}-\frac{1}{4}\partial_i\partial_jh^{(2)}_{ij}.\label{eq:h4toh2}
\end{align}
We can follow appendix \ref{section:EVd3} and \ref{section:EVd5} to obtain
\begin{align}
	\delta S^{ren}_B =\frac{dR^d\Omega_{d-2}}{8(d-1)(d+1)G_{d+1}}h^{(d)}_{ii}\label{eq:dSrenoddd}
\end{align}
Here by working with the renormalized quantities we recover the first law of entanglement entropy for general linearized variations of the metric, including both non-normalizable and normalizable modes. 

\subsection{Curvature Invariants Formula}
\label{section:CIF}
The first variation of the entanglement entropy around spherical entangling regions in AdS$_{d+1}$ with $d$ odd can be expressed in a particularly simple and elegant geometric form, using the expression for the renormalized entanglement entropy in terms of curvature and topological invariants \eqref{structure}. Since such variations do not change the topology of the entangling surface, the topological Euler invariant contribution does not change. All contributions from the extrinsic curvature are quadratic or higher order; since the extrinsic curvature vanishes to leading order, and this means the contributions ${\cal H}_p$ do not contribute to first variations (but do contribute to second variations). By analogous reasoning, the only contribution from the Weyl terms ${\cal W}_r$ comes from the term that is linear in the Weyl tensor. Thus we arrive at 
\begin{align}
\delta S^{ren} \propto - \frac{1}{4 G_{2n}} \delta {\cal W}\label{eq:dSdW}
\end{align}
where $G_{2n}$ is the Newton constant (with $2 n = d +1 $) and
\begin{align}
	\delta {\cal W} = \int_{\Sigma} d^{2 (n-1)} x \sqrt{g} \; \delta W_{1212} - \int_{\partial \Sigma} d^{2n -3} x \sqrt{h}\; \delta W_{1212} + \cdots  \label{eq:dWintW}
\end{align}
where $\delta W_{1212}$ is the pullback of the normal components of the bulk linearized Weyl curvature in an orthonormal frame and
$\delta W_{1212}$  is the pullback of the normal components of the boundary linearized Weyl curvature in an orthonormal frame. The boundary terms are such that $\delta {\cal W}$ is a finite conformal invariant for a generic non-normalizable metric perturbation. Note that the boundary term vanishes 
for AdS$_4$. The ellipses denote additional boundary terms expressed in terms of higher powers of the boundary Weyl curvature that are required for $n > 3$. 

The variation of renormalized entanglement entropy for $d=3,5$ is
\begin{eqnarray}
	d=3:&\quad\quad\quad\quad \delta S^{ren}(\tilde{B}) =  - \frac{1}{4G_{4}}\delta {\cal W}(\tilde{B}) \label{eq:dSdWd3}\\
	d=5:&\quad\quad\quad\quad \delta S^{ren}(\tilde{B}) =  - \frac{1}{12G_{6}}\delta {\cal W}(\tilde{B})\label{eq:dSdWd5}
\end{eqnarray}
In Poincar\'{e} coordinates the linear variation of the Weyl tensor $\delta W_{abcd}$ is,
\begin{align}
	\delta W_{\mu\nu\rho\sigma}&=\frac{1}{z^2}\mathcal{R}[\eta+h]_{\mu\nu\rho\sigma}+\frac{1}{2z^3}(h_{\mu\rho}'\eta_{\nu\sigma}+h_{\nu\sigma}'\eta_{\mu\rho}-h_{\mu\sigma}'\eta_{\nu\rho}-h_{\nu\rho}'\eta_{\mu\sigma})\label{eq:dW1}\\
	\delta W_{\mu\nu\rho z}&=\frac{1}{2z^2}[\partial_{\mu}h_{\nu\rho}'-\partial_{\nu}h_{\mu\rho}']\label{eq:dW2}\\
	\delta W_{\mu z\nu z}&=-\frac{1}{2z^2}h_{\mu\nu}''+\frac{1}{2z^3}h_{\mu\nu}'\label{eq:dW3}
\end{align}
where $\mathcal{R}[\eta+h]_{\mu\nu\rho\sigma}$ is the Riemann tensor for boundary metric $\eta_{\mu\nu}+h_{\mu\nu}$. In Poincar\'{e} coordinates, the two unit normals are
\begin{align}
	n_1&=z\frac{\partial}{\partial z}\\
	n_2&=\frac{z}{\sqrt{r^2+z^2}}\left(z\frac{\partial}{\partial z}+r\hat{x}^i\frac{\partial}{\partial x^i}\right)
\end{align}
Then the projection of Weyl tensor onto $N\tilde{B}$, $\delta W_{1212}$, is
\begin{align}
	\delta {W}_{1212}=\frac{z^4}{r^2+z^2}\left(z^2\delta W_{tztz}+r^2\hat{x}^i\hat{x}^j\delta W_{titj}+2zr\hat{x}^i\delta W_{tzti}\right)
\end{align}
The bulk Weyl integral becomes
\begin{align}
	&\int_{\tilde{B}}d^{d-1}x\sqrt{\gamma}\;W_{1212}=\int^{\frac{\pi}{2}}_{\xi}du\int_{S^{d-2}}d\Omega_{d-2} \frac{R^{2}\cos^{d-2}u}{\sin^{d-5}u}\left(z^2W_{tztz}+r^2\hat{x}^i\hat{x}^jW_{titj}+2zr\hat{x}^iW_{tzti}\right)
\end{align}
and the boundary Weyl integral is
\begin{align}
	&\int_{\partial\tilde{B}}d^{d-2}x\sqrt{\widetilde{\gamma}}\;W_{1212}=\int_{S^{d-2}}d\Omega_{d-2} \frac{R^{2}\cos^{d-2}u}{\sin^{d-6}u}\left(z^2W_{tztz}+r^2\hat{x}^i\hat{x}^jW_{titj}+2zr\hat{x}^iW_{tzti}\right)
\end{align}
Substituting $(\ref{eq:dW1})-(\ref{eq:dW3})$ into the above integrals
\begin{align}
\int_{\tilde{B}}d^{d-1}x\sqrt{\gamma}\;\delta W_{1212}=&\int^{\frac{\pi}{2}}_{\xi}du\int_{S^{d-2}}d\Omega_{d-2}R^2\bigg( \frac{\cos^{d-2}u}{\sin^{d-5}u}\big[-\frac{1}{2}h_{tt}''+\frac{1}{2R\sin u}h_{tt}'\big]\label{eq:IntW}\\
	&+\frac{\cos^{d}u}{\sin^{d-3}u}\hat{x}^i\hat{x}^j\Big[\mathcal{R}_{titj}+\frac{1}{2R\sin u}(h_{tt}'\eta_{ij}+h_{ij}'\eta_{tt})\Big]\nonumber\\
	&+\frac{\cos^{d-1}u}{\sin^{d-4}u}\hat{x}^i[\partial_{t}h_{ti}'-\partial_{i}h_{tt}']\bigg)\nonumber
\end{align}
and
\begin{align}
	\int_{\tilde{B}}d^{d-2}x\sqrt{\widetilde{\gamma}}\;\delta{W}_{1212}=&\int_{S^{d-2}}d\Omega_{d-2}R^2\bigg( \frac{\cos^{d-2}u}{\sin^{d-6}u}\big[-\frac{1}{2}h_{tt}''+\frac{1}{2R\sin u}h_{tt}'\big]\label{eq:IntdW}\\
	&+\frac{\cos^{d}u}{\sin^{d-4}u}\hat{x}^i\hat{x}^j\Big[\mathcal{R}_{titj}+\frac{1}{2R\sin u}(h_{tt}'\eta_{ij}+h_{ij}'\eta_{tt})\Big]\nonumber\\
	&+\frac{\cos^{d-1}u}{\sin^{d-5}u}\hat{x}^i[\partial_{t}h_{ti}'-\partial_{i}h_{tt}']\bigg)\nonumber
\end{align}
where ${}'=\frac{\partial}{\partial z}$ Up to order $R^{d}$, the relevant components of the integrand are obtained by Taylor expanding about the origin and eliminating the odd components as the it is integrated over $S^{d-2}$. In appendix \ref{section:RWI}, we expand $\mathcal{R}_{titj}$ into linear perturbation $h_{\mu\nu}$ then further relate the higher order non-normalizable modes $h_{\mu\nu}^{(n<d)}$ to the lower order non-normalizable modes via the Einstein equation. Finally, we can see all the lower order non-normalizable modes perturbation are cancelled and the renormalized Weyl integral is
\begin{align}
		d=3:&\quad\quad\quad\quad \delta\mathcal{W}=-\frac{3R^3\Omega_{1}}{16G_{4}}h^{(3)}_{tt}\label{eq:dWintd3}\\
		d=5:&\quad\quad\quad\quad \delta\mathcal{W}=-\frac{5R^5\Omega_{3}}{16G_{6}}h^{(5)}_{tt}.\label{eq:dWintd5}
\end{align}
Then substituting $(\ref{eq:dWintd3}),(\ref{eq:dWintd5})$ into  $(\ref{eq:dSdWd3}),(\ref{eq:dSdWd5})$ to get the renormalized entanglement entropy. We recovered the infinitesimal first law of entanglement entropy in $(\ref{eq:hddSren})$ for variation that includes perturbation of non-normalizable modes,
\begin{align}
	d=3:&\quad\quad\quad\quad \delta S^{ren}_B=\frac{3R^3\Omega_{1}}{48G_{4}}h^{(3)}_{tt}\\
	d=5:&\quad\quad\quad\quad \delta S^{ren}_B=\frac{5R^5\Omega_{3}}{192G_{6}}h^{(5)}_{tt}
\end{align}

\subsection{Cancellation of Divergences in $d=4$}
\label{section:CDd4}
We now turn from odd dimensional boundaries to even dimensions and show how the cancellation of divergences of the renormalized entanglement entropy works in $d=4$. A general perturbation of the boundary metric $h_{\mu\nu}$ can be expanded around the boundary $z=0$
\begin{equation}
	h_{\mu\nu}=h^{(0)}_{\mu\nu}(r,\theta,\phi)+z^2h^{(2)}_{\mu\nu}(r,\theta,\phi)+\dotsi
\end{equation}
Since on $\tilde{B}$ the coordinate $r$ is a function of $z$. The coefficient in the expansion of the metric perturbation can be further expanded around $r=R$. For $h^{(0)}_{\mu\nu}(r,\theta,\phi)$ the expansion is
\begin{align}
	h^{(0)}_{\mu\nu}(r,\theta,\phi)&=h^{(0)}_{\mu\nu}(R,\theta,\phi)+(r-R)\partial_r h^{(0)}_{\mu\nu}(R,\theta,\phi)+\cdots\label{eq:polayRexp}
	\\
	&=h^{(0)}_{\mu\nu}(R,\theta,\phi)-\frac{z^2}{2R}\partial_r h^{(0)}_{\mu\nu}(R,\theta,\phi)+\cdots
\end{align}
So the variation of the regularized entanglement entropy in polar coordinates for $d=4$ is,
\begin{align}
	\delta S^{reg}_B=&\frac{1}{8G_5}\int^R_{\epsilon}dz\int_{S^2}d\Omega_2\bigg[\frac{1}{z^3}\Big(h^{(0)}_{\theta\theta}+\frac{1}{\sin^2{\theta}}h^{(0)}_{\Phi\Phi}\Big)+\frac{1}{z}\Big(-\frac{1}{2R^2}h^{(0)}_{\theta\theta}-\frac{1}{2R^2\sin^2\theta}h^{(0)}_{\phi\phi}\nonumber\\
	&+h^{(0)}_{rr}+\frac{1}{R^2}h^{(0)}_{\theta\theta}+\frac{1}{R^2\sin^2\theta}h^{(0)}_{\phi\phi}-\frac{1}{2R}\partial_r h^{(0)}_{\theta\theta}-\frac{1}{2R\sin^2\theta}\partial_r h^{(0)}_{\phi\phi}+h^{(2)}_{\theta\theta}+\frac{1}{\sin^2\theta}h^{(2)}_{\phi\phi}\Big)\bigg]\label{eq:dSregd4}
\end{align}
Evaluating the $z$ integral and the divergent terms are,
\begin{align}
	(\delta S^{reg}_B)^{div}=&\frac{1}{8G_5}\int_{S^2}d\Omega_2\bigg[\frac{1}{2\epsilon^2}\Big(h^{(0)}_{\theta\theta}+\frac{1}{\sin^2{\theta}}h^{(0)}_{\Phi\Phi}\Big)-\ln\epsilon\Big(-\frac{1}{2R^2}h^{(0)}_{\theta\theta}-\frac{1}{2R^2\sin^2\theta}h^{(0)}_{\phi\phi}\nonumber\\
	&+h^{(0)}_{rr}+\frac{1}{R^2}h^{(0)}_{\theta\theta}+\frac{1}{R^2\sin^2\theta}h^{(0)}_{\phi\phi}-\frac{1}{2R}\partial_r h^{(0)}_{\theta\theta}-\frac{1}{2R\sin^2\theta}\partial_r h^{(0)}_{\phi\phi}+h^{(2)}_{\theta\theta}+\frac{1}{\sin^2\theta}h^{(2)}_{\phi\phi}\Big)\bigg]\label{eq:dSregdivd4}
\end{align}
We can see that $(\ref{eq:dSregdivd4})$ is identical to $(\ref{eq:dSctdivd4})$ so the divergences of the regularized entanglement entropy will be removed by the counterterms in the renormalization procedure.
\begin{align}
	(\delta S^{reg}_B)^{div}=(\delta S^{ct}_B)^{div}.
\end{align}
More explicitly, in Cartesian coordinate, the set of counterterms from $(\ref{eq:renhee})$ is
\begin{align}
	\delta S^{ct}_B&=\frac{1}{16G_{5}}\int_{S^2}d\Omega_2\bigg[ (h_{ii}-\hat{x}^i\hat{x}^jh_{ij})(\frac{r^2}{\epsilon^2}+\ln \epsilon) \nonumber\\
	&+\ln \epsilon\Big(-\delta\mathcal{R}_{tt}+\delta\mathcal{R}_{rr}-\frac{2}{3}\delta\mathcal{R}\Big)-\ln \epsilon\Big(-2h_{ii}+\frac{1}{r}\partial_r(r^2h_{ii}-x^ix^jh_{ij})\Big)\bigg].
\end{align}
Following section \ref{section:EVCs} we get
\begin{align}
	\delta S^{ct}_B&=\frac{1}{16G_{5}}\int_{S^2}d\Omega_2 \frac{r^2}{\epsilon^2}\big(h_{ii}-\hat{x}^i\hat{x}^jh_{ij}\big)+\ln\epsilon\bigg[ \big(h^{(0)}_{ii}-\hat{x}^i\hat{x}^jh^{(0)}_{ij}\big)-2R^2\big(h^{(2)}_{ii}-\hat{x}^i\hat{x}^jh^{(2)}_{ij}\big)\nonumber\\
	&-\big(-r\hat{x}^k\partial_kh^{(0)}_{ii}+2\hat{x}^i\hat{x}^jh^{(0)}_{ij}+r\hat{x}^i\hat{x}^j\hat{x}^k\partial_kh^{(0)}_{ij}\big)\bigg]\label{eq:ExdSCtd4}.
\end{align} 
Note that there are finite contributions from the first term in $(\ref{eq:ExdSCtd4})$.

\section{Integral Renormalized First Law}\label{section:IRFL}

Under general variations of the boundary metric where both the non-normalizable and normalizable modes are not fixed, we need to modify the relation between the conserved charges $(\ref{eq:Lemma4.1})$ and the associated first law. Since the spatial slice $\Sigma$ where the charges are defined has a boundary, we cannot neglect the total derivative terms. In fact the boundary terms capture all the divergent behaviour of the Noether charge and act as counterterms. 

As mentioned in the section \ref{section:Ham}, the asymptotic conformal Killing vector used to define the Noether charges in \cite{Papadimitriou:2005ii} has to follow the fall off condition $(\ref{eq:falloff})$ which our modular flow generator $\xi_B$ in $(\ref{eq:xiB})$ does not satisfy. We shall see later that all these extra terms are essential to match the universal divergences of the entanglement entropy. 

Charges defined on asymptotic boundary $\tilde{B}$ and entangling surface $B$ have asymptotic behaviours analogous to the entanglement entropy. In $(d+1)$ even spacetime dimensions, the finite charges are universal. For $(d+1)$ odd spacetime dimensions, the finite parts are scheme dependent, and change covariantly under changes of scheme. Hence, the first law of entanglement entropy in odd bulk dimensions requires appropriate finite counterterms. 

\subsection{Charges in the Entangling Region}\label{section:CER}
The Noether charge form $\boldsymbol{Q}[\xi]$ is the exact term in the conserved current form induced by the vector $\xi$. For pure Einstein gravity (with or without cosmological constant), it can be expressed as
\begin{align}
	\boldsymbol{Q}[\xi] &= - \frac{1}{16\pi G_N}\star d\xi\nonumber\\
& = - \frac{1}{16\pi G_N}\boldsymbol{\varepsilon}_{ab}\nabla^a\xi^b, \label{eq:NCQ}
\end{align}
up to exact terms. Since the extra exact terms will introduce boundary terms in the integral over $B$ and $\tilde{B}$ respectively, we will treat $(\ref{eq:NCQ})$ as the definition of $\boldsymbol{Q}[\xi]$ to avoid confusion. In asymptotically locally AdS, the full expression for the 
Noether charge form is then written as
\begin{align}
	\boldsymbol{Q}^{full}[\xi]=\boldsymbol{Q}[\xi]-\iota_{\xi}\boldsymbol{B}
\end{align}
with $\boldsymbol{B}$ defined as
\begin{align}
	\boldsymbol{B}=- \frac{1}{8\pi G_N}\boldsymbol{\varepsilon}_{a}n^a\left(K_{(d)}+\lambda_{ct}\right)
\end{align}
where $n$ is the radial unit normal pointing outwards from the asymptotic boundary $\partial \mathcal{M}$.

\bigskip

The holographic charge form $\boldsymbol{\mathcal{Q}}[\xi]$ is defined in terms of the $d^{th}$ term in the dilatation eigenfunction expansion of the canonical momentum, $\pi_{(d)}^{bc}$, through the following expression
\begin{align}
	\boldsymbol{\mathcal{Q}}[\xi]=-\boldsymbol{\varepsilon}_{ab}n^a2\pi_{(d)}^{bc}\xi_c.
\end{align}
In our setup, the full Noether current form $\boldsymbol{J}^{full}[\xi_B]$ is induced by the bulk modular flow of a bulk Killing vector $\xi_B$. The full Noether charge on the spatial slice $\Sigma_{\epsilon}$ can be thought of as the charge captured by the surface from the current:
\begin{align}
	{Q}^{full}[\xi_B]&=\int_{\Sigma_{\epsilon}}\boldsymbol{J}^{full}[\xi_B]\nonumber\\
	&=\int_{\Sigma_{\epsilon}}\boldsymbol{\Theta}[\delta_{\xi_B}\phi]-\iota_{\xi_B}\boldsymbol{L}^{onshell}\nonumber\\
&=- \int_{\Sigma_{\epsilon}} \iota_{\xi_B}\boldsymbol{L}^{onshell}
\end{align}
As shown in $(\ref{eq:dQ})$, the onshell Noether current form is exact. 

In $(\ref{eq:EntropyInt},\ref{eq:ModEInt})$, we defined the bulk entanglement entropy by an integral of the Noether charge form over $\tilde{B}_{\epsilon}$  and the modular energy through an integral of the holographic charge form over $B_{\epsilon}$. In order to relate the two we need to generalize $(\ref{eq:Lemma4.1})$ to
\begin{align}
	\int_{B_{\epsilon}}\boldsymbol{Q}^{full}[\xi] - \boldsymbol{\Delta}[\xi] =- \int_{B_{\epsilon}}\boldsymbol{\mathcal{Q}}[\xi].\label{eq:New4.1}
\end{align}
Here $\boldsymbol\Delta$ captures the counterterms associated with renormalizing the divergences of $\boldsymbol{Q}^{full}$; this term is needed as the quantity on the righthandside, ${\boldsymbol{\mathcal{Q}}}$, is renormalized. We could redefine the Noether charge on the lefthandside to include these counterterms, but in what follows we keep track of the contributions separately to emphasise how the counterterm contributions arise. 

The counterterms need to be included here because of our more general falloff conditions for the perturbations. This contribution vanishes in \cite{Papadimitriou:2005ii} because of the stricter fall-off condition of $\xi$ which makes the radial derivative of $\xi$ vanishes and the counterterms integrate to zero as the integral is over a surface with no boundary. In \cite{Faulkner:2013ica} this term vanishes due to the falloff conditions imposed on the metric perturbations. 

The conserved charge forms $\boldsymbol{Q}^{full}$ and $\boldsymbol{\mathcal{Q}}$ can be interpreted as Hamiltonian potentials, as explained in detail in appendix $\ref{section:CPSH}$. $\boldsymbol{\Delta}$ in this context is the difference of the counterterm contributions of the two Hamiltonian potentials. In the covariant phase space formalism, given an action with boundary terms, one can obtain the presymplectic current through variation of the Lagrangian and boundary terms. The presymplectic current maps the vector field in the configuration space to the Hamiltonian potential. 

We are interested in renormalized quantities and there are two ways to see how the counterterms arise in the Hamiltonian potential. The first approach is to use the renormalized action that includes the counterterms, and then obtain the presymplectic current and Hamiltonian potential. We denote this as the full Hamiltonian potential because it is equal to the full Noether charge form when $\pi_{(d)}^{\mu\nu}=0$
\begin{align}
	\boldsymbol{H}^{full}[\xi]&=\boldsymbol{Q}[\xi]+\boldsymbol{b}^{GH}[\xi]-\boldsymbol{b}^{ct}[\xi]\\
	&=\boldsymbol{H}^{GH}[\xi]-\boldsymbol{b}^{ct}[\xi]\\
	&=\boldsymbol{Q}^{full}[\xi],
\end{align}
where $\boldsymbol{b}^{GH}$ and $\boldsymbol{b}^{ct}$ represent the Gibbon-Hawking boundary term and counterterm contribution. Here $\boldsymbol{H}^{GH}$ is Hamiltonian potential obtained from the action that only includes the Gibbon-Hawking boundary term. 

The second way to see how the counterterms arise is to renormalize the Gibbon-Hawking Hamiltonian potential $\boldsymbol{H}^{GH}$ directly by subtracting the lower order terms in the dilatation eigenvalue expansion. The renormalized Gibbon-Hawking Hamiltonian potential is given in terms of the 
holographic charge form when $\pi_{(d)}^{\mu\nu}=0$
\begin{align}
	\boldsymbol{H}^{GH}_{(d)}[\xi]&=\boldsymbol{H}^{GH}[\xi]-\boldsymbol{H}^{GH}_{ct}[\xi]\\
	&=-\boldsymbol{\mathcal{Q}}[\xi].
\end{align}
Hence we can interpret $\boldsymbol{\Delta}$ as the difference of the two aforementioned Hamiltonian potentials
\begin{align}
\boldsymbol{\Delta}[\xi]=\boldsymbol{H}^{GH}_{ct}[\xi]-\boldsymbol{b}^{ct}[\xi]\label{eq:D2HGHbct}
\end{align}
As we will see this term in the perturbed case this is exact and represents the counterterms for the entanglement entropy.

\bigskip

On $B_{\epsilon}$, for bulk Killing vector, $\xi_B$ the full Noether charge form is
\begin{align}
		\boldsymbol{Q}^{full}[\xi_B]\vert_{B{\epsilon}}=-\frac{1}{8\pi G_N}\boldsymbol{\varepsilon}_{zt}n^z\left(-z\partial_z\xi^t_B+K_{\mu}^{t}\xi^{\mu}_B-K_{(d)}\xi_B^t-\lambda_{ct}\xi^t_B\right)
\end{align}
where the first term on the right hand side was neglected in \cite{Papadimitriou:2005ii} due to the falloff condition $(\ref{eq:falloff})$. The holographic charge form is
\begin{align}
	\boldsymbol{\mathcal{Q}}[\xi_B]\vert_{B{\epsilon}}&=-\boldsymbol{\varepsilon}_{zt}n^z2\pi_{(d)}^{tc}\xi_{B\;c}\nonumber\\
	 &=\frac{1}{8\pi G_N}\boldsymbol{\varepsilon}_{zt}n^z\left(K^{ta}_{(d)}-K_{(d)}\gamma^{ta}\right)\xi_{B\;a}\label{eq:MQinK}
\end{align}
where we used $(\ref{eq:pin})$ to express the holographic charge in terms of the $d^{th}$ term in the dilatation eigenfunction expansion of the extrinsic curvature. The difference in the charges $\boldsymbol{\Delta}[\xi_B]$ is
\begin{align}
	\boldsymbol{\Delta}[\xi_B]\vert_{B{\epsilon}}=-\frac{1}{8\pi G_N}\boldsymbol{\varepsilon}_{zt}n^z\left(-z\partial_z\xi^t_B+K_{ct\,\mu}^{\;\;t}\xi^{\mu}_B-\lambda_{ct}\xi^t_B\right). \label{eq:Deltaex}
\end{align}
It is important to remember that this expression is only valid when $\xi_B$ is Killing. We shall see later the perturbed difference of the charges $\delta\Delta[\xi_B]$ admits an extra term as $\xi_B$ is no longer Killing. In appendix \ref{section:DEE}, we follow \cite{Papadimitriou:2004ap} and derive the explicit dilatation eigenfunction expansion for $K^{\mu}_{\nu}$ and $\lambda$. In the unperturbed setting, the boundary metric is $d$-dimensional Minkowski metric $\eta_{\mu\nu}$. Only the zeroth term in the dilatation eigenfunction expansion is non-vanishing,
\begin{align}
	K^{\;\;\mu}_{(0)\,\nu}=\delta^{\mu}_{\nu}, && \lambda_{(0)}=1.
\end{align} 
From $(\ref{eq:MQinK})$ we know the holographic charge is zero 
\begin{align}
	\boldsymbol{\mathcal{Q}}[\xi_B]\vert_{B{\epsilon}}=0.
\end{align}
Then $\boldsymbol{\Delta}[\xi_B]$ simply equals the Noether charge
\begin{align}
	\boldsymbol{\Delta}[\xi_B]\vert_{B{\epsilon}}=\boldsymbol{Q}[\xi_B]\vert_{B{\epsilon}}&=-\frac{1}{8\pi G_N}\boldsymbol{\varepsilon}_{zt}n^z\left(-z\partial_z\xi^t_B\right)\nonumber\\
	&=\frac{z^3}{4  G_N}\boldsymbol{\varepsilon}_{zt} \label{eq:DeltaequalQ}
\end{align}
We can now turn our attention to the Noether charge form on the entangling surface $\tilde{B}{\epsilon}$. As explained in section \ref{section:1stLawIntro}, the integral of the Noether charge form over $\tilde{B}_{\epsilon}$ can be interpreted as both the entropy of the Rindler black hole and the entanglement entropy of boundary region $B_{\epsilon}$. Since $\xi_B$ vanishes on $\tilde{B}_{\epsilon}$, 
\begin{align}
	\boldsymbol{Q}^{full}[\xi_B]\vert_{\tilde{B}_{\epsilon}}&=\boldsymbol{Q}[\xi_B]\vert_{\tilde{B}_{\epsilon}}\nonumber\\
	&=\frac{1}{8\pi G_N}\boldsymbol{\varepsilon}_{wt}\partial^w\xi_B^t
\end{align} 
where we used the $w$ coordinate in $(\ref{eq:wcoord})$ and the Killing condition. Integrating over $\tilde{B}_{\epsilon}$,
\begin{align}
	\int_{B_{\epsilon}}	\boldsymbol{Q}^{full}[\xi_B]&=\frac{1}{8\pi G_N}\int_{B_{\epsilon}}	du\,d\Omega_{d-2}\frac{\cos^{d-2}u}{\sin^{d-1}u}\frac{2\pi}{w}\left(z^2+r^2\right)\nonumber\\
	&=\frac{R}{4 G_N}\int_{B_{\epsilon}}	du\,d\Omega_{d-2}\frac{\cos^{d-2}u}{\sin^{d-1}u}\nonumber\\
	&=S^{reg}_B
\end{align}
where we use $(\ref{eq:SIntu})$ to identify the second line with the regulated entanglement entropy.

\subsection{Variation of Charges}\label{section:VoC}

The variations of $\boldsymbol{Q}^{full}$ and $\boldsymbol{\mathcal{Q}}$ differ from the previous literature \cite{Faulkner:2013ica} when we allow variations of the non-normalizable modes. For general perturbations of $\gamma_{\mu\nu}$, the linear variation of the Noether charge form is
\begin{align}
	\delta \boldsymbol{Q}[\xi_{B}]&=\frac{-1}{16\pi G_N}\delta\left[\boldsymbol{\varepsilon}_{ab}\nabla^a\xi_{B}^b\right]\nonumber\\
	&=\frac{-1}{16\pi G_N}\boldsymbol{\varepsilon}_{ab}\left[\frac{\delta\gamma}{2}\nabla^a\xi_{B}^b-\delta g^{ac}\nabla_c\xi_{B}^b+g^{ac}\delta\Gamma^b_{dc}\xi_{B}^d\right]\nonumber\\
	&=\frac{-z^2}{8RG_N}\bigg[\boldsymbol{\varepsilon}_{ti}\left(x^ih_{kk} -x^jh_{ij}-(R^2-z^2-\vec{x}^2)\partial_th_{it}\right)\\
	&\quad\quad\quad\quad+\boldsymbol{\varepsilon}_{tz}\left(zh_{kk}+(R^2-z^2-\vec{x}^2)(-\frac{2}{z}h_{tt}+\partial_zh_{tt})\right)\bigg]\nonumber
\end{align}
Using the coordinates $(\ref{eq:wcoord})$ on $\tilde{B}_{\epsilon}$ we get the integral
\begin{align}
	\int_{\tilde{B}_{\epsilon}}\delta \boldsymbol{Q}[\xi_{B}]&=\frac{1}{8RG_N}\int_{\tilde{B}_{\epsilon}}d^{d-1}x\frac{1}{z^{d}}(R^2h_{kk}-x^ix^jh_{ij})\nonumber\\
	&=\frac{1}{8RG_N}\int_{\tilde{B}_{\epsilon}}d^{d-1}x(R^2-\vec{x}^2)^{-\frac{d}{2}}(R^2h_{kk}-x^ix^jh_{ij})
\end{align}
which is equal to the linear variation of the holographic entanglement entropy
\begin{align}
	\int_{\tilde{B}_{\epsilon}}\delta \boldsymbol{Q}[\xi_{B}]&=\delta S^{reg}_B \label{eq:dEntropyInt}
\end{align}
For the variation of the full Noether charge form, we need to evaluate the boundary term $\delta\boldsymbol{B}$. This term is related to the presymplectic form $\boldsymbol{\Theta}[\delta\phi]$ by
\begin{align}
	\boldsymbol{\Theta}[\delta\phi]&=\frac{\boldsymbol{d^dx}}{8\pi G_N}\delta\left(-\sqrt{\gamma}\lambda\right) \\
	&=\boldsymbol{d^dx}\left[\frac{1}{8\pi G_N}\delta\left(-\sqrt{\gamma}K\right)+\pi^{\mu\nu}\delta\gamma_{\mu\nu}\right]\nonumber\\
	&=\boldsymbol{d^dx}\left[\frac{1}{8\pi G_N}\delta\left(-\sqrt{\gamma}\left(K_{(d)}+\lambda_{ct}\right)\right)+\pi_{(d)}^{\mu\nu}\delta\gamma_{\mu\nu}\right] \nonumber\\
	&=\delta\boldsymbol{B}+\boldsymbol{\varepsilon}_{\scriptscriptstyle\partial \mathcal{M}_{\epsilon}}\pi_{(d)}^{\mu\nu}\delta\gamma_{\mu\nu}. \nonumber
\end{align}
where the d-form $\boldsymbol{d^dx}$ is
\begin{align}
	\boldsymbol{d^dx}=\frac{1}{d!}dx^0\wedge\cdots\wedge dx^{d-1}.
\end{align}
The variation of the full Noether charge form is then
\begin{align}
	\delta \boldsymbol{Q}^{full}[\xi_{B}]&=\delta \boldsymbol{Q}[\xi_{B}]-\iota_{\xi_B}\delta\boldsymbol{B} \\
	&=\delta \boldsymbol{Q}[\xi_{B}]-\iota_{\xi_B}\boldsymbol{\Theta}[\delta\phi]-\iota_{\xi_B}\boldsymbol{\varepsilon}_{\scriptscriptstyle\partial \mathcal{M}_{\epsilon}}\pi_{(d)}^{\mu\nu}\delta\gamma_{\mu\nu}. \nonumber 
\end{align}
The linear variation of the holographic charge is
\begin{align}
	\delta\boldsymbol{\mathcal{Q}}[\xi]\vert_{B_{\epsilon}}&=-\boldsymbol{\varepsilon}_{zt}n^z2\delta\pi_{(d)\,t}^{\;\;t}\xi_{B}^t.\label{eq:dHCQ}
\end{align}
This is related to the renormalized boundary energy momentum tensor via
\begin{align}
	2\delta\pi_{(d)}^{\mu\nu}=-\delta T_{ren}^{\mu\nu}.
\end{align}
Substituting this expression into $(\ref{eq:dHCQ})$ the integral of the variation of holographic charge form $\delta\boldsymbol{\mathcal{Q}}[\xi_{B}]$ on the boundary ball region $B_{\epsilon}$ is equal to the variation of modular energy
\begin{align}
	- \int_{{B}_{\epsilon}}\delta\boldsymbol{\mathcal{Q}}[\xi_B]=\delta E_B.\label{eq:dModE}
\end{align}
To express the variation of modular energy in terms of dilatation eigenfunction expansion of extrinsic curvature we vary $(\ref{eq:ConMom})$ to obtain
\begin{align}
	\delta\pi_{(d)\,\nu}^{\;\;\mu}&=\frac{-1}{16\pi G_N}\left(\delta K^{\;\;\mu}_{(d)\,\nu}-\delta K_{(d)} \delta^{\mu}_{\nu}\right)\label{eq:dConMom}\\
	\delta\pi_{(d)\, t}^{\;\; t}&=\frac{1}{16\pi G_N}\delta K_{(d)\,i}^{\;\;i}.
\end{align}
Using the tracelessness of $\delta K_{(d)\,\nu}^{\;\;\mu}$ at the linear level we can write $\delta\boldsymbol{\mathcal{Q}}[\xi]$ on $B_{\epsilon}$ as
\begin{align}
	\delta\boldsymbol{\mathcal{Q}}[\xi_B]\vert_{B_{\epsilon}}&=\frac{1}{8\pi G_N}\boldsymbol{\varepsilon}_{zt}n^z\delta K_{(d)\,t}^{\;\;t}\xi_{B}^t.
\end{align}
(This expression holds for all $d$, with conformal anomalies present if we write out $K_{(d)\,t}^{\;\;t}$ in terms of $g^{(n)}_{\mu\nu}$ and $\tilde{g}^{(d)}_{\mu\nu}$, see for example $(\ref{eq:K4ind4})$.)
The variation of the full Noether charge form $\delta\boldsymbol{Q}^{full}$ on $B_{\epsilon}$ is
\begin{align}
	\delta\boldsymbol{Q}^{full}[\xi_B]\vert_{B{\epsilon}}=\frac{z}{8\pi G_N}\boldsymbol{\varepsilon}_{z t}\bigg[&\left(\frac{\delta\gamma}{2} (K^t_{\mu}\xi_B^{\mu}-\frac{1}{z}\partial^z\xi_B^t)-\frac{1}{2z}\partial_t\xi_B^z\delta\gamma^{tt}+\xi_B^t\delta K^t_t\right)\\
	&-\xi_B^t\left(\frac{\delta\gamma}{2}+\delta K_{(d)}+\delta\lambda_{ct}\right)\bigg]. \nonumber
\end{align}
By using $(\ref{eq:dConMom})$, we can obtain the relation between $\delta\boldsymbol{Q}^{full}$ and $\delta\boldsymbol{\mathcal{Q}}$. 

Similarly to $(\ref{eq:New4.1})$, this revised version of $(\ref{eq:dLemma4.1})$ receives a contribution $\delta\boldsymbol{\Delta}[\xi_B]$. 
We get
\begin{align}
	\int_{B_{\epsilon}}\delta\boldsymbol{Q}^{full}[\xi_B]=\int_{B_{\epsilon}}-\delta\boldsymbol{\mathcal{Q}}[\xi_B]+\delta\boldsymbol{\Delta}[\xi_B].\label{eq:dNew4.1}
\end{align}
The latter term takes the form
\begin{align}
	\delta\boldsymbol{\Delta}[\xi_B]=\frac{z}{8\pi G_N}\boldsymbol{\varepsilon}_{z t}\left[-\frac{1}{2z} \partial^z\xi_B^t\delta\gamma-\frac{1}{2z}\partial_t\xi_B^z\delta\gamma^{tt}+\xi_B^t\left(\delta K^t_t-\delta\lambda\right)_{ct}\right].\label{eq:dDxiB}
\end{align}
Note that we can understand why this term arises for two reasons. Firstly, $\xi_B$ is no longer Killing with respect to the perturbed metric and secondly $\xi_{B}$ has a weaker falloff condition $(\ref{eq:newfalloff})$ instead of $(\ref{eq:falloff})$. Here we use an abbreviated notation:
\begin{align}
	\delta\gamma=\gamma^{\mu\nu}\delta\gamma_{\mu\nu},&&\delta K^{\mu}_{\nu}=\delta\left(\gamma^{\mu\sigma}K_{\sigma\nu}\right).
\end{align}
 In terms of Hamiltonian potentials, the $\delta\boldsymbol{\Delta}$ term is
\begin{align}
	\delta\boldsymbol{\Delta}[\xi_B]=\delta\boldsymbol{H}^{GH}_{ct}[\xi_B]-\delta\boldsymbol{b}^{ct}[\xi_B].
\end{align}
We further describe the origin of each term in the appendix $\ref{section:CPSH}$ and expressed $\delta\boldsymbol{\Delta}$ in $(\ref{eq:dDeltaexp})$ using the formalism of \cite{2020HarlowWu}.

Substituting the unperturbed flat boundary metric and the bulk Killing vector we obtain
\begin{align}
	\delta\boldsymbol{\Delta}[\xi_B]&=\frac{d^{d-1}x\,\sqrt{-\gamma}}{8\pi G_N}\left[\frac{\pi z^2}{R}(-h_{tt}+h_{ii})+\frac{\pi z^2}{R}h_{tt}+\frac{\pi (R^2-z^2-\vec{x}^2)}{R}\left(\delta K^t_t-\delta\lambda\right)_{ct}\right]\nonumber\\
	\delta\boldsymbol{\Delta}[\xi_B]&=\frac{d^{d-1}x\,z^{-d}}{8R G_N}\left[z^2h_{ii}+(R^2-z^2-\vec{x}^2)\left(\delta K^t_t-\delta\lambda\right)_{ct}\right].\label{eq:dDelta}
\end{align}
The variation of the onshell boundary Lagrangian, $\delta\lambda$, is related to the variation of the extrinsic curvature, $\delta K$, via the canonical momentum in $(\ref{eq:ConMom})$
\begin{align}
	-\frac{1}{8\pi G_N}\delta\left[\sqrt{\gamma}\lambda\right]&=-\frac{1}{8\pi G_N}\delta\left[\sqrt{\gamma}K\right]+\pi^{\mu\nu}\delta\gamma_{\mu\nu} \\
	-\frac{\sqrt{\gamma}}{8\pi G_N}\left(\frac{\lambda}{2}\delta\gamma+\delta\lambda\right)&=-\frac{\sqrt{\gamma}}{8\pi G_N}\left(\frac{K}{2}\delta\gamma+\delta K+\frac{K^{\mu\nu}-K\gamma^{\mu\nu}}{2}\delta\gamma_{\mu\nu}\right)\nonumber\\
	\frac{\lambda}{2}\delta\gamma+\delta\lambda&=\delta K+\frac{K^{\mu\nu}}{2}\delta\gamma_{\mu\nu}. \nonumber
\end{align}
For flat boundary metrics we have
\begin{align}
	\delta\lambda=\delta K.
\end{align}
We then get the following simplified expression for all dimension
\begin{align}	
	\delta\boldsymbol{\Delta}[\xi_B]&=\frac{z^{-d}}{8R G_N}\left[z^2h_{ii}+(R^2-z^2-\vec{x}^2)\left(\delta K^t_t-\delta K\right)_{ct}\right]\nonumber\\
 &=\frac{z^{-d}}{8R G_N}\left[z^2h_{ii}-(R^2-z^2-\vec{x}^2)\delta K_{ct\,i}^{\;\;i}\right].
\end{align}
The  extrinsic curvature counterterm means that all the terms appear earlier in the dilatation eigenfunction expansion, i.e.
\begin{align}
	\delta K_{\mu\nu}&=\delta K_{(0)\,\mu\nu}+\delta K_{(2)\,\mu\nu}+\cdots+\log z^2 \delta\tilde{K}_{(d)\,\mu\nu}+\delta K_{(d)\,\mu\nu}+\cdots\nonumber\\
	\delta K_{\mu\nu}&=\delta K_{ct\,\mu\nu}+\delta K_{(d)\,\mu\nu}+\cdots
\end{align}
From $(\ref{eq:dNew4.1})$ we can deduce that the divergence of the full Noether charge integral is equal to the divergence of the correction term integral,
\begin{align}
	\left(\int_{B_{\epsilon}}\delta\boldsymbol{Q}^{full}[\xi_B]\right)^{div}=\int_{B_{\epsilon}}\delta\boldsymbol{\Delta}^{div}[\xi_{B}].
\end{align}
In order to see how this divergence is equivalent to the divergence in the entanglement entropy, we need to use the Stoke's theorem of the full Noether charge form on $\Sigma_{\epsilon}$,
\begin{align}
	\int_{{B}_{\epsilon}}\delta \boldsymbol{Q}^{full}[\xi_{B}]-\int_{\tilde{B}_{\epsilon}}\delta \boldsymbol{Q}^{full}[\xi_{B}]=\int_{\Sigma_{\epsilon}}d\delta\boldsymbol{Q}^{full}[\xi_{B}].\label{eq:dQStokes}
\end{align}
The exterior derivative of the variation of Noether charge form can be deduced from $(\ref{eq:JTL})$ and $(\ref{eq:dQ})$,
\begin{align}
	d\delta\boldsymbol{Q}[\xi_B]&=\delta \boldsymbol{J}-\delta\boldsymbol{N} \\
	&=\delta\boldsymbol{\Theta}[\delta_{\xi_B}\psi]-\iota_{\xi_B}\delta\boldsymbol{L}+\delta\boldsymbol{N}\nonumber\\
	&=\delta\boldsymbol{\Theta}[\delta_{\xi_B}\psi]-\iota_{\xi_B}d\delta\boldsymbol{\Theta}[\delta\psi]-\iota_{\xi_B}\boldsymbol{E}^{\psi}\delta\psi+\delta\boldsymbol{N}\nonumber\\
	&=\delta\boldsymbol{\Theta}[\delta_{\xi_B}\psi]-\mathcal{L}_{\xi_B}\boldsymbol{\Theta}[\delta_{\xi_B}\phi]+d\iota_{\xi_B}\delta\boldsymbol{\Theta}[\delta\phi]-\iota_{\xi_B}\boldsymbol{E}^{\psi}\delta\phi+\delta\boldsymbol{N}\nonumber\\
	&=\boldsymbol{\omega}(\delta\psi,\delta_{\xi_B}\psi)+d\iota_{\xi_B}\delta\boldsymbol{\Theta}[\delta\psi]-\iota_{\xi_B}\boldsymbol{E}^{\phi}\delta\psi+\delta\boldsymbol{N}\nonumber\\
	&=d\iota_{\xi_B}\delta\boldsymbol{\Theta}[\delta\psi]-\iota_{\xi_B}\boldsymbol{E}^{\phi}\delta\psi+\delta\boldsymbol{N}\label{eq:ddQ}  
	\end{align}
where $\boldsymbol{\omega}(\delta_1\psi,\delta_2\psi)$ is the symplectic form 
\begin{align}
	\boldsymbol{\omega}(\delta_1\psi,\delta_2\psi)=\delta_2\boldsymbol{\Theta}[\delta_1\psi]-\delta_1\boldsymbol{\Theta}[\delta_2\psi]
\end{align}
and it vanishes when $\xi_B$ is Killing. Note the last two terms are off-shell terms. We first write out  $(\ref{eq:dNew4.1})$ as
\begin{align}
	\int_{{B}_{\epsilon}}\delta \boldsymbol{Q}[\xi_{B}]-\iota_{\xi_B}\boldsymbol{\Theta}[\delta\phi]=\int_{B_{\epsilon}}-\delta\boldsymbol{\mathcal{Q}}[\xi_B]+\delta\boldsymbol{\Delta}[\xi_B].\label{eq:dNewNew4.1}
\end{align}
Now substitute $(\ref{eq:ddQ})$ and $(\ref{eq:dNewNew4.1})$ into $(\ref{eq:dQStokes})$,
\begin{align}
	\int_{{B}_{\epsilon}}\delta \boldsymbol{Q}^{full}[\xi_{B}]-\int_{\tilde{B}_{\epsilon}}\delta \boldsymbol{Q}^{full}[\xi_{B}]&=\int_{\Sigma_{\epsilon}}d\iota_{\xi_B}\delta\boldsymbol{\Theta}[\delta\phi]+\delta\boldsymbol{N}-\iota_{\xi_B}\boldsymbol{E}^{\phi}\delta\phi-d\iota_{\xi_B}\delta\boldsymbol{B}\nonumber\\
	\int_{{B}_{\epsilon}}\delta \boldsymbol{Q}[\xi_{B}]-\iota_{\xi_B}\boldsymbol{\Theta}[\delta\phi]-\int_{\tilde{B}_{\epsilon}}\delta \boldsymbol{Q}[\xi_{B}]&=\int_{\Sigma_{\epsilon}}-\iota_{\xi_B}\boldsymbol{E}^{\phi}\delta\phi+\delta\boldsymbol{N}\nonumber\\
	\int_{{B}_{\epsilon}}-\delta \boldsymbol{\mathcal{Q}}[\xi_{B}]+\delta\boldsymbol{\Delta}[\xi_B]&=\int_{\tilde{B}_{\epsilon}}\delta \boldsymbol{Q}[\xi_{B}]+\int_{\Sigma_{\epsilon}}-\iota_{\xi_B}\boldsymbol{E}^{\phi}\delta\phi+\delta\boldsymbol{N}.\label{eq:full1stlawoff}
\end{align}
Onshell we get 
\begin{align}
		\int_{{B}_{\epsilon}}-\delta \boldsymbol{\mathcal{Q}}[\xi_{B}]&=\int_{\tilde{B}_{\epsilon}}\delta \boldsymbol{Q}[\xi_{B}]-\int_{{B}_{\epsilon}}\delta\boldsymbol{\Delta}[\xi_B].
\end{align}
Since the left hand side is manifestly finite we have
\begin{align}
	\left(\int_{\tilde{B}_{\epsilon}}\delta\boldsymbol{Q}[\xi_B]\right)^{div}&=\int_{B_{\epsilon}}\delta\boldsymbol{\Delta}^{div}[\xi_{B}]\\
	\delta S_B^{div}&=\int_{B_{\epsilon}}\delta\boldsymbol{\Delta}^{div}[\xi_{B}]\label{eq:dSdiv}
\end{align}
Therefore the integral of $\delta\boldsymbol{\Delta}$ on the boundary ball region can be thought of as the counterterm of the entanglement entropy. In the next section we will show that the finite part of $\delta\boldsymbol{\Delta}$  matches with the counterterm of the entanglement entropy as well. Hence we get the integral first law of entanglement entropy:
\begin{align}
	\delta E_B&=\delta S_B^{ren}.\label{eq:New1stLaw}
\end{align}
Finite counterterms contribute only when the CFT dimension is even. This is an expected result as the finite part of the entanglement entropy is scheme dependent in even $d$. Similarly, the left hand side is related to the renormalized energy momentum tensor which is also scheme dependent for even $d$. For odd $d$, the finite part of the renormalized entanglement entropy is universal We will see explicit examples in the following section.

{The implication of $\delta\boldsymbol{\Delta}$ acting as the density of the entanglement entropy counterterms is that $\delta\boldsymbol{\Delta}$ is exact,
\begin{align}
	\delta\boldsymbol{\Delta}[\xi_{B}]=d\delta \boldsymbol{S}^{ct}_B,
\end{align}
where $\delta \boldsymbol{S}^{ct}_B$ is the $(d-2)$-form that integrates to the entanglement entropy counterterm. This means the full Hamiltonian potential $\delta\boldsymbol{H}^{full}$ and the renormalized Gibbon-Hawking Hamiltonian potential $\delta\boldsymbol{H}^{GH}_{(d)}$ is equal up to an exact term. In the usual context of conserved quantities, this is attributed to the exact term ambiguity. Since the potential is integrated over a boundary manifold, which itself does not have boundary, the exact term ambiguity will not contribute to the conserved charge. For us, both the entanglement entropy and modular energy is defined as the integral of a manifold that does have boundary, so the exact term difference is no longer an ambiguity. Note the counterterm of the entanglement entropy is obtained systemically from the renormalized action through the replica trick, this indicates this exact term difference can be calculated from the renormalized action directly.  In the Hamiltonian holographic renormalization framework, we show how to obtain $\delta\boldsymbol{\Delta}$ from the counterterms contribution of the Hamiltonian potentials in appendix $\ref{section:CPSH}$. }

\subsection{Examples: Generalized First Law in $AlAdS_{d+1}$}\label{section:EXs}

We have shown that the variation of the modular energy $\delta E_B$ is equal to the integral of the holographic charge form over the boundary ball region $B_{\epsilon}$ in $(\ref{eq:dModE})$ and the variation of the entanglement entropy $\delta S_B$ is equal to the integral of the Noether charge form over the bulk entangling surface $\tilde{B}_{\epsilon}$ in $(\ref{eq:dEntropyInt})$. To complete the generalized first law of entanglement entropy $(\ref{eq:New1stLaw})$ for generic variations of the boundary metric $\delta\gamma_{\mu\nu}$ in $AlAdS_{d+1}$, we only need to check that the integral of the term $\delta\boldsymbol{\Delta}[\xi_B]$ over $B_{\epsilon}$ is the counterterm of the entanglement entropy,
\begin{align}
	\int_{{B}_{\epsilon}}\delta\boldsymbol{\Delta}[\xi_{B}]=\delta S^{ct}_B.\label{eq:IntDdSct}
\end{align}
In the following subsections, we will demonstrate this equality up to dimension $d=5$, thus implying the renormalized first law $(\ref{eq:New1stLaw})$, with scheme dependence of renormalized entropy and energy systematically matched.

\subsubsection{$d=3$}\label{section:AlAdS4ex}

The terms in the dilatation eigenfunction expansion of the extrinsic curvature variation are related to the Fefferman-Graham expansion of the boundary metric variation. For $d=3$, we only need to include terms up to $O(z^4)$ as higher order terms will not contribute to calculations 
in the limit $\epsilon\rightarrow0$:
\begin{align}
	\delta K_{(0)\,\nu}^{\;\;\mu}&=0\\
	\delta K_{(2)\,\nu}^{\;\;\mu}&=-z^2\eta^{\mu\sigma}h_{(2)\,\sigma\nu}+O(z^4)\\
	\delta{K}_{(3)\,\nu}^{\;\;\mu}&=-\frac{3}{2}z^3\eta^{\mu\sigma}h_{(3)\,\sigma\nu}+O(z^4).
\end{align}
In this case, the counterterm $\delta K_{ct\,\nu}^{\;\;\mu}$ is just the second term in the dilatation eigenfunction expansion $	\delta K_{(2)\,\nu}^{\;\;\mu}$. Hence the counterterm from $(\ref{eq:dDelta})$ gives
\begin{align}
	\delta\boldsymbol{\Delta}[\xi_B]&=\frac{\boldsymbol{d^2x}z^{-3}}{8R G_N}\left[z^2h_{ii}-(R^2-z^2-\vec{x}^2)\left(-z^2h_{(2)\,ii}\right)\right].
\end{align}
Keeping the terms up to $O(z)$ we have,
\begin{align}
	\delta\boldsymbol{\Delta}[\xi_B]&=\frac{\boldsymbol{d^2x}}{8R G_N}\left[\frac{1}{z}\left(h_{(0)\,ii}+(R^2-\vec{x}^2)h_{(2)\,ii}\right)\right].
\end{align}
We see that for this example in odd dimensions, $\delta\boldsymbol{\Delta}[\xi_B]$ has no term of order $z^0$ and  there is as expected no finite counterterm contribution to the entanglement entropy. 

To see the identification of the integral of $\delta\boldsymbol{\Delta}[\xi_B]$ over $B_{\epsilon}$ with the ordinary entanglement entropy counterterm in $(\ref{eq:ExdSCt})$, we need to use $(\ref{eq:Inth0h2})$ with the result
\begin{align}
	\int_{{B}_{\epsilon}}\delta\boldsymbol{\Delta}[\xi_{B}]&=\frac{1}{8RG_N}\int_{S^1}d\Omega_{1}\frac{r}{\epsilon}\left(h_{(0)\,ii}-\hat{x}^i\hat{x}^jh_{(0)\,ij}\right).
\end{align}
This matches with the variation in the counterterm $(\ref{eq:ExdSCt})$ exactly. Hence we have satisfied $(\ref{eq:IntDdSct})$ confirming that the general variation of the modular energy is the variation of the renormalized entanglement entropy. 

\subsubsection{$d=4$}\label{section:AlAdS5ex}

For $d=4$, in addition to including the logarithmic term in dilatation eigenfunction expansion, we also have to include terms up to $O(z^6)$ to evaluate both the divergent and finite contributions:
\begin{align}
	\delta K_{(0)\,\nu}^{\;\;\mu}&=0\\
	\delta K_{(2)\,\nu}^{\;\;\mu}&=-z^2\eta^{\mu\sigma}h_{(2)\,\sigma\nu}-z^2\eta^{\mu\sigma}\delta D_{(2)\,\sigma\nu}+O(z^6)\\
	\delta \tilde{K}_{(4)\,\nu}^{\;\;\mu}&=-2z^4\eta^{\mu\sigma}\tilde{h}_{(4)\,\sigma\nu}+O(z^6)\\
	\delta{K}_{(4)\,\nu}^{\;\;\mu}&=-2z^4\eta^{\mu\sigma}h_{(4)\,\sigma\nu}-z^4\eta^{\mu\sigma}\tilde{h}_{(4)\,\sigma\nu}+z^4\eta^{\mu\sigma}\delta D_{(2)\,\sigma\nu}+O(z^6) \label{eq:K4ind4}
\end{align}
where we use the notation
\begin{align}
	\delta D_{(n)\,\mu\nu}=\delta\left[\int g_{(n)\sigma\rho}\frac{g_{(n)\mu\nu}}{g_{(0)\sigma\rho}}\right].
\end{align}
It turns out at linear level that the second order term $\delta D_{(2)\,\mu\nu}$ is related to the coefficient of the logarithmic term in the Fefferman-Graham expansion as
\begin{align}
	\delta D_{(2)\,\mu\nu}=-2\tilde{h}_{(4)\,\mu\nu},
\end{align}
and hence it is also traceless. Then the relevant terms in the dilatation eigenfunction expansion for the extrinsic curvature are
\begin{align}
	\delta K_{(0)\,\nu}^{\;\;\mu}&=0\\
	\delta K_{(2)\,\nu}^{\;\;\mu}&=-z^2\eta^{\mu\sigma}h_{(2)\,\sigma\nu}+2z^4\eta^{\mu\sigma}\tilde{h}_{(4)\,\sigma\nu}+O(z^6)\\
	\delta \tilde{K}_{(4)\,\nu}^{\;\;\mu}&=-2z^4\eta^{\mu\sigma}\tilde{h}_{(4)\,\sigma\nu}+O(z^6)\\
	\delta{K}_{(4)\,\nu}^{\;\;\mu}&=-2z^4\eta^{\mu\sigma}h_{(4)\,\sigma\nu}-3z^4\eta^{\mu\sigma}\tilde{h}_{(4)\,\sigma\nu}+O(z^6).
\end{align}
The counterterm from $(\ref{eq:dDelta})$ is then
\begin{align}
\delta\boldsymbol{\Delta}[\xi_B]&=\frac{\boldsymbol{d^3}xz^{-4}}{8R G_N}\left[z^2h_{ii}-(R^2-z^2-\vec{x}^2)\left(-z^2h_{(2)\,ii}+2z^4\tilde{h}_{(4)\,ii}-2z^4\log z^2\tilde{h}_{(4)\,ii}\right)\right].
\end{align}
Neglecting the $O(z)$ terms as they vanish in the limit $\epsilon\rightarrow0$ we have
\begin{align}
	\delta\boldsymbol{\Delta}[\xi_B]&=\frac{\boldsymbol{d^3x}}{8R G_N}\left[\frac{1}{z^2}\left(h_{(0)\,ii}-(R^2-\vec{x}^2)h_{(2)\,ii}\right)
	+2(R^2-\vec{x}^2)(1-\log z^2)\tilde{h}_{(4)\,ii}\right].
\end{align}
Finally we need to transform this integral on boundary ball region $B_{\epsilon}$ into a surface integral on the sphere $\partial B_{\epsilon}$ via the manipulation of $h_{(n)\,\mu\nu}$ in appendix $\ref{section:Inth}$. First we use $(\ref{eq:Inth0h2})$ to turn the coefficient of $\epsilon^{-2}$ divergences into a surface integral
\begin{align}
	\int_{{B}_{\epsilon}}\delta\boldsymbol{\Delta}[\xi_{B}]&=\frac{1}{8RG_N}\bigg[\frac{(R^2-\epsilon^2)^{\frac{3}{2}}}{2\epsilon^2}\int_{S^2}d\Omega_{2}\,\left(h_{(0)\,ii}-\hat{x}^i\hat{x}^jh_{(0)\,ij}\right)\\\label{eq:IntBdDd4}
	&+\int_{{B}_{\epsilon}}d^3x\,h_{(2)\,ii}\nonumber\\
	&-2(1-\log\epsilon^2)\int_{{B}_{\epsilon}}d^3x\,(R^2-\vec{x}^2)\tilde{h}_{(4)\,ii}\bigg]\nonumber.
\end{align}
For the coefficient of the logarithmic divergence, we use $(\ref{eq:Inth4h2})$ to turn the $\tilde{h}_{(4)\,ii}$ integral into integrals of $h_{(2)\,ii}$ then use $(\ref{eq:Inth2h0})$ to turn the remaining volume integral into a surface integral of $h_{(0)\,ii}$. The final result is
\begin{align}
	\int_{{B}_{\epsilon}}\delta\boldsymbol{\Delta}[\xi_{B}]&=\frac{1}{8RG_N}\int_{S^2}d\Omega_{2}\,\bigg[\frac{r^3}{2\epsilon^2}\left(h_{(0)\,ii}-\hat{x}^i\hat{x}^jh_{(0)\,ij}\right)\\
    &+\log\epsilon^2\bigg(\frac{r^3}{2}(\hat{x}^i\hat{x}^jh_{(2)\,ij}-h_{(2)\,ii})\nonumber\\
    &+\frac{r}{4}(h_{(0)\,ii}-3\hat{x}^i\hat{x}^jh_{(0)\,ij}+x^j\partial_jh_{(0)\,ii}-\hat{x}^i\hat{x}^jx^k\partial_kh_{(0)\,ij})\bigg)\nonumber\\
	&-\frac{r^3}{2}(\hat{x}^i\hat{x}^jh_{(2)\,ij}-h_{(2)\,ii})\bigg]\nonumber 
\end{align}
Note that there are, as expected, finite contributions.  Comparing with $(\ref{eq:ExdSCtd4})$ we can see this term is exactly the counterterm for the entanglement entropy. Therefore in $AlAdS_5$ we have satisfied $(\ref{eq:IntDdSct})$. The renormalized stress tensor $T^{ren}_{\mu\nu}$ in $(\ref{eq:dTrend4})$ has a scheme dependent term proportional to $\tilde{h}_{(d)\,\mu\nu}$ that originates from the variation of anomaly term in the counterterm action. Therefore the finite counterterm in the entanglement entropy is necessary to match the contribution associated with the holographic conformal anomaly.

\subsubsection{$d=5$}\label{section:AlAdS6ex}
The $d=5$ case is very similar to the above example but without the logarithmic terms. The dilatation eigenfunction expansion for the variation of the extrinsic curvature is 
\begin{align}
	\delta K_{(0)\,\nu}^{\;\;\mu}&=0\\
	\delta K_{(2)\,\nu}^{\;\;\mu}&=-z^2\eta^{\mu\sigma}h_{(2)\,\sigma\nu}-z^2\eta^{\mu\sigma}\delta D_{(2)\,\sigma\nu}+O(z^6)\\
	\delta{K}_{(4)\,\nu}^{\;\;\mu}&=-2z^4\eta^{\mu\sigma}h_{(4)\,\sigma\nu}+z^4\eta^{\mu\sigma}\delta D_{(2)\,\sigma\nu}+O(z^6)
\end{align}
where at linear level we have
\begin{align}
	\delta D_{(2)\, ii}=\frac{2}{3}h_{(4)\,ii}.
\end{align}
Then the relevant dilatation eigenfunction expansion terms, up to $O(z^6)$, are
\begin{align}
	\delta K_{(0)\,i}^{\;\;i}&=0\\
	\delta K_{(2)\,i}^{\;\;i}&=-z^2h_{(2)\,ii}-\frac{2}{3}z^2h_{(4)\,ii}+O(z^6)\\
	\delta{K}_{(4)\,i}^{\;\;i}&=-\frac{4}{3}z^4h_{(4)\,ii}+O(z^6).
\end{align}
The counterterm from $(\ref{eq:dDelta})$ gives 
\begin{align}
	\delta\boldsymbol{\Delta}[\xi_B]&=\frac{\boldsymbol{d^4x}z^{-5}}{8R G_6}\left[z^2h_{ii}-(R^2-z^2-\vec{x}^2)\left(-z^2h_{(2)\,ii}-2z^4{h}_{(4)\,ii}\right)\right].
\end{align}
Neglecting the $O(z)$ terms as they vanish in the limit $\epsilon\rightarrow0$ we have,
\begin{align}
	\delta\boldsymbol{\Delta}[\xi_B]&=\frac{\boldsymbol{d^4x}}{8R G_6}\left[\frac{1}{z^3}\left(h_{(0)\,ii}-(R^2-\vec{x}^2)h_{(2)\,ii}\right)
	+\frac{2}{z}(R^2-\vec{x}^2){h}_{(4)\,ii}\right].\label{eq:IntBdDd5}
\end{align}
Now evaluate the integral of the correction following in appendix $\ref{section:Inth}$. We use $(\ref{eq:Inth0h2})$ and $(\ref{eq:Inh4h2})$ to get
\begin{align}
	\int_{{B}_{\epsilon}}\delta\boldsymbol{\Delta}[\xi_{B}]=&\frac{1}{8RG_6}\bigg[\frac{(R^2-\epsilon^2)^{2}}{3\epsilon^3}\int_{S^3}d\Omega_{3}\,\left(h_{(0)\,ii}-\hat{x}^i\hat{x}^jh_{(0)\,ij}\right)+\frac{1}{\epsilon}\int_{{B}_{\epsilon}}d^4x\,h_{(2)\,ii}\\
	&\quad\quad+\frac{(R^2-\epsilon^2)^{2}}{\epsilon}\int_{S^3}d\Omega_{3}\,\left(h_{(2)\,ii}-\hat{x}^i\hat{x}^jh_{(2)\,ij}\right)-\frac{3}{\epsilon}\int_{{B}_{\epsilon}}d^4x\,h_{(2)\,ii}\bigg]\nonumber.
\end{align}
The remaining volume integral of $h_{(2)\,ii}$ can be converted to surface integral via $(\ref{eq:Inh2h0})$,
\begin{align}
	\int_{{B}_{\epsilon}}\delta\boldsymbol{\Delta}[\xi_{B}]=&\frac{1}{8RG_6}\int_{S^3}d\Omega_{3}\bigg[\frac{R^4}{3\epsilon^3}\,\left(h_{(0)\,ii}-\hat{x}^i\hat{x}^jh_{(0)\,ij}\right)-\frac{2R^2}{3\epsilon}\,\left(h_{(0)\,ii}-\hat{x}^i\hat{x}^jh_{(0)\,ij}\right)\\
	&+\frac{R^4}{\epsilon}\,\left(h_{(2)\,ii}-\hat{x}^i\hat{x}^jh_{(2)\,ij}\right)-\frac{R^2}{3\epsilon}\left(h_{(0)\,ii}-4\hat{x}^i\hat{x}^jh_{(0)\,ij}+x^j\partial_jh_{(0)\,ii}-\hat{x}^i\hat{x}^jx^k\partial_kh_{(0)\,ij}\right)\bigg]\nonumber.
\end{align}
After rearranging we arrive at the final expression of the correction term
\begin{align}
	\int_{{B}_{\epsilon}}\delta\boldsymbol{\Delta}[\xi_{B}]=&\frac{1}{8RG_6}\int_{S^3}d\Omega_{3}\bigg[\frac{R^4}{3\epsilon^3}\,\left(h_{(0)\,ii}-\hat{x}^i\hat{x}^jh_{(0)\,ij}\right)-\frac{R^2}{\epsilon}\,\left(h_{(0)\,ii}-2\hat{x}^i\hat{x}^jh_{(0)\,ij}\right)\label{eq:FinaldDd5}\\
	&+\frac{R^4}{\epsilon}\,\left(h_{(2)\,ii}-\hat{x}^i\hat{x}^jh_{(2)\,ij}\right)-\frac{R^2}{3\epsilon}\left(x^j\partial_jh_{(0)\,ii}-\hat{x}^i\hat{x}^jx^k\partial_kh_{(0)\,ij}\right)\bigg]\nonumber.
\end{align}
This is in fact identical to the counterterm in $(\ref{eq:ExdSCt})$ when taking the limit $\epsilon\rightarrow 0$ and satisfying $(\ref{eq:IntDdSct})$. Note that in $(\ref{eq:ExdSCt})$ one has to expand $r=\sqrt{R^2-\epsilon^2}$ to arrive at $(\ref{eq:FinaldDd5})$. Since $d$ is odd, there is no finite counterterm and the renormalized first law is scheme independent.

\section{Conclusions and Outlook}

In this paper we have proven the renormalized first law of holographic entanglement entropy, in both infinitesimal and covariant versions, for generic variations of the 
metric. The original proofs of the first law of holographic entanglement entropy assumed that only normalisable modes of the metric were varied, corresponding to 
changing the state in the dual conformal field theory. Our proof extends to non-normalisable variations of the metric, corresponding to changing the background metric for  
the dual conformal theory. 

When the boundary dimension $d$ is odd, both the renormalized stressed tensor and renormalized area of the entangling surface are scheme independent and the holographic conformal anomaly is absent. When the boundary dimension $d$ is even, there are finite contributions from counterterms and one needs to ensure that the same renormalization scheme is used for the stress tensor and entanglement entropy; this follows immediately from the approach taken in \cite{Taylor:2016aoi} because the counterterms for the entanglement entropy are derived from the counterterms for the action given in \cite{deHaro2001} using the replica trick. In our setup the background about which we are perturbing is conformally flat and thus there are no explicit contributions from the conformal anomaly at linear order. 

The first law can also be derived using the covariant phase space approach, building on \cite{Faulkner:2013ica}, as well discussions of the covariant phase space formalism in the presence of boundaries \cite{2020HarlowWu} and boundary counterterm contributions to conserved charges \cite{Papadimitriou:2005ii}. The generalization to non-normalizable variations of the bulk metric, corresponding to deforming the background metric for the dual CFT, induces specific counterterms in the covariant phase space construction. We explain in detail how these relate to the boundary terms in \cite{2020HarlowWu}. Note that in the context of the laws of black holes one would fix the non-normalizable modes and therefore the our analysis differs from the renormalized black hole charge analysis of \cite{Papadimitriou:2005ii}.  

\bigskip

While the focus of this paper has been on proving the holographic first law of entanglement entropy for non-normalisable bulk metric variations, our methodology could be extended to many analyses within holographic information theory.  One could clearly explore perturbations of the surface itself, following 
\cite{Klebanov:2012yf,Allais:2014ata, Rosenhaus:2014zza, Mezei:2014zla}. The extension to higher derivative gravity theories would be straightforward in principle although 
one may need to resolve analogous technical ambiguities to those encountered in \cite{2014jc,2014xd}. Analyses of local reconstruction in the bulk from boundary entanglement such as \cite{Lashkari:2013koa, 2015LinOoguriStocia} assume normalizable fall offs of metric perturbations (corresponding to CFT states), but our approach facilitates the discussion of marginal and indeed even irrelevant deformations. To include the latter, one would simply add in the bulk field corresponding to the irrelevant operator, and compute renormalized quantities perturbatively in the irrelevant deformation. Other analyses where our methodology would be useful to extend the class of theories/states under consideration include discussions of subregion complexity and the first law of complexity \cite{Jang:2020cbm,Bernamonti2020} as well as analyses of the relation of holographic entanglement entropy to inverse mean curvature flow \cite{2017fw}. 

\bigskip

Finally, let us consider the expression for the variation of the entanglement entropy in terms of the Weyl tensor $(\ref{eq:dSdW})$. This relation could have been 
anticipated from the known relationship between the Einstein sector of conformal (Weyl) gravity and Einstein gravity \cite{maldacena2011einstein, 2016ao}. Up to a topological term the renormalized action for Einstein gravity is proportional to the Weyl squared term \cite{Anderson:2001vr, 2016ao,Anastasiou_2021}. Accordingly, 
the Wald entropy functional for the AdS Rindler black hole on the black hole horizon $\tilde{H}_{\epsilon}$ gives
\begin{align}
	S_{\scriptscriptstyle Wald} \propto \int_{\tilde{H}_{\epsilon}}W^{abcd}n_{ab}n_{cd}
\end{align}
where $n_{ab}$ is the binormal for the codimension two surface $\tilde{H}_{\epsilon}$. Using the standard Casini, Huerta and Myers approach \cite{Casini:2011kv} we can then
map this entropy to the entanglement entropy for a spherical region in a flat background. The computations in this paper relate to the first variation of this entropy under bulk metric variations and using the CHM map we immediately obtain the first term of the Weyl integral in $(\ref{eq:dWintW})$
 \begin{align}
 	\delta S^{\rm ren} \propto \int_{\tilde{B}_{\epsilon}}\delta W_{1212}.
 \end{align}
 This relation holds in all even bulk spacetime dimensions, even though the expressions for the renormalized entanglement entropy become increasingly complex expressions of the Euler characteristic and curvature invariants of the entangling surface in higher dimensions \cite{Taylor:2020uwf}. The variation manifestly simplifies to just this one term for linear variations of a spherical surface around a background with zero Weyl curvature. Working to higher order in the variations, and in more general setups, one should make use of the full form of the renormalized area in terms of Euler characteristic and curvature invariants in \cite{Taylor:2020uwf} to understand the underlying geometric structure.

\section*{Acknowledgements}
This work is funded by the STFC grant ST/P000711/1. This project has received funding and support from the European Union's Horizon 2020 research and innovation programme under the Marie Sklodowska-Curie grant agreement No 690575. LT would like to thank A. Poole and F. Capone for relevant discussions.

\appendix
\section{Infinitesimal first law}
\subsection{Useful Identities}
In this appendix, we provide some useful identities that are used in section $\ref{section:InfRFL}$. First we give angular integrals of the unit vectors, 
\begin{align}
	&\int_{S^d}d\Omega_d \hat{x}^{odd}=0\label{eq:AIntodd}\\
	&\int_{S^d}d\Omega_d \hat{x}^i\hat{x}^j=\frac{\Omega_d}{d+1}\delta^{ij}\label{eq:AInt2}\\
	&\int_{S^d}d\Omega_d \hat{x}^i\hat{x}^j\hat{x}^k\hat{x}^l=\frac{\Omega_d}{(d+3)(d+1)}(\delta^{ij}\delta^{kl}+\delta^{ik}\delta^{jl}+\delta^{il}\delta^{jk})\label{eq:AInt4}\\
	&\int_{S^d}d\Omega_d\, \hat{x}^i\hat{x}^j\hat{x}^k\hat{x}^l\hat{x}^p\hat{x}^q=\frac{15\Omega_d}{(d+5)(d+3)(d+1)}\delta^{(ij}\delta^{kl}\delta^{pq)}\label{eq:AInt6}\\
	&\int_{S^d}d\Omega_d\, \hat{x}^{i_1}\hat{x}^{i_2}\cdots\hat{x}^{i_{2n-1}}\hat{x}^{i_{2n}}=\Omega_d\prod_{r=1}^{n}\frac{2r-1}{(d+2r-1)}\delta^{(i_1i_2}\cdots\delta^{i_{2n-1}i_{2n})}\label{eq:OmegaxiInt}.
\end{align}
Since the angular integral of unit vectors is expressed as symmetrized Kronecker deltas, it is also useful to have the expression of the symmetrized Kronecker deltas contracted with derivatives of the metric perturbation:
\begin{align}
	&\delta^{(ij}\delta^{kl)}\partial_k\partial_lh_{ij}=\frac{1}{3}\partial_k\partial_kh_{ii}+\frac{2}{3}\partial_i\partial_jh_{ij}\\
	&\delta^{(ij}\delta^{kl}\delta^{pq)}\partial_k\partial_l\partial_p\partial_qh_{ij}=\frac{1}{5}\partial_k\partial_k\partial_l\partial_lh_{ii}+\frac{4}{5}\partial_k\partial_k\partial_i\partial_jh_{ij}\\
	&\delta^{(i_1i_2}\cdots\delta^{i_{2n-1}i_{2n})}\partial_{i_3}\partial_{i_4}\cdots\partial_{i_{2n-1}}\partial_{i_{2n}}h_{i_1i_2}=\frac{(\partial^2)^{n-2}}{2n-1}\big[\partial^2h_{i_1i_1}+(2n-2)\partial_{i_1}\partial_{i_2}h_{i_1i_2}\big]\label{eq:deltaspartialsh}.
\end{align}

\subsection{Explicit Variation in $d=3$}
\label{section:EVd3}
In this section we will show the procedure used to calculate the variation of the variation of regularized entanglement entropy and variation of the counterterms in $d=3$. Here we continue the calculation from $(\ref{eq:IntSregwcoord})$. First we consider the leading order in the Taylor expansion which has no derivatives and perform the angular integrals $(\ref{eq:AInt2})$ to get
\begin{align}
	\delta S^{reg}_B(\partial^0)&=\frac{\Omega_1}{8G_{4}}\int_{\xi}^{\pi/2}du\frac{\cos{u}}{\sin^2{u}}(1-\frac{\cos^2{u}}{2})h_{ii}(z,0,0),
\end{align}
where
\begin{align}
	h_{ii}(z)&=h^{(0)}_{ii}+R^2\sin^2{u}h^{(2)}_{ii}+R^3\sin^3{u}h^{(3)}_{ii}.
\end{align}
After performing the $u$ integrals we get
\begin{align}
	\delta S^{reg}_B(\partial^0)&=\frac{\Omega_1}{8G_4}\bigg[\frac{1}{2}\Big(\frac{1}{\sin{\xi}}-\sin{\xi}\Big)h^{(0)}_{ii}+\Big(\frac{2}{3}-\frac{\sin{\xi}}{2}-\frac{\sin^3{\xi}}{6}\Big)R^2h^{(2)}_{ii}\nonumber\\
	&\;+\Big(\frac{3}{8}-\frac{\sin^2{\xi}}{4}-\frac{\sin^4{\xi}}{8}\Big)R^3h^{(3)}_{ii}\bigg]
\end{align}
We also need to evaluate the higher derivative terms in the Taylor expansion. For our purposes we need only the Taylor expansion of $h^{(0)}_{ij}(x,y)$. The contribution of the one derivative term of the variation is
\begin{equation}
	\delta S^{reg}_B(\partial^1)=\frac{R}{8G_{4}}\int_{\xi}^{\pi/2}du\int^{2\pi}_{0}d\phi \frac{\cos^2{u}}{\sin^2{u}}(\delta^{ij}-\cos^2{u}\hat{x}^i\hat{x}^j)\hat{x}^k\partial_kh^{(0)}_{ij}.
\end{equation}
Using the angular integrals $(\ref{eq:AIntodd})$ we can deduce $\delta S^{reg}_B(\partial^1)$ vanishes. 

The contribution of the leading two derivative terms of the variation is
\begin{equation}
	\delta S^{reg}_B(\partial^2)=\frac{R^2}{8G_{4}}\int_{\xi}^{\pi/2}du\int^{2\pi}_{0}d\phi \frac{\cos^3{u}}{\sin^2{u}}(\delta^{ij}-\cos^2{u}\hat{x}^i\hat{x}^j)\frac{\hat{x}^k\hat{x}^l}{2!}\partial_k\partial_lh^{(0)}_{ij}.
\end{equation}
We evaluate the angular integrals by substituting the results from $(\ref{eq:AInt2})$ and $(\ref{eq:AInt4})$,
\begin{align}
	&\delta S^{reg}_B(\partial^2)=\frac{R^2\Omega_1}{8G_{4}}\int_{\xi}^{\pi/2}du \frac{\cos^3{u}}{\sin^2{u}}\frac{1}{4}\partial_j\partial_jh^{(0)}_{ii}-\frac{\cos^5{u}}{\sin^2{u}}\frac{1}{16}\Big(\partial_j\partial_jh^{(0)}_{ii}+2\partial_i\partial_jh^{(0)}_{ij}\Big).
\end{align}
Evaluating the $u$ integral and rearranging the derivatives of metric variation we get
\begin{align}
	&\delta S^{reg}_B(\partial^2)=\frac{R^2\Omega_1}{8G_{4}}\bigg[\frac{1}{4}\Big(-2+\frac{1}{\sin{\xi}}+\sin{\xi}\Big)\partial_j\partial_jh^{(0)}_{ii}\nonumber\\
	&\quad\quad\quad\quad-\frac{1}{16}\Big(-\frac{8}{3}+\frac{1}{\sin{\xi}}+2\sin{\xi}+\frac{\sin^3{\xi}}{3}\Big)\Big(\partial_j\partial_jh^{(0)}_{ii}+2\partial_i\partial_jh^{(0)}_{ij}\Big)\bigg]\nonumber\\
	&\delta S^{ref}_B(\partial^2)=\frac{R^2\Omega_1}{32G_{4}}\bigg[\Big(-\frac{4}{3}+\frac{3}{4\sin{\xi}}+\frac{\sin{\xi}}{2}-\frac{\sin^3{\xi}}{12}\Big)\partial_j\partial_jh^{(0)}_{ii}\nonumber\\
	&\;\;\;\;\;\;\;\;\;\;\;\;\;\;\;\;\;\Big(\frac{4}{3}-\frac{1}{2\sin{\xi}}-\sin{\xi}-\frac{\sin^3{\xi}}{6}\Big)\partial_i\partial_jh^{(0)}_{ij}\bigg]
\end{align}
We would like to take the limit of $\xi\rightarrow 0$ so we need to check the divergences are cancelled out by counterterms in $(\ref{eq:ExdSCt})$. We first evaluate the leading order terms in the Taylor series with no derivatives:
\begin{equation}
	\delta S^{ct}_B(\partial^0)=\frac{1}{8G_4}\int^{2\pi}_{0}dz\phi\frac{\cos{\xi}}{\sin{\xi}}(h_{ii}-\hat{x}^i\hat{x}^jh_{ij}).
\end{equation}
Evaluating the angular integral and expanding around $\xi=0$ we get
\begin{equation}
	\delta S^{ct}_B(\partial^0)=-\frac{\Omega_1}{16G_4}\bigg[\frac{1}{\sin{\xi}}-\frac{\sin{\xi}}{2}\bigg]\times\bigg[h^{(0)}_{ii}+R^2\sin^2{\xi}h^{(2)}_{ii}+R^3\sin^3{\xi}h^{(3)}_{ii}\bigg]
\end{equation}
We now evaluate contributions from the subleading one derivative terms in the Taylor expansion of $h^{(0)}_{ij}(z)$
\begin{equation}
	\delta S^{ct}_B(\partial^1)=-\frac{R}{8G_4}\int^{2\pi}_{0}\frac{\cos^2{\xi}}{\sin{\xi}}\bigg[\hat{x}^k\partial_kh^{(0)}_{ii}-\hat{x}^i\hat{x}^j\hat{x}^k\partial_kh^{(0)}_{ij}\bigg].
\end{equation}
Using the angular integrals $(\ref{eq:AIntodd})$ we can deduce $\delta S^{ct}_B(\partial^1)$ vanishes. The next leading two derivative 
contribution is
\begin{align}
	&\delta S^{ct}_B(\partial^2)=-\frac{R^2}{8G_4}\int^{2\pi}_{0}\frac{\cos^3{\xi}}{\sin{\xi}}\bigg[\hat{x}^k\hat{x}^l\partial_k\partial_lh^{(0)}_{ii}-\hat{x}^i\hat{x}^j\hat{x}^k\hat{x}^l\partial_k\partial_lh^{(0)}_{ij}\bigg]
\end{align}
After evaluating the angular integrals we obtain 
\begin{align}
	&\delta S^{ct}_B(\partial^2)=-\frac{R^2\Omega_1}{128G_4}\frac{\cos^3{\xi}}{\sin{\xi}}\bigg[3\partial_j\partial_jh^{(0)}_{ii}-2\partial_i\partial_jh^{(0)}_{ij}\bigg].
\end{align}
Combining the variations of the regularized entanglement entropy and the variation of the counterterms we get the following. For $\partial^0$ terms we have
\begin{align}
	\delta S^{reg}_B(\partial^0)+\delta S^{ct}_B(\partial^0)&=\frac{\Omega_1}{8G_4}\bigg[\frac{1}{2\sin{\xi}}h^{(0)}_{ii}+\frac{2}{3}R^2h^{(2)}_{ii}+\frac{3}{8}R^3h^{(3)}_{ii}\bigg]\nonumber\\
	&-\frac{\Omega_1}{16G_4}\frac{1}{\sin{\xi}}h^{(0)}_{ii},
\end{align}
and for $\partial^2$ terms have
\begin{align}
	\delta S^{reg}_B(\partial^2)+\delta S^{ct}_B\partial^2)&=\frac{R^2\Omega_1}{32G_{4}}\bigg[\Big(-\frac{4}{3}+\frac{3}{4\sin{\xi}}\Big)\partial_j\partial_jh^{(0)}_{ii}+\Big(\frac{4}{3}-\frac{1}{2\sin{\xi}}\Big)\partial_i\partial_jh^{(0)}_{ij}\bigg]\nonumber\\
	&-\frac{R^2\Omega_1}{128G_4}\frac{1}{\sin{\xi}}\bigg[3\partial_j\partial_jh^{(0)}_{ii}-2\partial_i\partial_jh^{(0)}_{ij}\bigg].
\end{align}
Gathering all terms together we obtain the variation of renormalized entanglement entropy,
\begin{align}
	\delta S^{ren}_B=\frac{\Omega_1}{8G_4}\bigg[-\frac{R^2}{3}\partial_j\partial_jh^{(0)}_{ii}+\frac{R^2}{3}\partial_i\partial_jh^{(0)}_{ij}+\frac{2}{3}R^2h^{(2)}_{ii}+\frac{3}{8}R^3h^{(3)}_{ii}\bigg].
\end{align}
Then from $(\ref{eq:h2toh0})$ we can express $h^{(2)}$ in terms of $h^{(0)}$ and the variation of the renormalized entanglement entropy becomes
\begin{align}
	\delta S^{ren}_B &=\frac{\Omega_1}{8G_4}\bigg[-\frac{R^2}{3}\partial_j\partial_jh^{(0)}_{ii}+\frac{R^2}{3}\partial_i\partial_jh^{(0)}_{ij}+\frac{2}{3}R^2\Big(\frac{1}{2}\partial_j\partial_jh^{(0)}_{ii}-\frac{1}{2}\partial_i\partial_jh^{(0)}_{ij}\Big)+\frac{3}{8}R^3h^{(3)}_{ii}\bigg] \nonumber \\
	&=\frac{3R^3\Omega_1}{64G_4}h^{(3)}_{ii},
\end{align}
which is the result stated in $(\ref{eq:dSrenoddd})$ for $d=3$.

\subsection{Explicit Variation in $d=5$}
\label{section:EVd5}
Following the same approaches as in the section above, we continue the calculation from $(\ref{eq:IntSregwcoord})$ for $d=5$. The variation of the regularized entanglement entropy to leading order of the near boundary approximation, the zero derivative terms in the Taylor expansion give
\begin{align}
	\delta S^{reg}_B(\partial^0)=\frac{\Omega_3}{8G_6}\int_{\xi}^{\pi/2}du \bigg[\frac{\cos^3{u}}{\sin^4{u}}\bigg(1-\frac{\cos^2{u}}{4}\bigg)h_{ii}\bigg].
\end{align}
Using the Fefferman Graham expansion and evaluating the $u$ integral we get
\begin{align}
	\delta S^{reg}_B(\partial^0)=&\frac{\Omega_3}{8G_6} \bigg[\bigg(\frac{1}{4\sin^3\xi}-\frac{1}{2\sin\xi}\bigg)h^{(0)}_{ii}+\bigg(\frac{3}{4\sin\xi}-\frac{4}{3}\bigg)R^2h^{(2)}_{ii}+\frac{8}{15}R^4h^{(4)}_{ii}+\frac{5}{24}R^5h^{(5)}_{ii}\bigg].
\end{align}
The two derivative terms give
\begin{align}
	\delta S^{reg}_B(\partial^2)=&\frac{R^2\Omega_3}{8G_6}\int_{\xi}^{\pi/2}du \frac{\cos^5u}{2!\sin^4u}\bigg[\frac{1}{4}\partial_k\partial_kh_{ii}-\frac{\cos^2u}{24}\big(\partial_k\partial_kh_{ii}+2\partial_i\partial_jh_{ij}\big)\bigg]
\end{align}
Using the Fefferman Graham expansion and evaluating the $u$ integral we get
\begin{align}
	\delta S^{reg}_B(\partial^2)=&\frac{\Omega_3}{8G_6} \bigg[\bigg(\frac{5}{144\sin^3\xi}-\frac{3}{16\sin\xi}+\frac{2}{9}\bigg)R^2\partial_k\partial_kh_{ii}^{(0)}+\bigg(-\frac{1}{72\sin^3\xi}+\frac{1}{8\sin\xi}-\frac{2}{9}\bigg)R^2\partial_i\partial_jh_{ij}^{(0)}\nonumber\\
	&-\bigg(\frac{5}{48\sin\xi}-\frac{4}{15}\bigg)R^4\partial_k\partial_kh_{ii}^{(2)}+\bigg(-\frac{1}{24\sin\xi}+\frac{2}{15}\bigg)R^4\partial_i\partial_jh_{ij}^{(2)}\bigg].
\end{align}
The four derivative terms are
\begin{align}
	\delta S^{reg}_B(\partial^4)=&\frac{R^4\Omega_3}{8G_6}\int_{\xi}^{\pi/2}du \frac{\cos^7u}{4!\sin^4u}\bigg[\frac{1}{8}\partial_k\partial_k\partial_l\partial_lh_{ii}-\frac{\cos^2u}{64}\big(\partial_k\partial_k\partial_l\partial_lh_{ii}+4\partial_k\partial_k\partial_i\partial_jh_{ij}\big)\bigg].
\end{align}
Using the Fefferman Graham expansion and evaluating the $u$ integral we get
\begin{align}
	\delta S^{reg}_B(\partial^4)=&\frac{\Omega_3}{8G_6} \bigg[\bigg(\frac{7}{4608\sin^3\xi}-\frac{5}{384\sin\xi}+\frac{1}{45}\bigg)R^4\partial_k\partial_k\partial_l\partial_lh_{ii}^{(0)}\nonumber\\
	&+\bigg(-\frac{1}{1152\sin^3\xi}+\frac{1}{96\sin\xi}-\frac{1}{45}\bigg)R^4\partial_k\partial_k\partial_i\partial_jh_{ij}^{(0)}\bigg].
\end{align}
We have thus obtained all the divergent and finite terms for the variation of regularized entanglement entropy up to $R^5$. Note that for the $\delta S^{reg}_B$ only even derivatives survive the angular integrals. This is no longer the case for the counterterms $\delta S^{ct}_B$ as some terms in $(\ref{eq:ExdSCt})$ contain an odd number of directional vectors $\hat{x}$. 

For the variation of the counterterms, $(\ref{eq:ExdSCt})$, in the near boundary approximation, the leading order zero derivative terms are
\begin{align}
	\delta S^{ct}_B(\partial^0)&=\frac{\Omega_3}{8G_6}\bigg[\bigg(-\frac{1}{4\sin^3\xi}+\frac{1}{2\sin\xi}\bigg)h_{ii}^{(0)}-\frac{3}{4\sin\xi}R^2h_{ii}^{(2)}\bigg].
\end{align}
The one derivative terms comes from the variation of the extrinsic curvature, corresponding to the last terms in $(\ref{eq:ExdSCt})$. Note that one derivative means up to and including the first derivative terms in the Taylor expansion:
\begin{align}
	\delta S^{ct}_B(\partial^1)&=\frac{1}{24G_6}\int_{S^3}d\Omega_3\frac{r^3}{z}\bigg[\hat{x}^k\hat{x}^l\partial_k\partial_lh_{ii}-\hat{x}^i\hat{x}^j\hat{x}^k\hat{x}^l\partial_k\partial_lh_{ij}\bigg].
\end{align}
After the integration only the following terms remain
\begin{align}
	\delta S^{ct}_B(\partial^1)&=\frac{\Omega_3}{8G_6}\bigg[\frac{5}{72\sin\xi}R^2\partial_k\partial_kh_{ii}^{(0)}-\frac{1}{36\sin\xi}R^2\partial_i\partial_jh_{ij}^{(0)}\bigg].
\end{align}
Similar procedures are used for higher derivative terms. The order two derivative terms are
\begin{align}
	\delta S^{ct}_B(\partial^2)=&\frac{\Omega_3}{8G_6} \bigg[\bigg(-\frac{5}{144\sin^3\xi}+\frac{17}{144\sin\xi}\bigg)R^2\partial_k\partial_kh_{ii}^{(0)} \\
	&+\bigg(+\frac{1}{72\sin^3\xi}-\frac{7}{72\sin\xi}\bigg)R^2\partial_i\partial_jh_{ij}^{(0)}-\frac{5}{48\sin\xi}R^4\partial_k\partial_kh_{ii}^{(2)}+\frac{1}{24\sin\xi}R^4\partial_i\partial_jh_{ij}^{(2)}\bigg]. \nonumber
\end{align}
The order three derivative terms are
\begin{align}
	\delta S^{ct}_B(\partial^3)&=\frac{1}{24G_6}\int_{S^3}d\Omega_3\frac{r^5}{3!z}\bigg[\hat{x}^k\hat{x}^l\hat{x}^p\hat{x}^q\partial_k\partial_l\partial_p\partial_qh_{ii}-\hat{x}^i\hat{x}^j\hat{x}^k\hat{x}^l\hat{x}^p\hat{x}^q\partial_k\partial_l\partial_p\partial_qh_{ij}\bigg]
\end{align}
Since only integrals with even directional vectors are non vanishing, there is no term of the form  $\partial^3h^{(0)}$. The remaining relevant terms are
\begin{align}
	\delta S^{ct}_B(\partial^3)&=\frac{\Omega_3}{8G_6}\bigg[\frac{7}{1152\sin\xi}R^4\partial_k\partial_k\partial_l\partial_lh_{ii}^{(0)}-\frac{1}{288\sin\xi}R^4\partial_k\partial_k\partial_i\partial_jh_{ij}^{(0)}\bigg].
\end{align}
and the four derivative terms
\begin{align}
	\delta S^{ct}_B(\partial^4)&=-\frac{\Omega_3}{8G_6} \bigg[\bigg(-\frac{7}{4608\sin^3\xi}+\frac{17}{4608\sin\xi}\bigg)R^4\partial_k\partial_k\partial_l\partial_lh_{ii}^{(0)}\nonumber\\
	&+\bigg(\frac{1}{1152\sin^3\xi}-\frac{1}{144\sin\xi}\bigg)R^4\partial_k\partial_k\partial_i\partial_jh_{ij}^{(0)}\bigg].
\end{align}
We have thence obtained all the relevant counterterms. 

For notational simplicity we express $h_{ii}=h$, $\partial_k\partial_k=\partial^2$ and $\partial_i\partial_jh_{ij}=h(\partial^2)$. To compute the renormalized entanglement entropy we arrange all the relevant terms at order $R^n$:

\subsubsection*{Order $R^0$}
\begin{align}
	\delta S_{ren}(R^0)=&\frac{\Omega_3}{8G_6} \bigg[\bigg(\frac{1}{4\sin^3\xi}-\frac{1}{2\sin\xi}\bigg)h^{(0)}+\bigg(-\frac{1}{4\sin^3\xi}+\frac{1}{2\sin\xi}\bigg)h^{(0)}\bigg]
\end{align}
\subsubsection*{Order $R^2$}
\begin{align}
	&\delta S_{ren}(R^2)=\frac{R^2\Omega_3}{8G_6} \bigg[\bigg(\frac{3}{4\sin\xi}-\frac{4}{3}\bigg)h^{(2)}+\bigg(\frac{5}{144\sin^3\xi}-\frac{3}{16\sin\xi}+\frac{2}{9}\bigg)\partial^2h^{(0)}\nonumber\\
	&+\bigg(-\frac{1}{72\sin^3\xi}+\frac{1}{8\sin\xi}-\frac{2}{9}\bigg)h^{(0)}(\partial^2)-\frac{3}{4\sin\xi}h^{(2)}+\frac{5}{72\sin\xi}\partial^2h^{(0)}-\frac{1}{36\sin\xi}h^{(0)}(\partial^2)\nonumber\\
	&+\bigg(-\frac{5}{144\sin^3\xi}+\frac{17}{144\sin\xi}\bigg)\partial^2h^{(0)}+\bigg(+\frac{1}{72\sin^3\xi}-\frac{7}{72\sin\xi}\bigg)h^{(0)}(\partial^2)\bigg]
\end{align}
\subsubsection*{Order $R^4$}
\begin{align}
	&\delta S_{ren}(R^4)=\frac{R^4\Omega_3}{8G_6} \bigg[\frac{8}{15}h^{(4)}+\bigg(\frac{5}{48\sin\xi}-\frac{4}{15}\bigg)\partial^2h^{(2)}+\bigg(-\frac{1}{24\sin\xi}+\frac{2}{15}\bigg)h^{(2)}(\partial^2)\nonumber\\
	&+\bigg(\frac{7}{4608\sin^3\xi}-\frac{5}{384\sin\xi}+\frac{1}{45}\bigg)\partial^2\partial^2h^{(0)}+\bigg(-\frac{1}{1152\sin^3\xi}+\frac{1}{96\sin\xi}-\frac{1}{45}\bigg)\partial^2h^{(0)}(\partial^2)\nonumber\\
	&-\frac{5}{48\sin\xi}\partial^2h^{(2)}+\frac{1}{24\sin\xi}h^{(2)}(\partial^2)+\frac{7}{1152\sin\xi}\partial^2\partial^2h^{(0)}-\frac{1}{288\sin\xi}\partial^2h^{(0)}(\partial^2)\nonumber\\
	&+\bigg(-\frac{7}{4608\sin^3\xi}+\frac{1}{44\sin\xi}\bigg)\partial^2\partial^2h^{(0)}+\bigg(\frac{1}{1152\sin^3\xi}-\frac{1}{144\sin\xi}\bigg)\partial^2h^{(0)}(\partial^2)\bigg]
\end{align}
\subsubsection*{Order $R^5$}
\begin{align}
	\delta S_{ren}(\partial^0)=&\frac{\Omega_3}{8G_6} \frac{5}{24}R^5h^{(5)}_{ii}
\end{align}
Using $(\ref{eq:h2toh0})$ and$(\ref{eq:h4toh2})$ to express all the higher order term in the Fefferman-Graham expansion in terms of lower order ones, we find that below order $R^5$ the variation of renormalized entanglement entropy is zero. More explicitly for each orders we have
\subsubsection*{Order $R^2$}
\begin{align}
	&\delta S_{ren}(R^2)=\frac{R^2\Omega_3}{8G_6} \bigg[\bigg(\frac{1}{8\sin\xi}-\frac{2}{9}\bigg)\big(\partial^2h^{(0)}-h^{(0)}(\partial^2)\big)+\bigg(\frac{5}{144\sin^3\xi}-\frac{3}{16\sin\xi}+\frac{2}{9}\bigg)\partial^2h^{(0)}\nonumber\\
	&+\bigg(-\frac{1}{72\sin^3\xi}+\frac{1}{8\sin\xi}-\frac{2}{9}\bigg)h^{(0)}(\partial^2)-\frac{1}{8\sin\xi}\big(\partial^2h^{(0)}-h^{(0)}(\partial^2)\big)+\frac{5}{72\sin\xi}\partial^2h^{(0)}\nonumber\\
	&-\frac{1}{36\sin\xi}h^{(0)}(\partial^2)+\bigg(-\frac{5}{144\sin^3\xi}+\frac{17}{144\sin\xi}\bigg)\partial^2h^{(0)}+\bigg(+\frac{1}{72\sin^3\xi}-\frac{7}{72\sin\xi}\bigg)h^{(0)}(\partial^2)\bigg]\nonumber\\
	&\delta S_{ren}(R^2)=0
\end{align}
\subsubsection*{Order $R^4$}
\begin{align}
	&\delta S_{ren}(R^4)=\frac{R^4\Omega_3}{8G_6} \bigg[\frac{2}{15}\big(\partial^2h^{(2)}-h^{(2)}(\partial^2)\big)-\bigg(\frac{5}{48\sin\xi}-\frac{4}{15}\bigg)\partial^2h^{(2)}+\bigg(-\frac{1}{24\sin\xi}+\frac{2}{15}\bigg)h^{(2)}(\partial^2)\nonumber\\
	&+\bigg(\frac{7}{4608\sin^3\xi}-\frac{5}{384\sin\xi}+\frac{1}{45}\bigg)\partial^2\partial^2h^{(0)}+\bigg(-\frac{1}{1152\sin^3\xi}+\frac{1}{96\sin\xi}-\frac{1}{45}\bigg)\partial^2h^{(0)}(\partial^2)\nonumber\\
	&-\frac{5}{48\sin\xi}\partial^2h^{(2)}+\frac{1}{24\sin\xi}h^{(2)}(\partial^2)+\frac{7}{1152\sin\xi}\partial^2\partial^2h^{(0)}-\frac{1}{288\sin\xi}\partial^2h^{(0)}(\partial^2)\nonumber\\
	&+\bigg(-\frac{7}{4608\sin^3\xi}+\frac{1}{144\sin\xi}\bigg)\partial^2\partial^2h^{(0)}+\bigg(\frac{1}{1152\sin^3\xi}-\frac{1}{144\sin\xi}\bigg)\partial^2h^{(0)}(\partial^2)\bigg],
\end{align}
gathering all the terms, it simplifies to
\begin{align}
	\delta S_{ren}(R^4)=&\frac{R^4\Omega_3}{8G_6} \bigg[-\frac{1}{45}\partial^2\big(\partial^2h^{(0)}-h^{(0)}(\partial^2)\big)\nonumber\\
	&\frac{1}{45}\partial^2\partial^2h^{(0)}-\frac{1}{45}\partial^2h^{(0)}(\partial^2)\bigg]\nonumber\\
	\delta S_{ren}(R^4)=&0
\end{align}
\subsubsection*{Order $R^5$}
This is the only order $\leq 5$ that is non-vanishing,
\begin{align}
	\delta S_{ren}(\partial^0)=&\frac{\Omega_3}{8G_6} \frac{5}{24}R^5h^{(5)}_{ii},
\end{align}
which matches with $(\ref{eq:dSrenoddd})$ for $d=5$.

\subsection{Renormalized Weyl Integrals}
\label{section:RWI}
This appendix provides the calculation details for section \ref{section:CIF}. In $(\ref{eq:IntW})$ and $(\ref{eq:IntdW})$, the Weyl integrals are given in terms of the Riemann tensor of the boundary of AdS, $\mathcal{R}_{\mu\nu\rho\sigma}$, and we need to expand $\mathcal{R}_{\mu\nu\rho\sigma}$ into linear perturbation $h_{\mu\nu}$. For $\mathcal{R}_{titj}$, we have the following expression
\begin{align}
	\mathcal{R}_{titj}&=\frac{1}{2}\left(\partial_t\partial_jh_{ti}+\partial_t\partial_ih_{tj}-\partial_t\partial_th_{ij}-\partial_i\partial_jh_{tt}\right)\\
	\mathcal{R}_{titj}&=\left(1+\frac{r^2\hat{x}^k\hat{x}^l}{2}\partial_k\partial_l\right)\frac{1}{2}\left(\partial_t\partial_jh^{(0)}_{ti}+\partial_t\partial_ih^{(0)}_{tj}-\partial_t\partial_th^{(0)}_{ij}-\partial_i\partial_jh^{(0)}_{tt}\right)\nonumber\\
	&+\frac{z^2}{2}\left(\partial_t\partial_jh^{(2)}_{ti}+\partial_t\partial_ih^{(2)}_{tj}-\partial_t\partial_th^{(2)}_{ij}-\partial_i\partial_jh^{(2)}_{tt}\right).\label{eq:Rbar}
\end{align}

\subsubsection*{The $d=3$ integral}
For $d=3$, we do not need the subleading term in the Taylor expansion of the metric perturbation as in $(\ref{eq:xi0exp})$. Also in $d=3$ the boundary integral $(\ref{eq:IntdW})$ is vanishing in the limit of $\xi\rightarrow 0$. After we substitute $(\ref{eq:Rbar})$ into $(\ref{eq:IntW})$ the renormalized Weyl integral $\mathcal{W}$ is mixed with different orders in the Fefferman-Graham expansion. Explicitly we have
\begin{align}
	\mathcal{W}&=\int_{\xi}^{\frac{\pi}{2}}du\int_{S^1}d\Omega_1\bigg[-\frac{3R^3\cos u\sin^3 u}{2}h^{(3)}_{tt}+\frac{3R^3\cos^3u\sin u\hat{x}^i\hat{x}^j}{2}\left(h^{(3)}_{tt}\eta_{ij}-h^{(3)}_{ij}\right)\\
	&+\frac{R^2\cos^3u\hat{x}^i\hat{x}^j}{2}\left(\partial_t\partial_jh^{(0)}_{ti}+\partial_t\partial_ih^{(0)}_{tj}-\partial_t\partial_th^{(0)}_{ij}-\partial_i\partial_jh^{(0)}_{tt}+2h^{(2)}_{tt}\eta_{ij}-2h^{(2)}_{ij}\right)\bigg]\nonumber.
\end{align}
After integrating over the circle we obtain
\begin{align}
	\mathcal{W}&=\Omega_1\bigg[-\frac{3R^3}{8}h^{(3)}_{tt}+\frac{3R^3}{16}\left(2h^{(3)}_{tt}-h^{(3)}_{ii}\right)\\
	&+\frac{R^2}{6}\left(2\partial_t\partial_ih^{(0)}_{ti}-\partial_t\partial_th^{(0)}_{ii}-\partial_i\partial_ih^{(0)}_{tt}+4h^{(2)}_{tt}-h^{(2)}_{ii}\right)\bigg]\nonumber.
\end{align}
By solving the Einstein equation order by order in the Fefferman-Graham expansion, we can deduced $h^{(n)}$ for $n<d$ from $h^{(0)}$. This gives
\begin{align}
	h^{(2)}_{ii}=&\frac{1}{2}\Big(\partial_k\partial_kh^{(0)}_{ii}-\partial_i\partial_jh^{(0)}_{ij}\Big)\\
	h^{(2)}_{tt}=&\frac{1}{4}\Big(\partial_k\partial_kh^{(0)}_{ii}-\partial_i\partial_jh^{(0)}_{ij}+\partial_t\partial_th^{(0)}_{ii}+\partial_k\partial_kh^{(0)}_{tt}-2\partial_t\partial_ih^{(0)}_{ti}\Big).
\end{align}
Using the above two expression sfor $h^{(2)}_{\mu\nu}$, we can easily simplify the renormalized Weyl integral as
\begin{align}
	\mathcal{W}&=-\frac{3R^3\Omega_{1}}{16}h^{(3)}_{tt}
\end{align}
which is the result stated in section \ref{section:CIF}.

\subsubsection*{The $d=5$ integral}
For $d=5$, we need the subleading term in the Taylor expansion of the metric perturbation as in $(\ref{eq:xi0exp})$. The relevant metric perturbation derivatives are
\begin{align}
	h_{tt}''&=\left(1+\frac{r^2\hat{x}^k\hat{x}^l}{2}\partial_k\partial_l\right)2h^{(2)}_{tt}+12z^2h^{(4)}_{tt}+20z^3h^{(5)}_{tt}\\
	h_{\mu\nu}'&=\left(1+\frac{r^2\hat{x}^k\hat{x}^l}{2}\partial_k\partial_l\right)2zh^{(2)}_{\mu\nu}+4z^3h^{(4)}_{\mu\nu}+5z^4h^{(5)}_{\mu\nu}\\
	\partial_{\mu}h_{\nu\rho}'&=2zr\hat{x}^k\partial_k\partial_{\mu}h_{\nu\rho}^{(2)},
\end{align}
where $'$ represent the radial derivative $\partial_z$. The renormalized Weyl integral $\mathcal{W}$ becomes
\begin{align}
	\mathcal{W}&=\int_{S^3}d\Omega_3\bigg[-\frac{8R^4}{15}h^{(4)}_{tt}-\frac{15R^5}{24}h^{(5)}_{tt}\\
	&+\left(-\frac{4R^2}{3}\hat{x}^i\hat{x}^j-\frac{4R^2}{5}\hat{x}^i\hat{x}^j\hat{x}^k\hat{x}^l\partial_k\partial_l\right)\nonumber\\
	&\quad\quad\times\left(\partial_t\partial_jh^{(0)}_{ti}+\partial_t\partial_ih^{(0)}_{tj}-\partial_t\partial_th^{(0)}_{ij}-\partial_i\partial_jh^{(0)}_{tt}+2h^{(2)}_{tt}\eta_{ij}+2h^{(2)}_{ij}\eta_{tt}\right)\nonumber\\
	&+\frac{4R^4}{15}\hat{x}^i\hat{x}^j\left(\partial_t\partial_jh^{(2)}_{ti}+\partial_t\partial_ih^{(2)}_{tj}-\partial_t\partial_th^{(2)}_{ij}-\partial_i\partial_jh^{(2)}_{tt}+4h^{(4)}_{tt}\eta_{ij}+4h^{(4)}_{ij}\eta_{tt}\right)\nonumber\\
	&+\frac{5R^5}{12}\hat{x}^i\hat{x}^j\left(h^{(5)}_{tt}\eta_{ij}+h^{(5)}_{ij}\eta_{tt}\right)\nonumber\\
	&+\frac{16R^4}{15}\hat{x}^i\hat{x}^k\partial_k\left(\partial_th^{(2)}_{ti}-\partial_ih^{(2)}_{tt}\right)\bigg].\nonumber
\end{align}
After integrating this over the $S^3$ using $(\ref{eq:OmegaxiInt})$ we get
\begin{align}
	\mathcal{W}=\Omega_3\bigg[\frac{R^2}{3}\Big(&-2\partial_t\partial_ih^{(0)}_{ti}+\partial_t\partial_th^{(0)}_{ii}+\partial_i\partial_ih^{(0)}_{tt}-8h^{(2)}_{tt}+2h^{(2)}_{ii}\Big)\\
	+\frac{R^4}{30}\Big(&-6\partial_t\partial_k\partial_k\partial_ih^{(0)}_{ti}+\partial_t\partial_t\partial_k\partial_kh^{(0)}_{ii}+2\partial_t\partial_t\partial_i\partial_jh^{(0)}_{ij}+3\partial_k\partial_k\partial_l\partial_lh^{(0)}_{tt}\nonumber\\
	&-22\partial_k\partial_kh^{(2)}_{tt}-2\partial_t\partial_th^{(2)}_{ii}+2\partial_k\partial_kh^{(2)}_{ii}+4\partial_i\partial_jh^{(2)}_{ij}+12\partial_t\partial_ih^{(2)}_{ti}\nonumber\\
	&+16h^{(4)}_{tt}-8h^{(4)}_{ii}\Big)\nonumber\\
	+\frac{R^5}{48}\Big(&-10h^{(5)}_{tt}-5h^{(5)}_{ii}\Big)\bigg]. \nonumber
\end{align}
Following the lower dimensional case, we need to related the terms of different orders in Fefferman-Grahm expansion to see the cancellation between divergent pieces. By solving the Einstein equations order by order in the Fefferman-Graham expansion, we can deduced $h^{(n)}$ for $n<d$ from $h^{(0)}$. Hence,
\begin{align}
	h^{(2)}_{ii}=&\frac{1}{6}\Big(\partial_k\partial_kh^{(0)}_{ii}-\partial_i\partial_jh^{(0)}_{ij}\Big)\\
	h^{(2)}_{tt}=&\frac{1}{24}\Big(\partial_k\partial_kh^{(0)}_{ii}-\partial_i\partial_jh^{(0)}_{ij}+3\partial_t\partial_th^{(0)}_{ii}+3\partial_k\partial_kh^{(0)}_{tt}-6\partial_t\partial_ih^{(0)}_{ti}\Big)\\
	h^{(2)}_{ti}=&\frac{1}{6}\Big(\partial_t\partial_ih^{(0)}_{jj}-\partial_t\partial_jh^{(0)}_{ij}-\partial_i\partial_jh^{(0)}_{tj}+\partial_k\partial_kh^{(0)}_{ti}\Big)\\
	h^{(4)}_{ii}=&\frac{1}{4}\Big(\partial_k\partial_kh^{(2)}_{ii}-\partial_i\partial_jh^{(2)}_{ij}\Big)\\
	h^{(4)}_{tt}=&\frac{1}{4}\Big(\partial_k\partial_kh^{(2)}_{tt}-\partial_t\partial_th^{(2)}_{ii}\Big)\\
	\partial_i\partial_jh^{(2)}_{ij}=&\frac{1}{24}\Big(3\partial_k\partial_k\partial_l\partial_lh^{(0)}_{ii}-3\partial_k\partial_k\partial_i\partial_jh^{(0)}_{ij}+\partial_t\partial_t\partial_k\partial_kh^{(0)}_{ii}\\
	&\quad-3\partial_k\partial_k\partial_l\partial_lh^{(0)}_{tt}+63\partial_t\partial_k\partial_k\partial_ih^{(0)}_{ti}-4\partial_t\partial_t\partial_i\partial_jh^{(0)}_{ij}\Big)\nonumber\\
	\partial_t\partial_ih^{(2)}_{it}=&\frac{1}{6}\Big(\partial_t\partial_t\partial_k\partial_kh^{(0)}_{ii}-\partial_t\partial_t\partial_i\partial_jh^{(0)}_{ij}\Big)
\end{align}
Substituting the above expressions for $h^{(n)}_{\mu\nu}$, we can easily simplify the renormalized Weyl integral as
\begin{align}
	\mathcal{W}&=-\frac{5R^5\Omega_{3}}{16}h^{(5)}_{tt}
\end{align}
which is the result stated in section \ref{section:CIF}.

\subsection{Variations in $AdS_5$}
\label{section:EVCs}
Here we will fill in the computational details of section \ref{section:CDd4} to show that the divergences of the variation of regularized entanglement entropy and variation of the counterterms match. In $(\ref{eq:dSregd4})$, the variation of regularized entanglement entropy was given in terms of both $h^{(0)}_{\mu\nu}$ and $h^{(2)}_{\mu\nu}$. In order to compare with the counterterm we will first express $h^{(2)}_{\mu\nu}$  as function of $h^{(0)}_{\mu\nu}$. 

Since the perturbed metric of $AdS_5$ satisfies the Einstein equation, the metric perturbation can be expanded and solved order by order in an asymptotic series. Using the results in \cite{deHaro2001},
\begin{equation}
	g^{(2)}_{\mu\nu}=-\frac{1}{2}\left(\mathcal{R}_{\mu\nu}[g^{(0)}]-\frac{1}{6}\mathcal{R}[g^{(0)}]g^{(0)}_{\mu\nu}\right).
\end{equation}
In $d=4$, we only need to consider terms of order up to $z^2$, hence we have
\begin{equation}
	h^{(2)}_{\mu\nu}=-\frac{1}{2}\left(\mathcal{R}_{\mu\nu}[\eta +h^{(0)}]-\frac{1}{6}\mathcal{R}[\eta +h^{(0)}](\eta_{\mu\nu}+h^{(0)}_{\mu\nu})\right) \label{eq:h2Rd4}.
\end{equation} 
Since the Ricci tensor of $\eta_{\mu\nu}$ vanishes, to first order of $h$ the Ricci tensor of $g^{(0)}$ is just the first order variation. For our interests the relevant terms then become
\begin{align}
	\frac{\sin\theta}{z}h^{(2)}_{\theta\theta} &=- \frac{\sin\theta}{2z}\delta\mathcal{R}_{\theta\theta}+\frac{R^2\sin\theta}{12z}\delta\mathcal{R}
	\\
	\frac{1}{\sin\theta z}h^{(2)}_{\phi\phi} &= -\frac{1}{2\sin\theta z}\delta\mathcal{R}_{\phi\phi}+\frac{R^2\sin\theta}{12z}\delta\mathcal{R}
\end{align} 
Using this expression, we can write the divergent term of the regularized entanglement entropy in $(\ref{eq:dSregdivd4})$ in terms of $h^{(0)}_{\mu\nu}$. 

Now we need to evaluate the variation of the counter terms and check all the divergences are cancelled. The induced metric $\widetilde\gamma$ of the regularised entangling surface $\partial\tilde{B}_{\epsilon}=\tilde{B}\vert_{z=\epsilon}$ is
\begin{equation}
	ds^2=\frac{R^2-\epsilon^2}{\epsilon^2}\big(d\theta^2+\sin^2\theta d\phi^2\big).
\end{equation}
Then the variation of the volume form is
\begin{align}
	\delta\sqrt{\widetilde\gamma}&= \frac{1}{2}\sqrt{\widetilde\gamma}\widetilde\gamma^{ij}\delta\widetilde\gamma_{ij}
	\\
	&=\frac{\sin\theta}{2}\bigg(\frac{1}{\epsilon^2}h^{(0)}_{\theta\theta}+\frac{1}{\epsilon^2\sin^2\theta}h^{(0)}_{\phi\phi}\bigg)  \nonumber
\end{align}
To calculate the variation of the counterterms we need to embed $\big(\partial\tilde{B},\widetilde\gamma\big)$ into $\big(AdS_5\vert_{z=\epsilon},\widetilde G\big)$ and find its unit normals which are
\begin{align}
	n_1=\frac{dt}{\epsilon},\quad\quad\quad\quad n_2=\frac{dr}{\epsilon}.
\end{align} 
The extrinsic curvature $K_{\mu\nu}$ is defined by $\frac{1}{2}\mathcal{L}_n\widetilde\gamma_{\mu\nu}$. The trace of the extrinsic curvature is then
\begin{align}
	K =\widetilde G^{\mu\nu}K_{\mu\nu}
	=\frac{1}{2}\widetilde\gamma^{ij}\mathcal{L}_n\widetilde\gamma_{ij}	=\mathcal{L}_n\ln \sqrt{\widetilde\gamma}
\end{align}
In time independent situations, $K_1$ vanishes. The extrinsic curvature corresponding to the radial normal is
\begin{align}
	K_2 =\epsilon\partial_r\ln \frac{r^2\sin\theta}{\epsilon^2}
		=\frac{2\epsilon}{r}
\end{align}
Although we are only taking linear order of metric variation which leaves the direction of the normals unchanged, the coefficients of unit normals $n_a$ vary. Specifically for $n_2$
\begin{align}
	\delta n_2= \delta (n^r\partial_r) = \delta \left  (\sqrt{\frac{1}{\widetilde G_{rr}}}\partial_r \right ) 
	= -\frac{\delta \widetilde G_{rr}}{2{\widetilde G_{rr}}^{\frac{3}{2}}}\partial_r
\end{align}
The variation of $K_2$ can be related to the variation of the metric $g$ as
\begin{align}
	\delta K_2 &= \delta n_2\big(\ln\sqrt{\widetilde\gamma}\big)+n_2\delta\big(\ln\sqrt{\widetilde\gamma}\big)
	\\
	&= -\frac{\delta \widetilde G_{rr}}{2{\widetilde G_{rr}}^{\frac{3}{2}}}\partial_r \ln\sqrt{\widetilde\gamma}+\epsilon\partial_r\bigg(\frac{1}{2}\widetilde\gamma^{ij}\delta\widetilde\gamma_{ij}\bigg)\nonumber
	\\
	&= -\frac{\epsilon}{r}h^{(0)}_{rr}-\frac{\epsilon}{r^3}h^{(0)}_{\theta\theta}-\frac{\epsilon}{r^3\sin^2\theta}h^{(0)}_{\phi\phi}+\frac{\epsilon}{2r^2}\partial_rh^{(0)}_{\theta\theta}+\frac{\epsilon}{2r^2\sin^2\theta}\partial_rh^{(0)}_{\phi\phi} \nonumber
\end{align}
Keeping only the divergence, the structure of the variation of the third term in $(\ref{eq:IntSct})$ is
\begin{align}
	\delta\big(\sqrt{\widetilde\gamma}k^2\big)&= \delta\big(\sqrt{\widetilde\gamma}\big)K_2^2+\sqrt{\widetilde\gamma}\delta\big(K_2^2\big)\label{eq:K226}
\end{align}
Separating the terms in $(\ref{eq:K226})$,
\begin{align}
	\delta\big(\sqrt{\widetilde\gamma}\big)K_2^2=\ &\sin\theta \Big(\frac{2}{R^2}h^{(0)}_{\theta\theta}+\frac{2}{R^2\sin^2\theta}h^{(0)}_{\phi\phi}\Big)
	\\
	2\sqrt{\widetilde\gamma}K_2\delta K_2=\ &\sin\theta \Big(-4h^{(0)}_{rr}-\frac{4}{R^2}h^{(0)}_{\theta\theta}-\frac{4}{R^2\sin^2\theta}h^{(0)}_{\phi\phi}\nonumber
	\\
	&+\frac{2}{R}\partial_rh^{(0)}_{\theta\theta}+\frac{2}{R\sin^2\theta}\partial_r h^{(0)}_{\phi\phi}\Big)
\end{align}
The remaining terms are the variation of Ricci scalar and projected Ricci tensor. Note that $\mathcal{R}_{aa}$ in \cite{Fursaev_1995,Fursaev_2013} was given in a Euclidean setting. After Wick rotating the normal direction back to Lorentzian signature, we obtain
\begin{align}
	\mathcal{R}_{aa}&=\mathcal{R}_{\mu\nu}(in^{\mu}_1)(in^{\nu}_1)+\mathcal{R}_{\mu\nu}n^{\mu}_2n^{\nu}_2\nonumber\\
	\mathcal{R}_{aa}&=z^2(-\mathcal{R}_{tt}+\mathcal{R}_{rr}).
\end{align} 
Again we use the fact that our unperturbed spacetime is flat so the variation of these terms is
\begin{align}
	\delta\big(\mathcal{R}_{aa}-\frac{2}{3}\mathcal{R}\big) &=\delta\mathcal{R}_{\mu\nu}n_a^{\mu}n_a^{\nu}-\frac{2}{3}\delta\mathcal{R}\nonumber\\
	&=-\frac{z^2}{3}\delta\mathcal{R}_{tt}+\frac{z^2}{3}\delta\mathcal{R}_{rr}-\frac{2z^2}{3}\bigg(\frac{1}{r^2}\delta\mathcal{R}_{\theta\theta}+\frac{1}{r^2\sin^2\theta}\delta\mathcal{R}_{\phi\phi}\bigg)\nonumber
	\\
	&= \frac{z^2}{3}\delta\mathcal{R}-\frac{z^2}{r^2}\delta\mathcal{R}_{\theta\theta}-\frac{z^2}{r^2\sin^2\theta}\delta\mathcal{R}_{\phi\phi}\label{eq:dRd4}
\end{align}
notice there is an abuse of notation where in the first line $\delta\mathcal{R}={\widetilde G}^{\mu\nu}\delta\mathcal{R}_{\mu\nu}$ and in the last line $\delta\mathcal{R}={g^{(0)}}^{\mu\nu}\delta\mathcal{R}_{\mu\nu}$. Using $(\ref{eq:h2Rd4})$ we can write $(\ref{eq:dRd4})$ in terms of $h^{(2)}$,
\begin{align}
	\delta\mathcal{R}_{\mu\nu}n_a^{\mu}n_a^{\nu}-\frac{2}{3}\delta\mathcal{R} = \frac{2z^2}{r^2}h^{(2)}_{\theta\theta}+\frac{2z^2}{r^2\sin^2\theta}h^{(2)}_{\phi\phi}
\end{align}
The divergent contributions to the counterterms are
\begin{align}
	(\delta S^{ct}_B)^{div}=&\frac{1}{8G_5}\int_{S^2} d\Omega_2 \bigg[\frac{1}{2\epsilon}\Big(h^{(0)}_{\theta\theta}+\frac{1}{\sin^2\theta}h^{(0)}_{\phi\phi}\Big)+\frac{\ln\epsilon}{2}\Big(\frac{1}{R^2}h^{(0)}_{\theta\theta}+\frac{1}{R^2\sin^2\theta}h^{(0)}_{\phi\phi} \nonumber\\
	&-2h^{(0)}_{rr}-\frac{2}{R^2}h^{(0)}_{\theta\theta}-\frac{2}{R^2\sin^2\theta}h^{(0)}_{\phi\phi}+\frac{1}{R}\partial_rh^{(0)}_{\theta\theta}+\frac{1}{R\sin^2\theta}\partial_r h^{(0)}_{\phi\phi}\nonumber\\
	& -2h^{(2)}_{\theta\theta}-\frac{2}{\sin^2\theta}h^{(2)}_{\phi\phi}\Big)\bigg]\label{eq:dSctdivd4}
\end{align}
which matches with $(\ref{eq:dSregdivd4})$.

\section{Asymptotic expansions and integrals}
	
	
\subsection{Dilatation Eigenfunction Expansion}
\label{section:DEE}
Under dilatation transformation $x^{\mu} \rightarrow \Omega x^{\mu}$, the boundary metric transforms as
\begin{align}
	\gamma_{\mu\nu}\rightarrow \Omega^2\gamma_{\mu\nu}.
\end{align}
In terms of infinitesimal operator 
\begin{align}
	\gamma_{\mu\nu}\rightarrow (1+\epsilon\delta_D)\gamma_{\mu\nu}
\end{align}
where $1+\epsilon=\Omega$. The dilatation operator for the boundary metric $\gamma$ is then
\begin{align}
\delta_D =2  \int d^d x \gamma_{\mu\nu}\frac{\delta}{\delta \gamma_{\mu\nu}}
\end{align}
which replaces $\gamma_{\mu\nu}$ with $2\gamma_{\mu\nu}$ as the dilatation weight of the metric is $2$. The dilatation operator in general contains all fields that transform non-trivially under dilatation. For our purposes we will actually only consider pure gravitational systems so the dilatation operator only contains the metric $\gamma_{\mu\nu}$. In the radial gauge the extrinsic curvature depends only on $\gamma_{\mu\nu}$ and it curvature can be expanded in Fefferman-Graham coefficients as
\begin{align}
	K_{\mu\nu}[\gamma]&=-\frac{z}{2}\partial_{z}\gamma_{\mu\nu} \nonumber \\
	&=z^{-2}g_{\mu\nu}^{(0)}-z^{2}g_{\mu\nu}^{(4)}+\cdots+\frac{2-d}{2}\tilde{g}_{\mu\nu}^{(d)}\log z^2-\tilde{g}_{\mu\nu}^{(d)}+\frac{2-d}{2}g_{\mu\nu}^{(d)}+\cdots\label{eq:KFGE}
\end{align} 
and in dilatation eigenfunction expansion 
\begin{align}
	K_{\mu\nu}[\gamma]=K_{(0)\;\mu\nu}[\gamma]+K_{(2)\;\mu\nu}[\gamma]+\cdots+\tilde{K}_{(d)\;\mu\nu}[\gamma]\log z^2+K_{(d)\;\mu\nu}[\gamma]+\cdots\label{eq:KDEE}
\end{align}
where the logarithmic terms are only present for even $d$. The dilatation eigenfunctions transform according to their order: we have homogenous transformations for $K_{(n<d)\;\mu\nu}$ and $\tilde{K}_{(d)\;\mu\nu}$ ,
\begin{align}
	\delta_DK_{(n)\;\mu\nu}=(2-n)K_{(n)\;\mu\nu}\label{eq:dKhom}
\end{align}
and inhomogenous transformations for $K_{(d)\;\mu\nu}$,
\begin{align}
	\delta_DK_{(d)\;\mu\nu}=(2-d)K_{(d)\;\mu\nu}-2\tilde{K}_{(d)\;\mu\nu}.\label{eq:dKinhom}
\end{align}
The origin of the inhomogenous transformation will become obvious when we relate the two expansions. To do that we need to express the radial derivative in terms of functional derivative of $\gamma_{\mu\nu}$
\begin{align}
	-z\partial_z=-z\partial_z\vert_{\gamma_{\mu\nu}=const}+\int d^dx 2K_{\mu\nu}[\gamma]\frac{\delta}{\delta \gamma_{\mu\nu}}.\label{eq:radderK}
\end{align}
Let us drop the first term as we are considering field that does not depend on $z$ explicitly. We know from $(\ref{eq:dKhom})$ that the zero$^{th}$ term in the dilatation eigenfunction expansion $K_{(0)\;\mu\nu}[\gamma]$ is proportional to $\gamma_{\mu\nu}$ then comparing with the leading term in $(\ref{eq:KFGE})$ we can deduce
\begin{align}
	K_{(0)\;\mu\nu}[\gamma]=\gamma_{\mu\nu}.
\end{align}
We see that expanding the extrinsic curvature in $(\ref{eq:radderK})$ the radial derivative is related to the dilatation operator by
\begin{align}
	-z\partial_z=\delta_D+\delta_{(2)}+\cdots
\end{align}
where
\begin{align}
	\delta_{(n)}=\int d^dx 2K_{(n)\;\mu\nu}[\gamma]\frac{\delta}{\delta \gamma_{\mu\nu}}.
\end{align}
Taylor expanding the $K_{(n)\;\mu\nu}[\gamma]$ about $z^{-2}g^{(0)}_{\mu\nu}$
\begin{align}
	K_{(n)\;\mu\nu}[\gamma]=K_{(n)\;\mu\nu}[z^{-2}g^{(0)}]+\int g^{(2)}_{\rho\sigma}\frac{\delta K_{(n)\;\mu\nu}}{\delta \gamma_{\mu\nu}}\vert_{\gamma=z^{-2}g^{(0)}}+\cdots\label{eq:KnTaylor}
\end{align}
Since $K_{(n)\mu\nu}[z^{-2}g^{(0)}]$ are also dilatation eigenfunctions, we can rescale the metric to get rid of the implicit $z$ dependence. Using the integrated transformation of $(\ref{eq:dKhom})$ for $K_{(n<d)\;\mu\nu}$ and $\tilde{K}_{(d)\;\mu\nu}$,
\begin{align}
	K_{(n)\;\mu\nu}[z^{-2}g^{(0)}]=z^{n-2}K_{(n)\;\mu\nu}[g^{(0)}].
\end{align}
Now that we know at the leading order we can write the dilatation operator in terms of the radial derivative $\delta_D\sim-z\partial_z$ for implicit $z$ dependence terms then,
\begin{align}
	-z\partial_z\left(\tilde{K}_{(d)\;\mu\nu}[\gamma]\log z^2+K_{(d)\;\mu\nu}[\gamma]\right)&\sim\delta_D\left(\tilde{K}_{(d)\;\mu\nu}[\gamma]\log z^2+K_{(d)\;\mu\nu}[\gamma]\right).
\end{align}
Note the bracket term depends on $z$ through $\gamma$ only because of the diffeomorphism invariance of the bulk action. Expanding the bracket we get
\begin{align}
	-z\partial_z\tilde{K}_{(d)\;\mu\nu}\log z^2-2\tilde{K}_{(d)\;\mu\nu}-z\partial_zK_{(d)\;\mu\nu} & \sim \delta_D\tilde{K}_{(d)\;\mu\nu}\log z^2+\delta_DK_{(d)\;\mu\nu}\label{eq:dzKddDKd}
\end{align}
and for all $n$ at leading order of $z$ we have
\begin{align}
	-z\partial_zK_{(n)\;\mu\nu}[z^{-2}g^{(0)}]\sim(2-n)K_{(n)\;\mu\nu}[z^{-2}g^{(0)}].
\end{align}
Hence matching the leading order terms in $(\ref{eq:dzKddDKd})$ we get back the inhomogenous transformation in $(\ref{eq:dKinhom})$. After all the steps above we arrive at the $z$ expansion of the dilatation eigenfunctions,
\begin{align}
	K_{(0)\;\mu\nu}[\gamma]&=z^{-2}g^{(0)}_{\mu\nu}+g^{(2)}_{\mu\nu}+\cdots\\
	K_{(2)\;\mu\nu}[\gamma]&=K_{(2)\;\mu\nu}[g^{(0)}]+z^2\int g^{(2)}_{\rho\sigma}\frac{\delta K_{(2)\;\mu\nu}}{\delta g^{(0)}_{\mu\nu}}+\cdots
\end{align}
and so on. The final steps to relate the Fefferman-Graham coefficients to the dilatation eigenfunctions is to express $K_{(n)\;\mu\nu}[g^{(0)}]$ in terms of $g^{(m)}_{\mu\nu}$. In general $K_{(n)\;\mu\nu}[g^{(0)}]$ are obtained by comparing with the $z^{n-2}$ in $(\ref{eq:KFGE})$, i.e. for $d>4$
\begin{align}
	z^0:\quad K_{\mu\nu}[\gamma]&=g^{(2)}_{\mu\nu}+K_{(2)\;\mu\nu}[g^{(0)}]\\
	&=0\nonumber\\
	z^2:\quad K_{\mu\nu}[\gamma]&=z^2g^{(4)}_{\mu\nu}+\int g^{(2)}_{\rho\sigma}\frac{\delta K_{(2)\;\mu\nu}}{\delta g^{(0)}_{\mu\nu}}+z^2K_{(4)\;\mu\nu}[g^{(0)}]\\
	&=-z^2g^{(4)}_{\mu\nu}\nonumber
\end{align}
so we get
\begin{align}
	K_{(2)\;\mu\nu}[g^{(0)}]&=-g^{(2)}_{\mu\nu}\\
K_{(4)\;\mu\nu}[g^{(0)}]&=-2g^{(4)}_{\mu\nu}+\int g^{(2)}_{\rho\sigma}\frac{\delta g^{(2)}_{\mu\nu}}{\delta g^{(0)}_{\mu\nu}}.
\end{align}
 For larger $n$, there will be functional derivative terms coming from the Taylor expansion in $(\ref{eq:KnTaylor})$ at the $z^n$ order, for example
 \begin{align}
 	z^{n-2}\int g^{(n-2)}\cdot\frac{\delta K_{(2)}}{\delta g^{(0)}},\cdots,z^{n-2}\left(\int\cdots\int\right)^m (g^{(p_1)}\cdots g^{(p_m)})\cdot(\frac{\delta }{\delta g^{(0)}}\cdots\frac{\delta }{\delta g^{(0)}})K_{(q)},\cdots
  \end{align}
where $q+p_1+\cdots+p_m=n$. Of course when onshell all $g^{n<d}_{\mu\nu}$ and $\tilde{g}^{(d)}_{\mu\nu}$ are functions of $g^{(0)}_{\mu\nu}$. Order by order, we can write all the dilatation eigenfunctions in terms of the terms Fefferman-Graham expansion.

	\subsection{Volume Integrals of $h_{(n)}$}
	\label{section:Inth}
	This appendix will address some technical steps omitted in section $\ref{section:EXs}$. In those examples, the integral term in $(\ref{eq:IntDdSct})$ is given by a volume integral over $B_{\epsilon}$. We know the counterterm is given by surface integral over the regulated boundary of the entangling surface $\partial\tilde{B}_{\epsilon}$. Since $\partial\tilde{B}_{\epsilon}=\partial{B}_{\epsilon}$, we need to express the integral term as a surface integral over $\partial{B}_{\epsilon}$. In the following we will show the relation between volume and surface integrals of the terms in the Fefferman-Graham expansion.  
	
	The leading term in the Fefferman-Graham expansion, $h_{(0)\,\mu\nu}$, is part of the boundary data hence should be treated as independent variable. Nonetheless, we can express them as combination of total derivatives and moment density of the derivatives of $h_{(0)\,\mu\nu}$. For the spatial trace $h_{(0)\,ii}$ we have
	\begin{align}
		(d-2)h_{(0)\,ii}&=\partial_i\left({x}^ih_{(0)\,jj}-{x}^jh_{(0)\,ij}-\frac{\vec{x}^2}{2}\left(\partial_ih_{(0)\,jj}-\partial_ih_{(0)\,ij}\right)\right)\\
		&+\frac{\vec{x}^2}{2}\left(\partial_i\partial_ih_{(0)\,jj}-\partial_i\partial_jh_{(0)\,ij}\right).\nonumber
	\end{align}
	From the Einstein equation the last bracket above is related to $h_{(2)\,ii}$ by  $(\ref{eq:h2toh0})$ and we get	
	\begin{align}
		h_{(0)\,ii}&=\frac{1}{d-2}\partial_i\left({x}^ih_{(0)\,jj}-{x}^jh_{(0)\,ij}-\frac{\vec{x}^2}{2}\left(\partial_ih_{(0)\,jj}-\partial_ih_{(0)\,ij}\right)\right)+\vec{x}^2h_{(2)\,ii}.
	\end{align}
	Integrating over $B_{\epsilon}$, we obtain a surface integral and a second moment of $h_{(2)\,ii}$ over $B_{\epsilon}$,
	\begin{align}
		\int_{{B}_{\epsilon}}d^{d-1}x\,h_{(0)\,ii}&=\frac{1}{d-2}\int_{\partial{B}_{\epsilon}}d^{d-2}x\hat{{x}}^i\left({x}^ih_{(0)\,jj}-{x}^jh_{(0)\,ij}-\frac{\vec{x}^2}{2}\left(\partial_ih_{(0)\,jj}-\partial_ih_{(0)\,ij}\right)\right)\\
		&+\int_{{B}_{\epsilon}}d^{d-1}x\,\vec{x}^2h_{(2)\,ii}.\nonumber
	\end{align}
	Since $\partial B_{\epsilon}$ is a sphere of radius $\vec{x}^2=R^2-\epsilon^2$, we can reverse the surface integral for the last terms in the first line to get back a volume integral of $h_{(2)\,ii}$ over $B_{\epsilon}$. 
	\begin{align}
		\int_{{B}_{\epsilon}}d^{d-1}x\,h_{(0)\,ii}&=\frac{(R^2-\epsilon^2)^{\frac{d-1}{2}}}{d-2}\int_{\partial{B}_{\epsilon}}d\Omega_{d-2}\,\left(h_{(0)\,ii}-\hat{x}^i\hat{x}^jh_{(0)\,ij}\right)\\
		&-\int_{{B}_{\epsilon}}d^{d-1}x\,(R^2-\epsilon^2-\vec{x}^2)h_{(2)\,ii}.\nonumber
	\end{align}
Gathering the terms that appear in the integral correction terms we get
\begin{align}
	\int_{{B}_{\epsilon}}d^{d-1}x\,\left(h_{(0)\,ii}+(R^2-\vec{x}^2)h_{(2)\,ii}\right)&=\frac{(R^2-\epsilon^2)^{\frac{d-1}{2}}}{d-2}\int_{S^{d-2}}d\Omega_{d-2}\,\left(h_{(0)\,ii}-\hat{x}^i\hat{x}^jh_{(0)\,ij}\right)\\\label{eq:Inth0h2}
	&+	\int_{{B}_{\epsilon}}d^{d-1}x\,\epsilon^2h_{(2)\,ii}.\nonumber
\end{align}
	For $d>3$, the integral correction term contains higher order terms in the Fefferman-Graham expansion. In general, the $n^{th}$ order terms are second derivative of $(n-2)^{th}$. The following expressions for evaluating volume integral of a generic second derivative of a tensor will be useful later on. First the second moment of such a derivative is
	\begin{align}
		\int_{{B}_{\epsilon}}d^{d-1}x\,\vec{x}^2\partial_i\partial_jA_{ij}=\int_{{B}_{\epsilon}}d^{d-1}x\,\partial_i\left(\vec{x}^2\partial_jA_{ij}-2\vec{x}^jA_{ij}\right)+2A_{ii}
	\end{align}
	then the shifted second moment is
	\begin{align}
	\int_{{B}_{\epsilon}}d^{d-1}x\,(R^2-\vec{x}^2)\partial_i\partial_jA_{ij}&=\int_{{B}_{\epsilon}}d^{d-1}x\,\partial_i\left(R^2\partial_jA_{ij}-\vec{x}^2\partial_jA_{ij}+2\vec{x}^jA_{ij}\right)-2A_{ii}\nonumber\\
	&=\epsilon^2\int_{\partial{B}_{\epsilon}}d^{d-2}x\,\hat{x}^i\partial_jA_{ij}+\int_{\partial{B}_{\epsilon}}d^{d-2}x\,2\hat{x}^ix^jA_{ij}-\int_{{B}_{\epsilon}}d^{d-1}x\,2A_{ii}\nonumber\\
	&=2r^{d-1}\int_{S^{d-2}}d\Omega_{d-2}\,\hat{x}^i\hat{x}^jA_{ij}-2\int_{{B}_{\epsilon}}d^{d-1}x\,A_{ii}\\\label{eq:IntddA}
	&+\epsilon^2r^{d-2}\int_{S^{d-2}}d\Omega_{d-2}\,\hat{x}^i\partial_jA_{ij}.\nonumber
   \end{align}	
	The $d=4$ examples in section $\ref{section:AlAdS5ex}$, we have integral of the form of $(\ref{eq:IntddA})$ where
	\begin{align}
		\tilde{h}_{(4)\,ii}=\partial_i\partial_jA_{ij}
	\end{align}
	and
	\begin{align}
		A_{ij}=\frac{1}{8}\left(h_{(2)\,ij}-\delta_{ij}h_{(2)\,kk}\right).
	\end{align}
	
	Neglecting the $O(\epsilon^2)$ term since they are irrelevant in $(\ref{eq:IntBdDd4})$ we get,
	\begin{align}
		\int_{{B}_{\epsilon}}d^3x\,(R^2-\vec{x}^2)\tilde{h}_{(4)\,ii}&=\frac{(R^2-\epsilon^2)^{\frac{3}{2}}}{4}\int_{S^2}d\Omega_{2}\,(\hat{x}^i\hat{x}^jh_{(2)\,ij}-h_{(2)\,ii})\label{eq:Inth4h2}\\
		&+\frac{1}{2}\int_{{B}_{\epsilon}}d^3x\,h_{(2)\,ii}.\nonumber
	\end{align}
	As seen in $(\ref{eq:h2toh0})$, $h_{(2)\,ii}$ is the second derivative of $h_{(0)\,ij}$, the last volume integral can be easily turned into surface integral,
	\begin{align}
		\int_{{B}_{\epsilon}}d^3x\, h_{(2)\,ii}&=\frac{(R^2-\epsilon^2)^{\frac{1}{2}}}{4}\int_{S^2}d\Omega_{2}\hat{x}^i(\partial_ih_{(0)\,jj}-\partial_jh_{(0)\,ij})\nonumber\\
		&=\frac{r}{4}\int_{S^2}d\Omega_{2}\hat{x}^i\bigg(\partial_ih_{(0)\,jj}-\partial_rh_{(0)\,ir}\nonumber\\
		&-\frac{1}{r^2}\partial_{\theta}h_{(0)\,\theta i}-\frac{1}{r^2\sin^2\theta}\partial_{\phi}h_{(0)\,\phi i}-\frac{\cos\theta}{r^2\sin\theta}h_{(0)\,\theta i}-\frac{2}{r}h_{(0)\,ri}\bigg)\nonumber\\
		&=\frac{r}{4}\int_{S^2}d\phi d\theta\sin\theta \bigg(\hat{x}^i\partial_ih_{(0)\,jj}-\hat{x}^i\hat{x}^j\hat{x}^k\partial_kh_{(0)\,ij}\label{eq:Inth2d4}\\
		&+\frac{1}{r^3}h_{(0)\,\theta \theta}+\frac{1}{r^3\sin^2\theta}h_{(0)\,\phi \phi}-\frac{2\hat{x}^i\hat{x}^j}{r}h_{(0)\,ij}\bigg)\nonumber
	\end{align}
	where we went from the first line to the second line by evaluating $\partial_jh_{(0)\,ij}$ in polar coordinates. From the second line to the third line we integrate by parts and we transform $r$ coordinate to Cartesian. Finally we can transform the angular coordinate into Cartesian coordinates,
	\begin{align}
		\int_{{B}_{\epsilon}}d^3x\, h_{(2)\,ii}=\frac{r}{4}\int_{S^2}d\Omega_{2}&\bigg[h_{(0)\,ii}-3\hat{x}^i\hat{x}^jh_{(0)\,ij}\label{eq:Inth2h0}\\
		&+x^j\partial_jh_{(0)\,ii}-\hat{x}^i\hat{x}^jx^k\partial_kh_{(0)\,ij}\bigg].\nonumber
	\end{align}
	For the $d=4$ example in section $\ref{section:AlAdS6ex}$, we have integrals of the form of $(\ref{eq:IntddA})$ where	
	\begin{align}
		{h}_{(4)\,ii}=\partial_i\partial_jB_{ij}
	\end{align}
	and
	\begin{align}
		B_{ij}=\frac{1}{4}\left(\delta_{ij}h_{(2)\,kk}-h_{(2)\,ij}\right).
	\end{align}
	Neglecting the $O(\epsilon^2)$ term since they are irrelevant in $(\ref{eq:IntBdDd5})$ we get 
	\begin{align}
	\int_{{B}_{\epsilon}}d^4x\,(R^2-\vec{x}^2){h}_{(4)\,ii}&=\frac{(R^2-\epsilon^2)^{\frac{5}{2}}}{2}\int_{S^3}d\Omega_{3}\,(h_{(2)\,ii}-\hat{x}^i\hat{x}^jh_{(2)\,ij})\label{eq:Inh4h2}\\
	&-\frac{3}{2}\int_{{B}_{\epsilon}}d^4x\,h_{(2)\,ii}\nonumber
    \end{align}	
Following the steps in $(\ref{eq:Inth2d4})$ we can evaluate the volume integral of $h_{(2)\,ii}$,
	\begin{align}
	\int_{{B}_{\epsilon}}d^4x\, h_{(2)\,ii}&=\frac{(R^2-\epsilon^2)^{\frac{3}{2}}}{6}\int_{S^3}d\Omega_{3}\hat{x}^i(\partial_ih_{(0)\,jj}-\partial_jh_{(0)\,ij})\\
	&=\frac{r^3}{6}\int_{S^3}d\Omega_{3}\hat{x}^i\bigg(\partial_ih_{(0)\,jj}-\hat{x}^j\hat{x}^k\partial_kh_{(0)\,ij}\nonumber\\
	&-\frac{1}{r^2}\partial_{\theta_1}h_{(0)\,\theta_1 i}-\frac{1}{r^2\sin^2\theta_1}\partial_{\theta_2}h_{(0)\,\theta_2 i}-\frac{1}{r^2\sin^2\theta_1\sin^2\theta_2}\partial_{\phi}h_{(0)\,\phi i}\nonumber\\
	&-\frac{2\cos\theta_1}{r^2\sin\theta_1}h_{(0)\,\theta_1 i}-\frac{\cos\theta_2}{r^2\sin^2\theta_1\sin\theta_2}h_{(0)\,\theta_2 i}-\frac{3}{r}h_{(0)\,ri}\bigg)\nonumber\\
	&=\frac{r^3}{6}\int_{S^3}d\phi d\theta_1d\theta_2\sin^2\theta_1\sin\theta_2 \bigg(\hat{x}^i\partial_ih_{(0)\,jj}-\hat{x}^i\hat{x}^j\hat{x}^k\partial_kh_{(0)\,ij}\nonumber\\
	&+\frac{1}{r^3}h_{(0)\,\theta_1 \theta_1}+\frac{1}{r^3\sin^2\theta_1}h_{(0)\,\theta_2 \theta_2}+\frac{1}{r^3\sin^2\theta_1\sin^2\theta_2}h_{(0)\,\phi \phi}-\frac{3\hat{x}^i\hat{x}^j}{r}h_{(0)\,ij}\bigg)\nonumber
\end{align}
Finally transforming into Cartesian coordinate we get
\begin{align}
	\int_{{B}_{\epsilon}}d^4x\, h_{(2)\,ii}=\frac{r^3}{6}\int_{S^3}d\Omega_{3}&\bigg[h_{(0)\,ii}-4\hat{x}^i\hat{x}^jh_{(0)\,ij}\label{eq:Inh2h0}\\
	&+x^j\partial_jh_{(0)\,ii}-\hat{x}^i\hat{x}^jx^k\partial_kh_{(0)\,ij}\bigg].\nonumber
\end{align}

\section{Covariant Phase Space Hamiltonian}\label{section:CPSH}
In this section we follow the formalism in \cite{2020HarlowWu} but here we consider the renormalized action, as well as different conditions on the vector. The variational problem of a Lagrangian theory with bulk and boundary terms requires the variation of both the bulk and boundary terms to be zero onshell. Therefore the sum of the presymplectic potential and the variation of the boundary terms should be exact on the boundary of the manifold
\begin{align}
	\boldsymbol{\Theta}[\delta \phi]-\delta\boldsymbol{B}=d\boldsymbol{C}[\delta \phi].
\end{align}
The presymplectic current can be expressed as 
\begin{align}
 \boldsymbol{\omega}[\delta_1\phi,\delta_2\phi]=\delta_1\left(\boldsymbol{\Theta}[\delta_2\phi]-d\boldsymbol{C}[\delta_2\phi]\right)
\end{align}
where $\delta$ is the exterior derivative on the configuration space. In Einstein gravity with cosmological constant and Gibbons-Hawking boundary term, without imposing any boundary condition, we get
\begin{align}
	\boldsymbol{\Theta}[\delta g]-\delta\boldsymbol{B}^{GH}=d\boldsymbol{C}^{GH}[\delta g]+\boldsymbol{\pi}\cdot\delta g.
\end{align}
The exact contribution $\boldsymbol{C}^{GH}[\delta g]$ captures the variation of the metric in the normal direction and the canonical momentum term captures the usual variation of the induced metric. Hence one can eliminate this term by imposing a radial gauge condition. However, as we will see later the variation of $\boldsymbol{C}^{GH}$ will have a non zero contribution. On $B_{\epsilon}$ we get
\begin{align}
	\boldsymbol{C}^{GH}[\delta g]&=-\frac{\boldsymbol{\varepsilon}_{\mu\nu}}{16\pi G_N}\gamma^{\nu\sigma}n^{\mu}n^{\rho}\delta g_{\sigma \rho}\label{eq:CGHdg}\\
	&=-\frac{\boldsymbol{\varepsilon}_{zt}}{16\pi G_N}\delta g^{tz}.
\end{align}
The variation of the Hamiltonian along the vector field $\xi$ can be constructed from the presymplectic form $\tilde{\Omega}$
\begin{align}
	\delta H[\xi]=-\iota_{X_{\xi}} \tilde{\Omega}
\end{align}
where $X_{\xi}$ is the configuration space vector that takes the one form in configuration space to the Lie derivative in configuration space
\begin{align}
 X_{\xi}(\delta\phi)=\mathcal{L}_{X_{\xi}}\phi.
\end{align}
The Lie derivative in configuration space only varies the dynamical fields along $\xi$ direction and the Lie derivative in spacetime varies both the dynamical fields and background fields along the $\xi$ direction. Any tensor is called covariant under the diffeomorphism induced by $\xi$ if the two Lie derivatives coincide
\begin{align}
	\mathcal{L}_{X_{\xi}}T=\mathcal{L}_{{\xi}}T.
\end{align}  
In general, the normal is constructed from a background function, 
\begin{align}
	n\propto df,
\end{align}
such that the level sets of the function define  a foliation. Anything that distinguishes the normal direction from other directions is not covariant unless we impose an extra condition on $\xi$,
\begin{align}
	\mathcal{L}_{\xi}f=\xi(f)=0
\end{align}
which implies the normal direction of $\xi$ vanishes. We label the difference between the two Lie derivatives of $\boldsymbol{C}$ along generic $\xi$ by
\begin{align}
		\boldsymbol{D}[\xi]=\mathcal{L}_{{\xi}}\boldsymbol{C}-\mathcal{L}_{X_{\xi}}\boldsymbol{C}.\label{eq:D2C}
\end{align}
Since the presymplectic form is given by the integral of the presymplectic form, $\omega$, on the Cauchy surface $\mathcal{C}$, we can express the variation of the Hamiltonian as
\begin{align}
		\delta H[\xi]=\int_{\mathcal{C}}-\iota_{X_{\xi}} \boldsymbol{\omega}.
\end{align}
Through some algebra in a generic theory onshell we get
\begin{align}
		-X_{\xi}\cdot \boldsymbol{\omega}=d\left(\delta \boldsymbol{Q}[\xi]-\iota_{\xi}\delta\boldsymbol{B}-\delta\iota_{X_{\xi}}\boldsymbol{C}-\iota_{\xi}\boldsymbol{\pi}\cdot\delta\phi-\boldsymbol{D}[\xi]+d\iota_{\xi}\boldsymbol{C}\right).
\end{align}
Let us define the Hamiltonian potential as the density over $\partial\mathcal{C}$ so
\begin{align}
	\delta \boldsymbol{H}[\xi]=\delta \boldsymbol{Q}[\xi]-\iota_{\xi}\delta\boldsymbol{B}-\delta\iota_{X_{\xi}}\boldsymbol{C}-\iota_{\xi}\boldsymbol{\pi}\cdot\delta\phi-\boldsymbol{D}[\xi]+d\iota_{\xi}\boldsymbol{C}.
\end{align}
We can see that the Hamiltonian potential has an exact term ambiguity because the Hamiltonian is defined to be the integral of the Hamiltonian form over a manifold with no boundary. We will now show that the full Noether charge form is a well defined Hamiltonian potential of the renormalized action. Since we have found that the holographic charge form is equal to the full Noether charge form up to an exact term, the Hamiltonian defined through holographic charge form is the full Noether charge. In the context of the first law of entanglement entropy, neither the entanglement entropy nor the modular energy is a Hamiltonian or a conserved charge, and hence the exact term difference matters. Here we will derive an expression for $\delta \boldsymbol{\Delta}[\xi_B]$ in terms of the quantities defined above. 
\\
\\
In \cite{2020HarlowWu}, the case of Einstein gravity with cosmological constant and Gibbons-Hawking boundary term was considered. The boundary condition imposed was
\begin{align}
	\boldsymbol{\pi}\cdot\delta g=0\label{eq:Pidg0}
\end{align}
and restricting normal direction of $\xi$ to be identically zero. Under these conditions, the variation of the Hamiltonian potential is
\begin{align}
	\delta\boldsymbol{H}^{BY}[\xi]&=\delta\left(\boldsymbol{\varepsilon}_{\mu\nu}n^{\mu}{T}^{\nu}_{\sigma}\xi^{\sigma}\right)\\
	&=\delta\left(\boldsymbol{\varepsilon}_{\mu\nu}n^{\mu}2{\pi}^{\nu}_{\sigma}\xi^{\sigma}\right)\\
	&=-\delta\boldsymbol{\mathcal{Q}}[\xi]
\end{align}
where $T_{\mu\nu}$ is the Brown York stress tensor given by
\begin{align}
	T^{\mu\nu}=-\frac{1}{8\pi G}\left(K^{\mu\nu}-\gamma^{\mu\nu}K\right).
\end{align} 
In our case, not only we do not impose the boundary condition $(\ref{eq:Pidg0})$, we also need to use the vector field $\xi_B$ which will introduce a term relating to the normal component of $\xi_B$. 

The Hamiltonian potential from Einstein gravity with Gibbons-Hawking boundary term is
\begin{align}
		\delta \boldsymbol{H}^{GH}[\xi_B]&=\delta \boldsymbol{Q}[\xi_B]-\iota_{\xi_B}\delta\boldsymbol{B}^{GH}-\delta\iota_{X_{\xi_B}}\boldsymbol{C}^{GH}-\iota_{\xi_B}\boldsymbol{\pi}\cdot\delta g-\boldsymbol{D}^{GH}[\xi_B]+d\iota_{\xi_B}\boldsymbol{C}^{GH}\nonumber\\
		&=\delta \boldsymbol{Q}[\xi_B]-\iota_{\xi_B}\delta\boldsymbol{B}^{GH}-\delta\iota_{X_{\xi_B}}\boldsymbol{C}^{GH}-\iota_{\xi_B}\boldsymbol{\pi}\cdot\delta g\nonumber\\
		&=\delta\boldsymbol{H}^{BY}[\xi_B]-\frac{\boldsymbol{\varepsilon}_{\mu\nu}n^{\mu}\tau^{\nu}}{16\pi G_N}\delta \gamma \tau^{\alpha}\partial_{\alpha}(\xi_B^{\beta}n_{\beta})-\iota_{\xi_B}\boldsymbol{\pi}\cdot\delta g\label{eq:dHGH}
\end{align}
where $\tau$ is the future pointing timelike normal vector. To get to the last line we also used the following properties for the Killing vector $\xi_B$ and in radial gauge,
\begin{align}
	\iota_{\xi_B}\boldsymbol{C}^{GH}=\boldsymbol{D}^{GH}[\xi_B]=0.
\end{align}
 When we consider the renormalized action there are additional counterterms in the full Hamiltonian potential
\begin{align}
	\delta \boldsymbol{H}^{full}[\xi]=\delta \boldsymbol{Q}[\xi]&-\iota_{\xi}\delta\boldsymbol{B}^{GH}-\delta\iota_{X_{\xi}}\boldsymbol{C}^{GH}-\iota_{\xi}\boldsymbol{\pi}\cdot\delta g-\boldsymbol{D}^{GH}[\xi]+d\iota_{\xi}\boldsymbol{C}^{GH}\label{eq:dHfull} \\
	&+\iota_{\xi}\delta\boldsymbol{B}^{ct}+\delta\iota_{X_{\xi}}\boldsymbol{C}^{ct}+\iota_{\xi}\boldsymbol{\pi}^{ct}\cdot\delta g+\boldsymbol{D}^{ct}[\xi]-d\iota_{\xi}\boldsymbol{C}^{ct}.\nonumber
\end{align}
Simplifying the above equation by gathering the boundary terms we get
\begin{align}
	\delta \boldsymbol{H}^{full}[\xi]=\delta \boldsymbol{Q}[\xi]+\delta \boldsymbol{b}^{GH}[\xi]-\delta \boldsymbol{b}^{ct}[\xi].\label{eq:Hfull2Q}
\end{align}
Hence the Gibbon-Hawking Hamiltonian potential is related to the full Hamiltonian potential by
\begin{align}
		\delta\boldsymbol{H}^{GH}[\xi]=\delta \boldsymbol{H}^{full}[\xi]+\delta\boldsymbol{b}^{ct}[\xi]. \label{eq:HGH2Hfull}
\end{align}
From $(\ref{eq:CGHdg})$  and $(\ref{eq:D2C})$ we can deduce 
\begin{align}
	\boldsymbol{C}^{GH}=\boldsymbol{C}^{ct},&&\boldsymbol{D}^{GH}=\boldsymbol{D}^{ct}
\end{align}
then we have
\begin{align}
	\delta \boldsymbol{H}^{full}[\xi]=\delta \boldsymbol{Q}^{full}[\xi]-\iota_{\xi}\boldsymbol{\pi}_{(d)}\cdot\delta g.
\end{align}
The full Hamiltonian potential is equal to the full Noether charge when the last term is zero. For a conformal Killing vector we can apply the tracelessness condition on $\boldsymbol{\pi}^{\mu\nu}_{(d)}$. In our case, the unperturbed $\boldsymbol{\pi}^{\mu\nu}_{(d)}$ is zero by itself, so we can relax all boundary condition on $\delta g_{\mu\nu}$. 

By inspecting the dilatation eigenvalue expansion of $(\ref{eq:dHGH})$, the renormalized Brown-York Hamiltonian potential $\delta\boldsymbol{H}^{BY}_{(d)}[\xi_B]$ can be expressed in terms of $\delta \boldsymbol{H}^{GH}[\xi_B]$ and its counterterm,
\begin{align}
\delta\boldsymbol{H}^{BY}_{(d)}[\xi_B]-\iota_{\xi_B}\boldsymbol{\pi}_{(d)}\cdot\delta g= \delta\boldsymbol{H}^{GH}[\xi_B]- \delta\boldsymbol{H}^{GH}_{ct}[\xi_B].
\end{align}
In our setting $\boldsymbol{\pi}^{\mu\nu}_{(d)}=0$ so the renormalized Brown-York Hamiltonian potential is obtained by subtracting the lower order terms in the dilatation eigenvalue expansion of the Gibbons-Hawking Hamiltonian potential. This should be distinguished from the full Hamiltonian that is constructed form the renormalized Lagrangian or action. These two procedures of obtaining the Hamiltonian are equivalent if the difference between Hamiltonian potentials is exact. We will see in the following how the two renormalization procedures differ in the context of entanglement entropy and modular energy. 

First we use $(\ref{eq:dHGH})$ and $(\ref{eq:dHfull})$, to relate the two Hamiltonian potentials, $\delta\boldsymbol{H}^{BY}[\xi_B]$ and $\delta \boldsymbol{H}^{full}[\xi_B]$, by
\begin{align}
	\delta\boldsymbol{H}^{BY}[\xi_B]=\delta \boldsymbol{H}^{full}[\xi_B]-\iota_{\xi_B}\delta\boldsymbol{B}^{ct}-\delta\iota_{X_{\xi_B}}\boldsymbol{C}^{ct}+\frac{\boldsymbol{\varepsilon}_{\mu\nu}n^{\mu}\tau^{\nu}}{16\pi G_N}\delta \gamma \tau^{\alpha}\partial_{\alpha}(\xi_B^{\beta}n_{\beta})
\end{align}
The difference between the two Hamiltonian potentials is not exact, and this implies $	\delta\boldsymbol{H}^{BY}[\xi_B]$ is not a proper Hamiltonian potential that integrates to give the Hamiltonian induced by $\xi_B$. 
However, we shall see that the renormalized Brown-York Hamiltonian potential or the holographic charge form is an appropriate Hamiltonian potential. Let us first express it in terms of the full Hamiltonian potential and all the counterterms,
\begin{align}
		\delta\boldsymbol{H}^{BY}_{(d)}[\xi_B]&=\delta \boldsymbol{H}^{full}[\xi_B]-\iota_{\xi_B}\delta\boldsymbol{B}^{ct}-\delta\iota_{X_{\xi_B}}\boldsymbol{C}^{ct}-\iota_{\xi_B}\boldsymbol{\pi}_{ct}\cdot\delta g-\delta\boldsymbol{H}^{GH}_{ct}[\xi_B]\\
		\delta\boldsymbol{H}^{BY}_{(d)}[\xi_B]&=\delta \boldsymbol{H}^{full}[\xi_B]-	\delta\boldsymbol{\Delta}[\xi_B].
\end{align}
The difference in the Hamiltonian potentials is non zero in general.

We can express the difference in Hamiltonian potentials as
\begin{align}
	\delta\boldsymbol{\Delta}[\xi_B]	&=\delta\boldsymbol{H}^{GH}_{ct}[\xi_B]+\iota_{\xi_B}\delta\boldsymbol{B}^{ct}+\delta\iota_{X_{\xi_B}}\boldsymbol{C}^{ct}+\iota_{\xi_B}\boldsymbol{\pi}_{ct}\cdot\delta g\\
	&=\delta\boldsymbol{H}^{GH}_{ct}[\xi_B]-\delta\boldsymbol{b}^{ct}[\xi_B].
\end{align}
Hence, the physical interpretation of $\delta\boldsymbol{\Delta}$ is the difference of counterterms in the two renormalization procedure where $\delta\boldsymbol{b}^{ct}$ is the counterterms contribution of the Hamiltonian potential derived from the renormalized action and $\delta\boldsymbol{H}^{GH}_{ct}$ is the counterterm of the Hamiltonian potential derived from the bare action. More explicitly the we have the expression that matches with $(\ref{eq:dDelta})$,
\begin{align}
	\delta\boldsymbol{\Delta}[\xi_B]=-\frac{\boldsymbol{\varepsilon}_{\mu\nu}n^{\mu}\tau^{\nu}}{16\pi G_N}\delta \gamma \tau^{\alpha}\partial_{\alpha}(\xi_B^{\beta}n_{\beta})+\delta\iota_{X_{\xi_B}}\boldsymbol{C}^{ct}+\iota_{\xi_B}\delta\boldsymbol{B}^{ct}+\delta\left({\boldsymbol{\varepsilon}_{\mu\nu}n^{\mu}}{2\pi}^{\nu}_{ct\,\sigma}\xi^{\sigma}\right),\label{eq:dDeltaexp}
\end{align}
with 
\begin{align}
	\delta\iota_{X_{\xi_B}}\boldsymbol{C}^{ct}=\frac{z\boldsymbol{\varepsilon}_{zt}}{16\pi G_N}\delta g^t_t\partial_z\xi_B^t.
\end{align}
Let us now dissect $(\ref{eq:dDeltaexp})$ term by term. 

The first term captures the non-covariant variation of the normal direction. In \cite{2020HarlowWu} this term is absent as they restrict the diffeomorphism generator to preserve covariance of the normal. In \cite{Papadimitriou:2005ii}, this term is absent as a stronger fall-off condition is imposed. 

The second term captures the variation of the diffeomorphism of the metric in the normal direction. This term is non-vanishing because $\xi_B$ is no longer Killing in the perturbed metric. Hence we do not see the equivalent of this term in the unperturbed $\boldsymbol{\Delta}[\xi_B]$ from $(\ref{eq:Deltaex})$. The last two terms are the standard counterterm contributions from the full Hamiltonian potential and Brown-York Hamiltonian potential. The non trivial result we found is that this difference is exact, the exterior derivative of the density of the entanglement entropy counterterms, 
\begin{align}
	\delta\boldsymbol{\Delta}[\xi_B]=d\delta\boldsymbol{S}^{ct}_B.
\end{align}
Then the Hamiltonian defined by the renormalized Brown-York Hamiltonian potential is the same as the full Hamiltonian potential,
\begin{align}
	\delta \mathcal{H}[\xi_B]&=\int_{\partial\mathcal{C}}	\delta\boldsymbol{H}^{BY}_{(d)}[\xi_B]\\
	&=\int_{\partial\mathcal{C}}	\delta \boldsymbol{H}^{full}[\xi_B]-d\delta\boldsymbol{S}^{ct}_B\nonumber\\
		&=\int_{\partial\mathcal{C}}	\delta \boldsymbol{H}^{full}[\xi_B]\nonumber\\
		\delta \mathcal{H}[\xi_B]&=\delta H^{full}[\xi_B].
\end{align}
For entanglement entropy and modular energy this difference matters because the integral is over a manifold with boundary that turns the exact term into the appropriate counterterm for the entanglement entropy. This analysis establishes the first law of renormalized entanglement entropy.

\bibliography{refs}

\providecommand{\href}[2]{#2}\begingroup\raggedright\begin{thebibliography}{10}

\bibitem{Faulkner:2013ica}
T.~Faulkner, M.~Guica, T.~Hartman, R.~C. Myers and M.~Van~Raamsdonk,
  \emph{{Gravitation from Entanglement in Holographic CFTs}},
  \href{http://dx.doi.org/10.1007/JHEP03(2014)051}{\emph{JHEP} {\bf 03} (2014)
  051}, [\href{https://arxiv.org/abs/1312.7856}{{\tt 1312.7856}}].

\bibitem{deHaro2001}
S.~de~Haro, K.~Skenderis and S.~N. Solodukhin, \emph{Holographic reconstruction
  of spacetime{\textparagraph}and renormalization in the ads/cft
  correspondence},
  \href{http://dx.doi.org/10.1007/s002200100381}{\emph{Communications in
  Mathematical Physics} {\bf 217} (Mar, 2001) 595--622}.

\bibitem{Taylor:2016aoi}
M.~Taylor and W.~Woodhead, \emph{{Renormalized entanglement entropy}},
  \href{http://dx.doi.org/10.1007/JHEP08(2016)165}{\emph{JHEP} {\bf 08} (2016)
  165}, [\href{https://arxiv.org/abs/1604.06808}{{\tt 1604.06808}}].

\bibitem{Taylor:2017zzo}
M.~Taylor and W.~Woodhead, \emph{{Non-conformal entanglement entropy}},
  \href{http://dx.doi.org/10.1007/JHEP01(2018)004}{\emph{JHEP} {\bf 01} (2018)
  004}, [\href{https://arxiv.org/abs/1704.08269}{{\tt 1704.08269}}].

\bibitem{Anastasiou:2018mfk}
G.~Anastasiou, I.~J. Araya, C.~Arias and R.~Olea, \emph{{Einstein-AdS action,
  renormalized volume/area and holographic Rényi entropies}},
  \href{http://dx.doi.org/10.1007/JHEP08(2018)136}{\emph{JHEP} {\bf 08} (2018)
  136}, [\href{https://arxiv.org/abs/1806.10708}{{\tt 1806.10708}}].

\bibitem{2018aao2}
G.~Anastasiou, I.~J. Araya, C.~Arias and R.~Olea, \emph{Einstein-ads action,
  renormalized volume/area and holographic rényi entropies},
  \href{http://dx.doi.org/10.1007/jhep08(2018)136}{\emph{Journal of High Energy
  Physics} {\bf 2018} (Aug, 2018) }.

\bibitem{Anastasiou:2019ldc}
G.~Anastasiou, I.~J. Araya, A.~Guijosa and R.~Olea, \emph{{Renormalized AdS
  gravity and holographic entanglement entropy of even-dimensional CFTs}},
  \href{http://dx.doi.org/10.1007/JHEP10(2019)221}{\emph{JHEP} {\bf 10} (2019)
  221}, [\href{https://arxiv.org/abs/1908.11447}{{\tt 1908.11447}}].

\bibitem{Anastasiou:2020smm}
G.~Anastasiou, J.~Moreno, R.~Olea and D.~Rivera-Betancour, \emph{{Shape
  dependence of renormalized holographic entanglement entropy}},
  \href{https://arxiv.org/abs/2002.06111}{{\tt 2002.06111}}.

\bibitem{Ryu:2006bv}
S.~Ryu and T.~Takayanagi, \emph{{Holographic derivation of entanglement entropy
  from AdS/CFT}},
  \href{http://dx.doi.org/10.1103/PhysRevLett.96.181602}{\emph{Phys. Rev.
  Lett.} {\bf 96} (2006) 181602},
  [\href{https://arxiv.org/abs/hep-th/0603001}{{\tt hep-th/0603001}}].

\bibitem{Papadimitriou:2004ap}
I.~Papadimitriou and K.~Skenderis, \emph{{AdS / CFT correspondence and
  geometry}}, \href{http://dx.doi.org/10.4171/013-1/4}{\emph{IRMA Lect. Math.
  Theor. Phys.} {\bf 8} (2005) 73--101},
  [\href{https://arxiv.org/abs/hep-th/0404176}{{\tt hep-th/0404176}}].

\bibitem{Papadimitriou:2005ii}
I.~Papadimitriou and K.~Skenderis, \emph{{Thermodynamics of asymptotically
  locally AdS spacetimes}},
  \href{http://dx.doi.org/10.1088/1126-6708/2005/08/004}{\emph{JHEP} {\bf 08}
  (2005) 004}, [\href{https://arxiv.org/abs/hep-th/0505190}{{\tt
  hep-th/0505190}}].

\bibitem{2020HarlowWu}
D.~Harlow and J.-q. Wu, \emph{Covariant phase space with boundaries},
  \href{http://dx.doi.org/10.1007/jhep10(2020)146}{\emph{Journal of High Energy
  Physics} {\bf 2020} (Oct, 2020) }.

\bibitem{Fursaev_1995}
D.~V. Fursaev and S.~N. Solodukhin, \emph{Description of the riemannian
  geometry in the presence of conical defects},
  \href{http://dx.doi.org/10.1103/physrevd.52.2133}{\emph{Physical Review D}
  {\bf 52} (Aug, 1995) 2133–2143}.

\bibitem{Fursaev_2013}
D.~V. Fursaev, A.~Patrushev and S.~N. Solodukhin, \emph{Distributional geometry
  of squashed cones},
  \href{http://dx.doi.org/10.1103/physrevd.88.044054}{\emph{Physical Review D}
  {\bf 88} (Aug, 2013) }.

\bibitem{Taylor:2020uwf}
M.~Taylor and L.~Too, \emph{{Renormalized entanglement entropy and curvature
  invariants}}, \href{http://dx.doi.org/10.1007/JHEP12(2020)050}{\emph{JHEP}
  {\bf 12} (2020) 050}, [\href{https://arxiv.org/abs/2004.09568}{{\tt
  2004.09568}}].

\bibitem{Lee:1990nz}
J.~Lee and R.~M. Wald, \emph{{Local symmetries and constraints}},
  \href{http://dx.doi.org/10.1063/1.528801}{\emph{J. Math. Phys.} {\bf 31}
  (1990) 725--743}.

\bibitem{Wald:1993nt}
R.~M. Wald, \emph{{Black hole entropy is the Noether charge}},
  \href{http://dx.doi.org/10.1103/PhysRevD.48.R3427}{\emph{Phys. Rev.} {\bf
  D48} (1993) R3427--R3431}, [\href{https://arxiv.org/abs/gr-qc/9307038}{{\tt
  gr-qc/9307038}}].

\bibitem{Iyer:1994ys}
V.~Iyer and R.~M. Wald, \emph{{Some properties of Noether charge and a proposal
  for dynamical black hole entropy}},
  \href{http://dx.doi.org/10.1103/PhysRevD.50.846}{\emph{Phys. Rev.} {\bf D50}
  (1994) 846--864}, [\href{https://arxiv.org/abs/gr-qc/9403028}{{\tt
  gr-qc/9403028}}].

\bibitem{Iyer:1995kg}
V.~Iyer and R.~M. Wald, \emph{Comparison of the noether charge and euclidean
  methods for computing the entropy of stationary black holes},
  \href{http://dx.doi.org/10.1103/physrevd.52.4430}{\emph{Physical Review D}
  {\bf 52} (Oct, 1995) 4430–4439}.

\bibitem{Wald:1999wa}
R.~M. Wald and A.~Zoupas, \emph{General definition of “conserved
  quantities” in general relativity and other theories of gravity},
  \href{http://dx.doi.org/10.1103/physrevd.61.084027}{\emph{Physical Review D}
  {\bf 61} (Mar, 2000) }.

\bibitem{Casini:2011kv}
H.~Casini, M.~Huerta and R.~C. Myers, \emph{{Towards a derivation of
  holographic entanglement entropy}},
  \href{http://dx.doi.org/10.1007/JHEP05(2011)036}{\emph{JHEP} {\bf 05} (2011)
  036}, [\href{https://arxiv.org/abs/1102.0440}{{\tt 1102.0440}}].

\bibitem{Klebanov:2012yf}
I.~R. Klebanov, T.~Nishioka, S.~S. Pufu and B.~R. Safdi, \emph{{On Shape
  Dependence and RG Flow of Entanglement Entropy}},
  \href{http://dx.doi.org/10.1007/JHEP07(2012)001}{\emph{JHEP} {\bf 07} (2012)
  001}, [\href{https://arxiv.org/abs/1204.4160}{{\tt 1204.4160}}].

\bibitem{Allais:2014ata}
A.~Allais and M.~Mezei, \emph{{Some results on the shape dependence of
  entanglement and R\'enyi entropies}},
  \href{http://dx.doi.org/10.1103/PhysRevD.91.046002}{\emph{Phys. Rev. D} {\bf
  91} (2015) 046002}, [\href{https://arxiv.org/abs/1407.7249}{{\tt
  1407.7249}}].

\bibitem{Rosenhaus:2014zza}
V.~Rosenhaus and M.~Smolkin, \emph{{Entanglement Entropy for Relevant and
  Geometric Perturbations}},
  \href{http://dx.doi.org/10.1007/JHEP02(2015)015}{\emph{JHEP} {\bf 02} (2015)
  015}, [\href{https://arxiv.org/abs/1410.6530}{{\tt 1410.6530}}].

\bibitem{Mezei:2014zla}
M.~Mezei, \emph{{Entanglement entropy across a deformed sphere}},
  \href{http://dx.doi.org/10.1103/PhysRevD.91.045038}{\emph{Phys. Rev. D} {\bf
  91} (2015) 045038}, [\href{https://arxiv.org/abs/1411.7011}{{\tt
  1411.7011}}].

\bibitem{2014jc}
J.~Camps, \emph{Generalized entropy and higher derivative gravity},
  \href{http://dx.doi.org/10.1007/jhep03(2014)070}{\emph{Journal of High Energy
  Physics} {\bf 2014} (Mar, 2014) }.

\bibitem{2014xd}
X.~Dong, \emph{Holographic entanglement entropy for general higher derivative
  gravity}, \href{http://dx.doi.org/10.1007/jhep01(2014)044}{\emph{Journal of
  High Energy Physics} {\bf 2014} (Jan, 2014) }.

\bibitem{Lashkari:2013koa}
N.~Lashkari, M.~B. McDermott and M.~Van~Raamsdonk, \emph{{Gravitational
  dynamics from entanglement 'thermodynamics'}},
  \href{http://dx.doi.org/10.1007/JHEP04(2014)195}{\emph{JHEP} {\bf 04} (2014)
  195}, [\href{https://arxiv.org/abs/1308.3716}{{\tt 1308.3716}}].

\bibitem{2015LinOoguriStocia}
J.~Lin, M.~Marcolli, H.~Ooguri and B.~Stoica, \emph{Locality of gravitational
  systems from entanglement of conformal field theories},
  \href{http://dx.doi.org/10.1103/physrevlett.114.221601}{\emph{Physical Review
  Letters} {\bf 114} (Jun, 2015) }.

\bibitem{Jang:2020cbm}
D.~Jang, Y.~Kim, O.-K. Kwon and D.~D. Tolla, \emph{{Renormalized Holographic
  Subregion Complexity under Relevant Perturbations}},
  \href{http://dx.doi.org/10.1007/JHEP07(2020)137}{\emph{JHEP} {\bf 07} (2020)
  137}, [\href{https://arxiv.org/abs/2001.10937}{{\tt 2001.10937}}].

\bibitem{Bernamonti2020}
A.~Bernamonti, F.~Galli, J.~Hernandez, R.~C. Myers, S.-M. Ruan and J.~Simón,
  \emph{Aspects of the first law of complexity},
  \href{http://dx.doi.org/10.1088/1751-8121/ab8e66}{\emph{Journal of Physics A:
  Mathematical and Theoretical} {\bf 53} (Jul, 2020) 294002}.

\bibitem{2017fw}
S.~Fischetti and T.~Wiseman, \emph{A bound on holographic entanglement entropy
  from inverse mean curvature flow},
  \href{http://dx.doi.org/10.1088/1361-6382/aa6ad0}{\emph{Classical and Quantum
  Gravity} {\bf 34} (May, 2017) 125005}.

\bibitem{maldacena2011einstein}
J.~Maldacena, \emph{Einstein gravity from conformal gravity},  2011.

\bibitem{2016ao}
G.~Anastasiou and R.~Olea, \emph{From conformal to einstein gravity},
  \href{http://dx.doi.org/10.1103/physrevd.94.086008}{\emph{Physical Review D}
  {\bf 94} (Oct, 2016) }.

\bibitem{Anderson:2001vr}
M.~T. Anderson, \emph{L${}^2$ curvature and volume renormalization of the ahe
  metrics on 4 manifolds},
  \href{http://dx.doi.org/10.4310/MRL.2001.v8.n2.a6}{\emph{Math. Res. Lett.}
  {\bf 8} (2001) 171}, [\href{https://arxiv.org/abs/math/0011051}{{\tt
  math/0011051}}].

\bibitem{Anastasiou_2021}
G.~Anastasiou, I.~J. Araya and R.~Olea, \emph{Einstein gravity from conformal
  gravity in 6d},
  \href{http://dx.doi.org/10.1007/jhep01(2021)134}{\emph{Journal of High Energy
  Physics} {\bf 2021} (Jan, 2021) }.

\end{thebibliography}\endgroup


\end{document}